# Wide band gap chalcogenide semiconductors

Rachel Woods-Robinson[1,2,3$], Yanbing Han[4,1$], Hanyu Zhang[1], Tursun Ablekim[1], Imran Khan[1], Kristin Persson[2,3], Andriy Zakutayev[1*]

1. Materials Science Center, National Renewable Energy Laboratory, Golden, CO 80401, USA
2. Department of Applied Science and Technology, University of California, Berkeley, CA 94720, USA
3. Materials Sciences Division, Lawrence Berkley National Laboratory, Berkeley, CA 94720, USA
4. School of Physics and Microelectronics, Zhengzhou University, Zhengzhou 450052, China

* Corresponding Author: andriy.zakutayev@nrel.gov
$ Rachel Woods-Robinson and Yanbing Han contributed equally to this review

# Abstract

Wide band gap semiconductors are essential for today's electronic devices and energy applications due to their high optical transparency, controllable carrier concentration, and tunable electrical conductivity. The most intensively investigated wide band gap semiconductors are transparent conductive oxides (TCOs) such as tin-doped indium oxide (ITO) and amorphous In-Ga-Zn-O (IGZO) used in displays and solar cells, carbides (e.g. SiC) and nitrides (e.g. GaN) used in power electronics, as well as emerging halides (e.g. γ-CuI) and 2D electronic materials (e.g. graphene) used in various optoelectronic devices. Compared to these prominent materials families, chalcogen-based (*Ch* = S, Se, Te) wide band gap semiconductors are less heavily investigated, but stand out due to their propensity for p-type doping, high mobilities, high valence band positions (i.e. low ionization potentials), and broad applications in electronic devices such as CdTe solar cells. This manuscript provides a review of wide band gap chalcogenide semiconductors. First, we outline general materials design parameters of high performing transparent semiconductors, as well as the theoretical and experimental underpinnings of the corresponding research methods. We proceed to summarize progress in wide band gap ($E_G$ > 2 eV) chalcogenide materials—namely II-VI *MCh* binaries, $CuMCh_2$ chalcopyrites, $Cu_3MCh_4$ sulvanites, mixed-anion layered Cu*MCh*(O,F), and 2D materials—and discuss computational predictions of potential new candidates in this family, highlighting their optical and electrical properties. We finally review applications of chalcogenide wide band gap semiconductors—e.g. photovoltaic and photoelectrochemical solar cells, transparent transistors, and light emitting diodes—that employ wide band gap chalcogenides as either an active or passive layer. By examining, categorizing, and discussing prospective directions in wide band gap chalcogenides, this review aims to inspire continued research on this emerging class of transparent semiconductors and thereby enable future innovations for optoelectronic devices.

# Table of Contents

# 1. Introduction

Wide band gap (WBG, or "wide-gap") semiconductors are critical to various electronic devices such as transparent contacts, p-n junctions, and thin film transistors.[1] Since the 1950s, oxide wide band gap semiconductors have been intensively investigated, in particular for their contradicting properties of high transparency and high conductivity. The transparent conducting oxide (TCO) Sn-doped $In_2O_3$, known as ITO, has been paramount to a variety of commercial devices in the past decades.[2] Possible alternatives such as F-doped $SnO_2$ (FTO)[3] and Al-doped ZnO (AZO)[4] have been explored in-depth and implemented into commercial devices. In the 21$^{st}$ century, multinary transparent amorphous oxide semiconductors (TAOS) such as indium zinc gallium oxide (IGZO) have also been heavily investigated as channel layers in thin film transistors (TFT) due to their high transparency, high mobility, and good uniformity,[5] leading to their commercial use in liquid crystal displays (LCDs). Although these n-type TCOs show excellent performance, p-type doping and high hole mobilities in oxides has proven much more difficult to achieve in practice. This is due primarily to (1) intrinsic limitations in dispersion that localize the valence band holes, and (2) challenges in introducing the holes by p-type doping and minimizing compensating defects. A breakthrough was achieved with the proposed strategy of "chemical modulation of the valence band" (CMVB) was proposed, using delafossite $CuAlO_2$ as a preliminary example.[6] This approach uses hybridization of O 2*p* states with metal 3*d* states at the valence band maximum (VBM), increasing dispersion. Subsequently, the use of this strategy and other approaches has led to a variety of Cu-based p-type TCOs, though optoelectronic properties of p-type TCOs still do not compare with n-type TCOs. Due to these challenges, only a narrow subset of p-type TCO materials have been incorporated into commercial device applications, e.g. $Cu_2O$ and SnO, which are mainly used in thin film transistors (TFTs).[7]

Despite the focus of research and device integration on wide-gap oxide materials, transparent semiconductors are not limited to oxides. To date (2020), there are several classes of non-oxide semiconductors that have been experimentally demonstrated to be transparent and conducting. Of interest to this review, the following group VI chalcogenide ("*Ch*") semiconductors have been investigated: (1) binary II-VI *MCh* semiconductors (e.g. ZnS,[8] CdS,[9] $Zn_xCd_{1-x}S$[10]), as well as other binary $M_xCh_y$ semiconductors (e.g. $SnS_2$, $In_2S_3$, where *M* = metal); (2) ternary chalcopyrite I-III-$Ch_2$ semiconductors (mostly Cu-based), represented by $CuAlS_2$,[11] and other ternaries (e.g. $\alpha$-$BaCu_2S_2$[12] and $Cu_3TaS_4$[13]); (3) multinary layered mixed-anion compounds, such as LaCuO*Ch*,[14] BaCuSF, and CuSCN;[15] and (4) 2D chalcogenides, such as $MoS_2$, including both binary and ternary materials. Additionally, significant research has been done on wide band gap semiconductors (also called WBG semiconductors or WBGSs) with anions from groups IV (e.g. graphene,[16] and carbon nanotubes,[17] SiC[18]), V (e.g. GaN,[19,20] zinc nitrides[21]), and VII (e.g. CuI[22]), which are not discussed here. A network diagram schematic of these material classes is depicted in **Figure 1**, with chalcogenides highlighted in red. Many review papers and book chapters address various aspects of this broad field of WBG semiconductors. Some examine oxide TCOs, focusing on intrinsic material properties,[23,24] while others highlight applications such as transparent electronics,[1] TFTs,[7] and photovoltaics.[25] Others summarize nitrides,[26] halides,[22] and carbides,[27] including graphene.[28] A few reviews have briefly mentioned wide-gap chalcogenide semiconductors, but limit their focus to oxides or a narrow subset of chalcogenides.[29] Thus, here we focus on these wide band gap chalcogenides, referring to chalcogen anions as "*Ch*" (with *Ch* = S, Se, Te) and not including O for the purposes of this article.

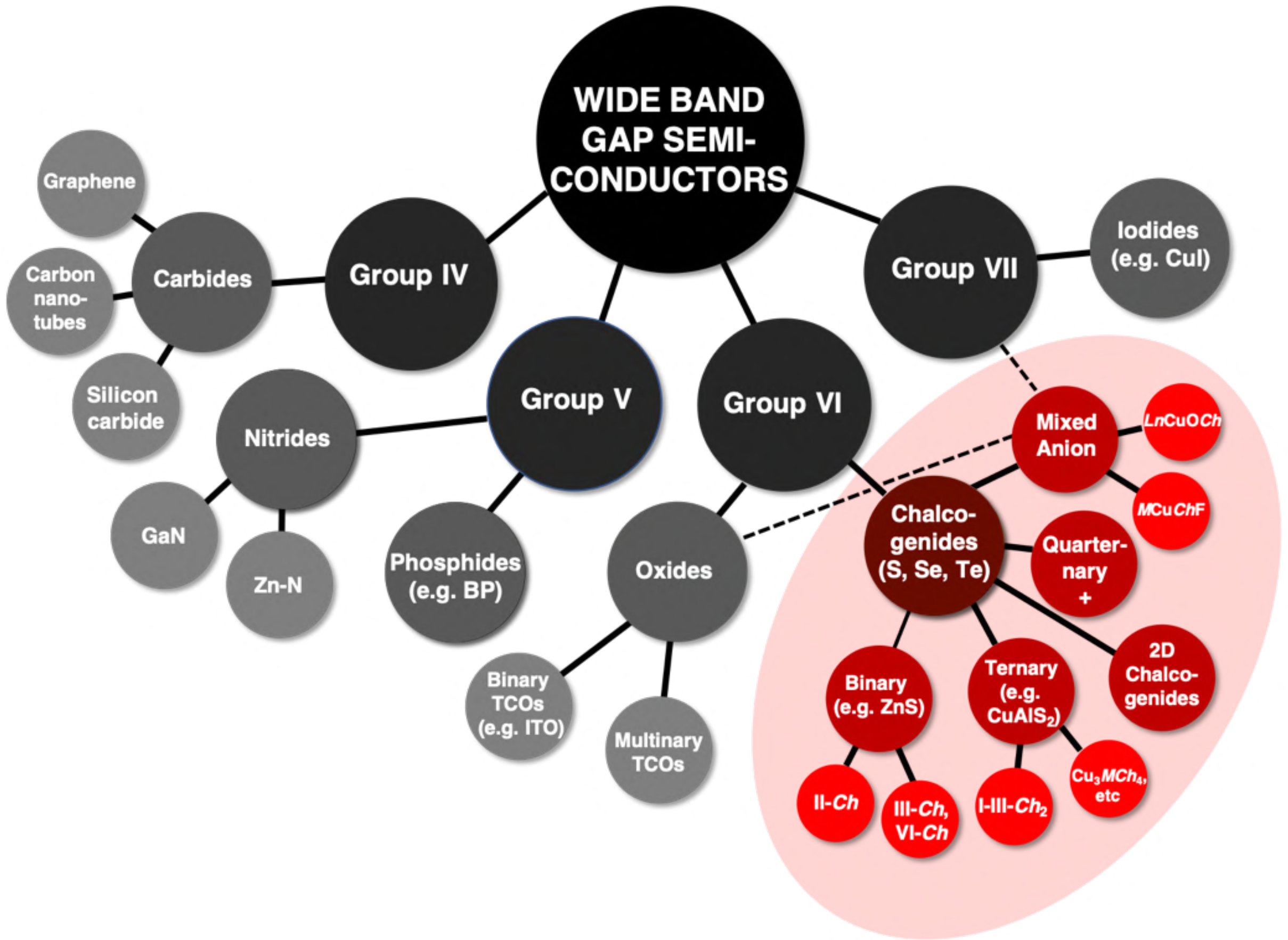

**Figure 1**. Schematic network diagram depicting various material classes of wide band gap semiconductors. Classes are sorted by anion group. Chalcogenide wide-gap semiconductors, which are the focus of this review article, are highlighted in red (with *Ch* = S, Se, Te).

As evident from the literature, wide band gap chalcogenide semiconductors present several distinctions in comparison to oxide counterparts, with particular advantages for optoelectronic applications and as p-type semiconductors. The covalency of a metal-VI (*M*-VI) bond tends to increase going down group VI, leading to larger orbital overlaps and lower hole effective masses. Thus, heavier chalcogenides have higher-lying *p* orbitals (S 3*p*, Se 4*p*, Te 5*p*) that can hybridize with *M* (e.g. Cu 3*d*) to form more disperse, delocalized valence bands than in oxides.[6] This can result in higher achievable mobilities in p-type chalcogenides, e.g. up to 20 $cm^2 V^{-1} s^{-1}$ for Cu-based compounds,[12,30,31] compared to <1 $cm^2 V^{-1} s^{-1}$ for Cu-based p-type wide-gap oxides (with some exceptions,[32] up to 10 $cm^2 V^{-1} s^{-1}$ in single crystal $CuAlO_2$). However, the band gap $E_G$ (and optical transparency window) tends to be smaller for chalcogenides than for oxides, though this is not always the case (e.g. ZnS has a wider gap than ZnO, see Section 3). There is also evidence that chalcogenides are easier to dope p-type than oxides due to higher-lying *p*-derived valence bands and smaller ionization energies, a postulated design principle for high p-type doping.[33] Additionally, weaker *M*-*Ch* bonds compared to *M*-O bonds in these materials allow for ease of synthesis, but may lead to increased degradation and stability concerns in chalcogenides. We note fewer stable chalcogenides exist in nature and in the experimental literature than oxides (see below), partially due to decomposition tendencies in air and water, but implies there are likely many unexplored chalcogenides. Considering these factors, it is of value to review wide-gap chalcogenide semiconductors.

To further demonstrate the need for this investigation, **Figure 2**a displays the extent to which the material space of wide band gap chalcogenides is underexplored by plotting the distribution of computed band gaps for all known thermodynamically stable oxides and chalcogenides in the Materials Project database.[34] Most of these materials correspond to experimentally synthesized compounds from the Inorganic Crystal Structure Database (ICSD).[35,36] We show all compounds with predicted gaps greater than 1 eV to represent wide-gap semiconductors, because density functional theory (DFT) tends to *underestimate* band gap by a factor of 1–2 with GGA and GGA+U functionals,[37,] unless U is empirically fit to recreate the experimental gap (see Section 2.2.1).[38] The resulting plot shows a similar statistical distribution, but for about 1.5 times more oxides (~15,456) than chalcogenides (~10,611), which are further subdivide into 3,367 sulfides, 2,330 selenides, and 1,767 tellurides (see **Figure 2b**). This can be in part attributed to the generally tendency of oxides to be more stable and earth abundant, which is why so many have been discovered in nature and/or realized synthetically. However recent research confirms synthesis of many new computationally predicted compounds unlikely to occur in nature, e.g. in the space of ternary nitrides,[39,40] suggesting possible realization of many new chalcogenides with chemical compositions and stoichiometries that have never been synthesized (and thus are missing from **Figure 2**). Moreover, many computationally "metastable" structures can be stabilized through techniques such as heterostructural alloying.[41] Thus, exciting and challenging research lies ahead to uncover new wide band gap chalcogenide materials with potential uses in energy applications and beyond.

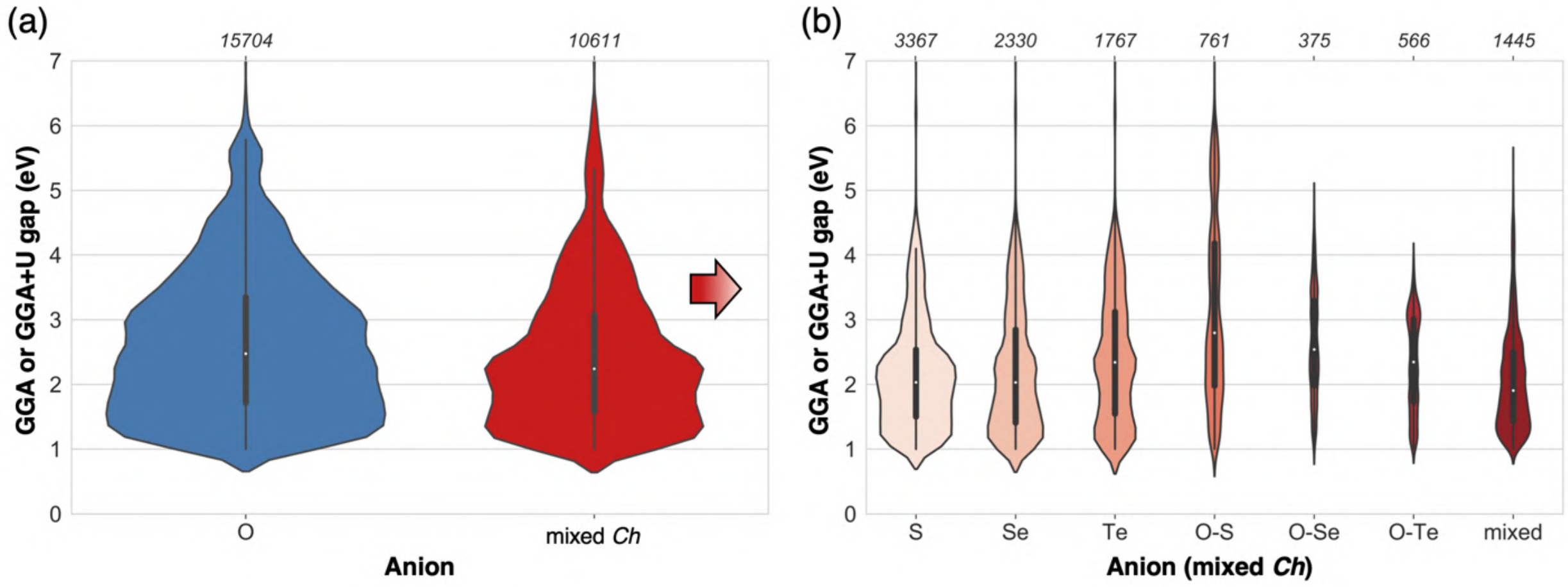

**Figure 2**. (a) Violin plot of the smoothed distribution of computed GGA or GGA+U band gaps within all known oxide materials (blue) and within all chalcogenide and mixed chalcogenides (red). (b) Chalcogenides and mixed chalcogenides are further categorized by anion type, with "mixed" referring to any mixed-anion chalcogenide. Data is only plotted for materials with band gaps greater than 1 eV. Note that the GGA+U functional significantly underestimates the band gap, so this is a reasonable cutoff for materials screening of wide-gap chalcogenides ($E_G > 2$ eV in this review). Violin plots are binned by count, and all data is from the open source Materials Project database.[34] We note that some of the mixed compounds listed may in fact have a chalcogenide *cation* rather than anion (e.g. sulfates).

In this review, following the introduction (Section 1), we highlight the functional materials properties of interest in wide band gap chalcogenide semiconductors (Section 2), and discuss computational and experimental methods to calculate, synthesize, characterize, and optimize such materials. Next, we discuss the prominent classes of chalcogenide semiconductors classified by cations, stoichiometry, and crystal structure (Section 3), as shown in red in **Figure 1**. We highlight materials that have been realized experimentally (Tables 1–3), and identify promising theoretical predictions (Table 4) to inform and inspire continued research in this field. Lastly, we summarize the most prominent applications of wide-gap chalcogenide semiconductors in electronic devices (Section 4), and discuss new potential device architectures. Since there is no clear boundary between transparent and non-transparent, we define "wide band gap semiconductors" for the purposes of this review as semiconductors whose band-to-band absorption edge appears larger than 2 eV (wavelengths less than 620 nm). Among wide band gap chalcogenide semiconductors, we focus on electronic n- or p-type dopable materials, which excludes highly insulating and amorphous chalcogenides, as well as energy storage and nanogenerator applications of chalcogenides from this review, referring the reader elsewhere.[42,43] Due to the multidisciplinary nature of these materials, this review is intended for a wide audience, from computational scientists to experimental materials scientists to engineers focusing on particular optoelectronic technologies. Thus, we have tailored each section to be either read independently or together as a comprehensive review.

# 2. Materials properties and research methods

The primary requirements of functional wide band gap chalcogenide semiconductors discussed in this text differ by application, but most share four basic criteria: (1) synthesizability and thermochemical stability, (2) a wide enough band gap to ensure transparency to light of a particular wavelength range, (3) high enough mobility to enable optoelectronic device integration, (4) n- or p-dopability to ensure a sufficient concentration of charge carriers. Here, we discuss both the theoretical and experimental considerations of each of these properties, which are pertinent to designing high performance wide-gap chalcogenide materials. We will focus on properties relevant to photovoltaics and transparent electronics, since they are desirable and rare, though we also mention properties important in other applications such as photocathodes for solar water splitting and light emitting diodes (LEDs) (see Section 4). We also provide evidence why chalcogenides in particular could be an important, underexplored chemical space to look for materials with these properties, and we highlight some unique challenges of chalcogenide materials.

Up until recently, material candidates were selected for synthesis based primarily on chemical intuition and theoretical approximations. Historically, it was easier to grow materials one-by-one and measure their conductivities and transparencies than to perform expensive *ab initio* computation to estimate such properties, so most experimental materials in Section 3 were developed in this manner. Recent advances in high-performance computing and efforts such as the Materials Genome Initiative (MGI) have driven development of computational platforms such as the Materials Project,[34] AFLOWLIB,[44] NRELMatDB,[45–47] and Open Quantum Materials Database,[48] that compile data from density functional theory (DFT) calculations and allow facile exploration and screening of material candidates for desired properties. This accessibility of data has fundamentally changed how experimental materials scientists select candidate materials for synthesis. In Section 3 we will discuss some key findings from *ab initio* computational screenings of wide-gap chalcogenides using the properties and descriptors discussed here.

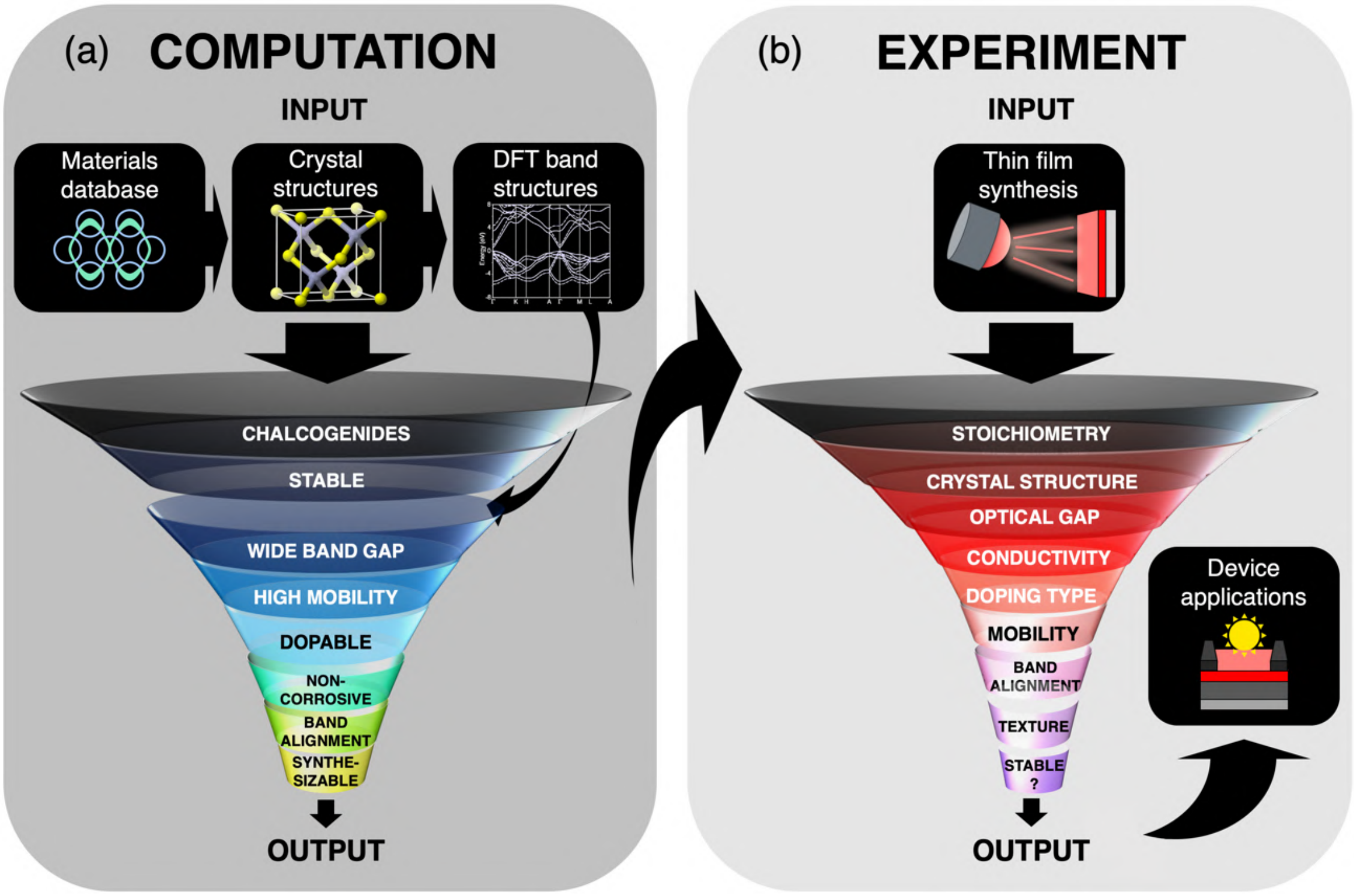

**Figure 3**. (a) Computational and (b) experimental screening funnels to search for desirable properties of wide-gap semiconductors within a computational materials database and using experimental methods. The computational figure is adapted with permission from Woods-Robinson et al.[49]

Computational materials screenings often make use of a "funnel" approach to filter out materials of particular interest, depicted by **Figure 3**a. Usually the more computationally "inexpensive" steps are performed first, and are followed by more in-depth steps that require more computer resources. In some studies "inputs" are drawn from materials property databases, while others calculate novel classes of compounds that have never been synthesized (e.g. via substitutions, random structure searches, or machine learning). As shown, a screening methodology typically requires (1) an initially data subset in a selected range of stoichiometries or structures (e.g. inorganic metal chalcogenides), (2) a proxy for thermodynamic stability, and then (3) a series of descriptors calculated from the electronic band structure or other methods to estimate a property of interest (e.g. effective mass to estimate mobility). Experimental materials discovery is illustrated by an analogous "experimental funnel" in Figure 3b. This process starts with synthesis of material candidates sometimes from the computational funnel (Figure 3a), and then a series of characterization experiments are performed to measure composition, structure, and various properties, with an end goal often to incorporate the characterized material into a device. We will discuss the "rungs" of the computational and experimental funnels next, some outputs in Section 3, and possible device applications in Section 4.

## 2.1 Synthesizability and stability

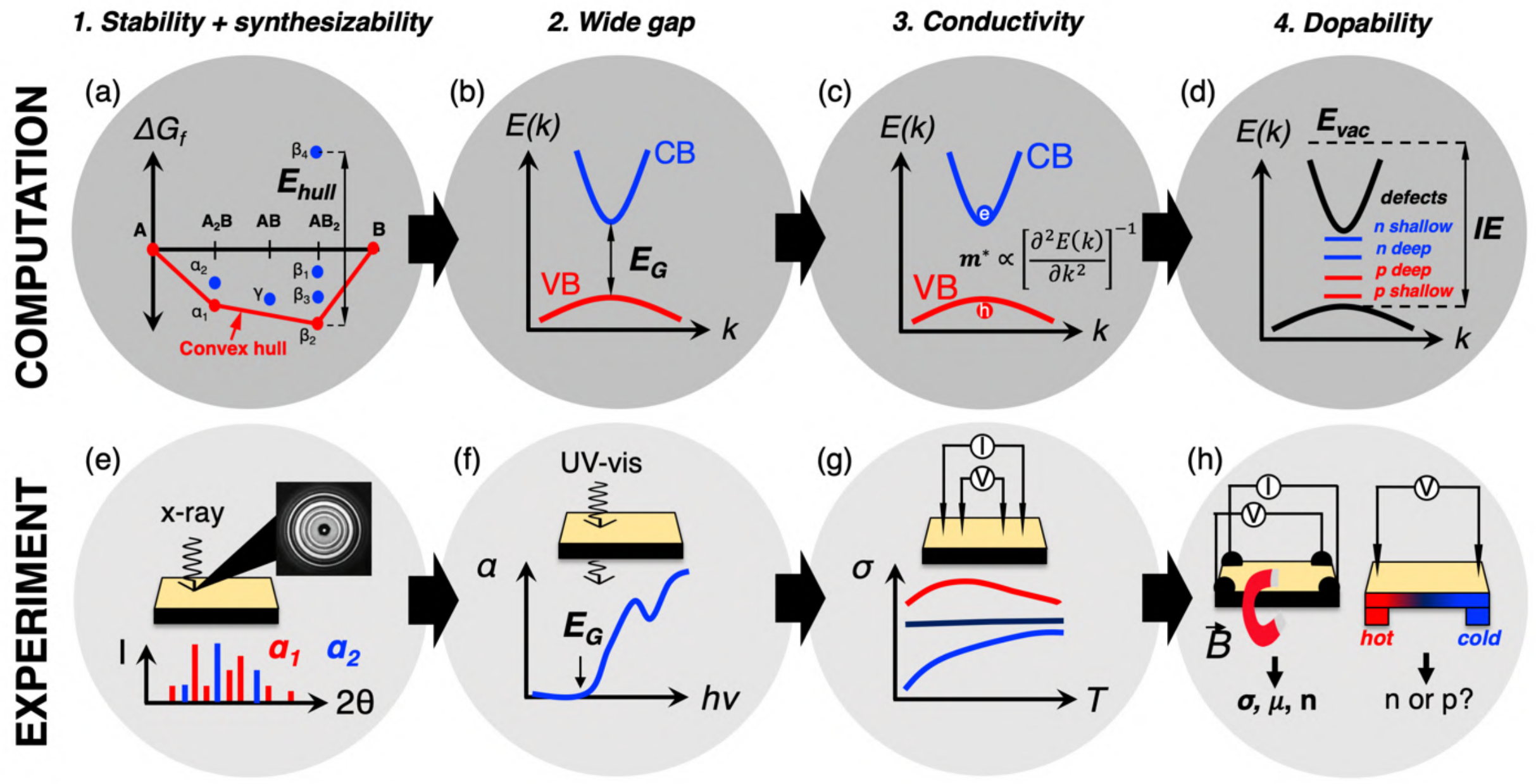

**Figure 4**. Schematic representations of the four computational descriptors highlighted in the text: (a) energy above the convex hull $E_{hull}$ as a descriptor for thermodynamic stability (b) band gap $E_G$ as a descriptor for transparency, (c) effective mass $m^*$ as a descriptor for mobility, (d) defect levels and ionization energy ($IE$) as descriptors for dopability. The corresponding experimental properties and tools are: (e) crystal structure and phase stability measured by x-ray diffraction, (f) absorption and optical band gap using UV-vis spectrophotometry, (g) conductivity $\sigma$ (temperature-dependent $\sigma$ shown here) using a four-point probe, and (h) quantification of mobility $\mu$, carrier concentration $n$, and carrier type using the Hall measurement probe (left) and a Seebeck thermocouple probe (right). Details are described in the text.

### *2.1.1 Thermodynamic stability calculations*

A wide-gap chalcogenide material must be both synthesizable and stable to be useful for device applications. A first-order computational descriptor for synthesizability is estimated by comparing formation energies of compounds within the same compositional space and by investigating their degradation pathways. Shown in **Figure 4**a, a first-order computational descriptor for synthesizability, "energy above the convex hull" ($E_{hull}$). For a given compound e.g. $AB_2$ polymorph $\beta_4$, $E_{hull}$ measures how far $\beta_4$'s formation energy ($\Delta G_f$) is from a convex hull described by the thermodynamic ground states within compositional space e.g. $A$-$B$. An $E_{hull}$ of 0 meV/atom indicates a ground state, and correlates strongly with a high probability of synthesizability. An $E_{hull}$ < 0.03 eV/atom is usually considered "stable" for oxides (within thermal energy $kT$), and "metastable" compounds tend to lie in the range $E_{hull}$ < 0.1 eV/atom,[49] though it is not established whether chalcogenides conform to these limits.[50] An upper $E_{hull}$ limit of synthesizability has recently been hypothesized by an amorphous "synthesizability skyline", and the skyline of chalcogenides was found to be significantly higher than that of oxides.[51] This suggests that thermodynamically metastable polymorphs may more achievable for chalcogenides than for oxides, which is a significant advantage for the number of possible materials that can be studied.

The $E_{hull}$ assessment can be applied early in a screening because it is usually computationally inexpensive. It does not require band structure calculations, only a DFT structure relaxation and cross comparison to other structures within a database. However, assessing stability and synthesizability by thermodynamics alone does not tell the full story. One issue chalcogenides face more than oxides is their stability in air and aqueous environments. Sulfides and selenides in general are more likely to oxidize and decompose under standard conditions than oxides, due to low energy reaction pathways with water or water vapor to form $H_2S$ or $H_2Se$, though this is not universally true and can be studied computationally for particular compounds of interest. For example, screenings for photoelectrochemical materials have used calculations of Pourbaix diagrams to assess the stability of a compound.[52,53] Additionally, p-type dopants must be stable within a host compound. For example, there have been several attempts of doping ZnO p-type with N and In/Ni, but these materials have later been shown to be unstable over time due to compensating donor defects.[54,55] It is also important to note that computationally promising "synthesizable" and "stable" materials still may not be possible to achieve in practice due to competing reactions, various decomposition pathways, degassing, or uncertainties in calculations, among other reasons.

### *2.1.2 Experimental synthesis and phase analysis*

One of the best ways to determine synthesizability and stability of a given material system is to *synthesize* it, but this is not necessarily straightforward. There is a huge variety of bulk and thin film growth techniques and numerous interconnected parameters to optimize in synthesis and post-processing, as covered in many texts.[56–60] Wide band gap semiconductors can be made in forms of bulk, thin film, 2D, or even quantum dots. Thin films are usually the most useful form for electronic device applications, and will be the focus in this review. Thin film transparent chalcogenides have been synthesized by (1) physical vapor deposition (PVD),[61] including sputtering,[62] pulsed laser deposition (PLD),[63] and thermal or electron-beam evaporation,[64,65] molecular beam epitaxy (MBE),[66,67] by (2) chemical vapor deposition, usually involving some reactions between precursors, such as atomic layer deposition (ALD),[68,69] metal organic chemical vapor deposition (MOCVD)[70] and plasma-enhanced chemical vapor deposition (PECVD),[71] or by (3) solution processes, including spray pyrolysis,[72] sol-gel,[73,74] and chemical bath deposition (CBD).[75] Post-deposition treatments can be applied to enhance crystallinity or introduce dopants, such as exposure to gas (e.g. sulfurization and selenization, common in chalcopyrite materials)[76] and rapid thermal annealing,[77] and films can be doped via ion implantation or diffusion. Synthesis methods for bulk semiconductors varies, spanning a wide range of material quality, including solid state reaction, spark plasma sintering[11], floating zone synthesis,[78] growth from melt techniques e.g. Czochralski and Bridgman growth,[79,80] etc. To achieve semiconducting materials in layer-controlled 2D forms, the two major directions are top-down methods, i.e. exfoliating materials from their bulk counterparts, and bottom-up methods, i.e. direct synthesis of 2D materials from constituent elements or precursors (see Section 3.6 for examples).[81] After synthesis, stoichiometry and elemental uniformity can be determined by x-ray fluorescence (XRF), Rutherford backscattering spectroscopy (RBS), x-ray photoelectron spectroscopy (XPS, limited to surface stoichiometry), energy dispersive x-ray spectroscopy (EDS or EDX), among other techniques. Sometimes off-stoichiometric synthesis is useful for achieving targeting properties and high conductivity.

Achieving the right stoichiometry of chemical elements does not guarantee that a crystal phase of interest has been made. Many of the low temperature growth techniques result in amorphous or nanocrystalline material, which has to be treated for crystallization. X-ray diffraction (XRD) techniques are often used to confirm whether a targeted crystal phase has been synthesized.[82] Measured diffraction peaks can be compared to standard reference patterns, which can be experimentally or computationally generated. Secondary phases may be present, but not visible with diffraction. Sometimes nanocrystalline phases can be detected by electron diffraction or Raman spectroscopy, among other methods, but Raman reference patterns for new materials are not as easily accessible. A cartoon schematic of XRD experiments is shown in **Figure 4e**.

Additional constraints exist when synthesizing chalcogenide materials for device applications, as discussed earlier. Most importantly, materials need to be stable in air and at the interface of relevant chemicals they are grown upon or below in a device stack, or exposed to during their lifecycle. To this end, it is useful to measure material properties after exposure in the air for a given time interval. Additional constraints include temperature stability (most devices have a "thermal budget"), stability to UV irradiation (for devices that will operate in the sun), Pourbaix stability (e.g. in PEC device applications, where materials are often exposed to extremely acidic or basic conditions), etc. It is also important to consider the lifetime of the device when designing materials to be incorporated, i.e. whether migration of dopants or segregation into secondary phases will occur over time. End-of-life decomposition and recyclability of materials is also becoming increasingly important in rational materials design, as researchers strive to implement principles of the circular economy.[83]

## 2.2 Band gap

### *2.2.1 Computational optical properties*

A material with a band gap $E_G$ higher than approximately 3.1 eV (corresponding to a band absorption onset at 400 nm), is typically considered "transparent" so long as other effects are negligible (e.g. free carrier absorption and defect absorption; see below). In this review we will also be discussing semi-transparent materials with band gaps greater than 2 eV (band absorption onset at 620 nm), hence the term "wide band gap semiconductors." Band gap and dominant band-to-band absorption can be estimated from electronic structure calculations from ground state density functional theory (DFT), as shown schematically in **Figure 4b** as the lowest energy difference between the conduction band minimum (CBM) and the valence band maximum (VBM). Different DFT functionals trade off band gap accuracy for computational efficiency. Fast, "cheap" DFT calculations using e.g. GGA+U have been performed and compiled for tens of thousands of experimentally known materials in computational databases (e.g. Materials Project[34]), but have been shown to significantly underestimate band gaps unless the U value is empirically fit to band gap. More expensive calculations such as hybrid functionals (e.g. HSE06) and GW can also be applied in a lower-throughput manner at a greater computational cost. These computational methods have been benchmarked to predict band gaps on a case-by-case basis within 25% uncertainty,[84] and are available in limited quantities (~1,000) in some computational databases (e.g. NRELMatDB[45–47]). In general oxides tend to exhibit a wider gap than counterpart chalcogenides (see Section 1), but as shown in **Figure 3** there are still many chalcogenides with wide band gaps. We note that a wide band gap alone does not necessarily guarantee optical transparency. Dominant transitions tend to occur at direct gaps, rather than indirect gaps that require phonon assistance for absorption. At high carrier doping, intraband scattering and free carrier absorption can cause loss of transparency within the near-infrared and visible spectrum that increases with doping concentration. Additionally, dopants can introduce defect levels within the gap that reduce optical transparency, and in some materials excitonic absorption may also become significant.[85] These effects are trickier to predict computationally.

### *2.2.2 Experimental optical properties*

Optical properties of transmittance, reflectance, and absorption are typically measured for thin films with UV-Vis-IR spectrophotometry as a function of wavelength, as depicted in **Figure 4f**. Transmittance and reflectance measurements can be used to calculate the absorption coefficient, $\alpha = -\ln\left(\frac{T}{1-R}\right)/d$, as a function of wavelength. From the absorption spectrum, the low-frequency cutoff is typically fit to estimate the band gap using the Tauc relation $(\alpha h\nu)^{1/r}$, where $r = 1/2$ is assumed for allowed direct transitions and $r = 2$ is assumed for allowed indirect transitions. Such fits are often ambiguous depending on the selected energy range, which is not standardized and depends on the discretion of the researcher. Band-to-band transitions typically occur in semiconductors for absorption coefficients of approximately $10^4$ $cm^{-1}$, so the cutoff is usually drawn around this value as an alternative method to estimate the optical absorption threshold without asserting the band gap. Thus, it can be very difficult to conclusively determine from Tauc plots whether a material has a direct or indirect band gap, and to accurately estimate indirect band gap for which the optical transition probability is usually low. In addition, researchers should be aware that transmittance and reflectance are *thickness dependent* measurements, leading to different accessible dynamic ranges of absorption coefficient measurements. It is important to keep this in mind when interpreting Tauc-derived experimental band gaps and comparing reported values for samples with very different thicknesses.

To illustrate some of these discrepancies, **Figure 5** shows the optical properties of sputtered zincblende ZnSe thin films, plotted as a function of film thickness. As evidenced from Figure 5a and b, transmittance values are a function of wavelength and are heavily influenced by Fabry-Perot interference fringes related to light scattering from the front and back of the film. Thus, it is ambiguous to report transmittance from a single wavelength number, yet this is still done routinely in the literature. One way to solve this issue is to use the reflectance-corrected transmittance, $T_{corr} = \frac{T}{1-R}$, as shown in Figure 5c. For a given thickness, this mathematical transformation results in a monotonically increasing $T_{corr}$ with wavelength, and thus a more reliable value. However, even $T_{corr}$ still depends somewhat on film thickness (Figure 5c), so thin films will appear "transparent" if thin enough. Considering these two factors, transmittance can be useful as a reference but is not an appropriate metric to compare various materials systems. Instead, it is recommended to report the $\alpha$ to avoid thickness and wavelength dependence. Figure 5d shows the absorption coefficients as a function of wavelength and thickness. In this case there is a fixed onset at ~490 nm where absorption reaches $10^4$ $cm^{-1}$ regardless of thickness, so we can interpret this value as a band gap of ~2.5 eV (compared to the literature reported gap of ~2.7 eV[120]).

Spectrophotometers can also be equipped with an integrating sphere to measure diffuse reflectance for rough samples. Other useful optical characterization techniques include Fourier transform infrared spectroscopy (FTIR) to probe the infrared, and ellipsometry to extract optical constants (i.e. refractive index *n* and extinction coefficient *k*), and there are many in-depth reviews and textbooks on these methods.[86–88] In this review, we will focus on reporting optical band gap because it is the standard value reported consistently in the literature, yet we will often use the "~" symbol to emphasize aforementioned experimental and analytical uncertainties. We also provide references that report absorption coefficient, transmittance, roughness, and other optical properties.

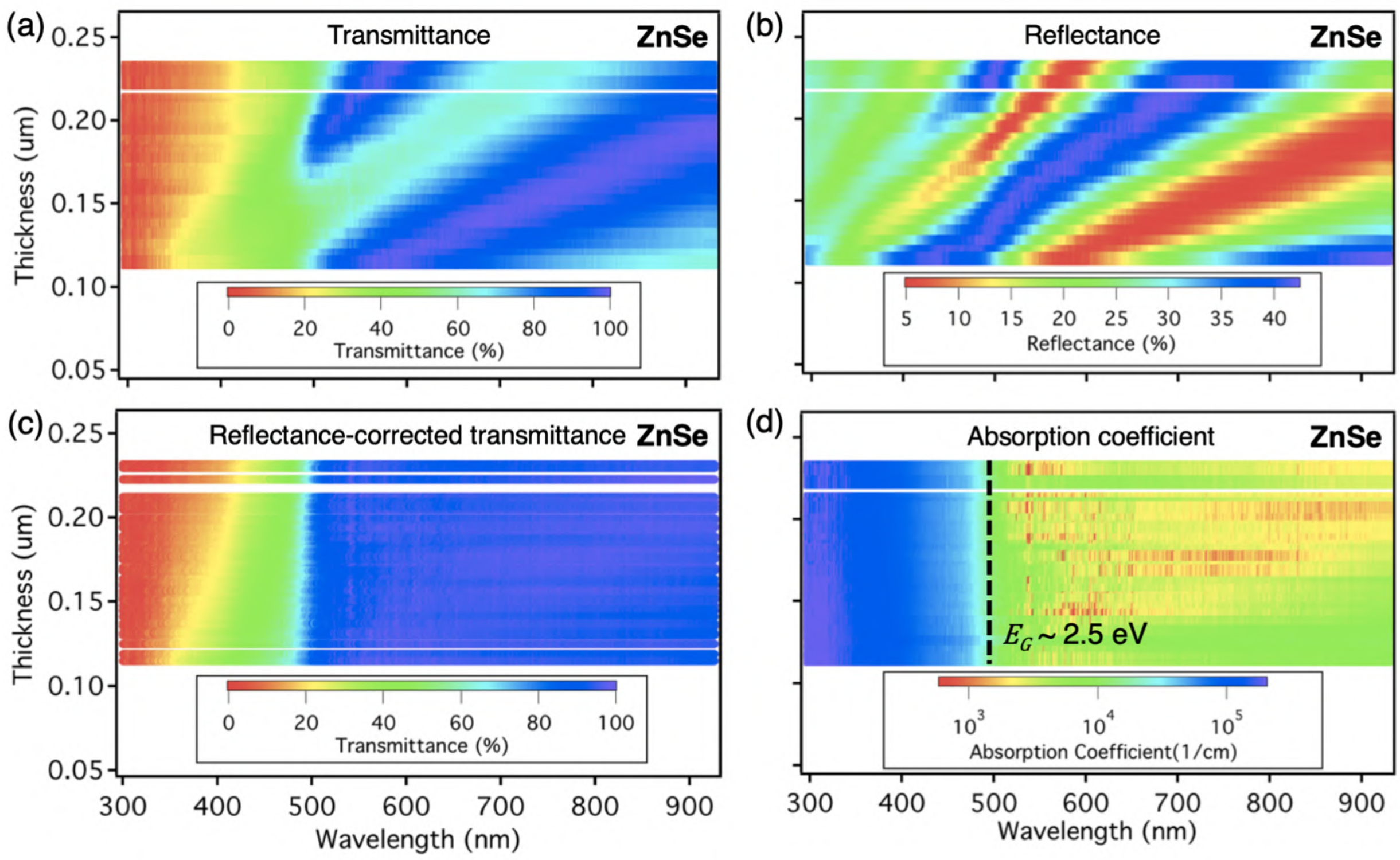

**Figure 5**. Optical properties of sputtered ZnSe thin films as a function of thickness: (a) transmittance, (b) reflectance, (c) reflectance-corrected transmittance, and (d) absorption coefficient, indicating how optical band gap $E_G$ is estimated experimentally.

## 2.3 Mobility and conductivity

The carrier mobility $\mu$ of a material is set by the intrinsic dispersion of the electronic bands as quantified by effective mass $m^*$, and is limited by the mean free path of carriers with charge $q$ (or the inverse of mean free path, scattering time $\tau$), such that:

$$\mu = \frac{q\tau}{m^*} \tag{1}$$

while electrical conductivity is proportional to both the carrier mobility and the carrier concentration $n$:

$$\sigma = qn\mu = \frac{q^2 n\tau}{m^*} \tag{2}$$

### *2.3.1 Effective mass and scattering time*

Conductivity can be estimated from DFT electronic band structures by first considering effective mass $m^*$ (electron effective mass $m_e^*$ for n-type conductivity, and hole effective mass $m_h^*$ for p-type conductivity). There are generally two types of effective mass reported in the literature. The first, density of states effective mass $m_{DOS}^*$, is estimated by integrating the entire density of states near the band edge and assuming non-degenerate doping. The second, band effective mass (or inertial effective mass) $m_b^*$, is band structure dependent and accounts for band anisotropy and degeneracies. This value is generally used for transport calculations, so we will refer to it herein. From band theory, the band effective mass is defined as the second partial derivative of band energy with respect to crystal momentum, and is represented by the effective mass tensor $\boldsymbol{m^*}_{ij}$:

$$\boldsymbol{m^*}_{ij} = \left[\frac{1}{\hbar^2}\frac{\partial^2 \boldsymbol{E}}{\partial k_i \partial k_j}\right]^{-1} \tag{3}$$

Diagonalizing and inverting this tensor results in an effective mass for three perpendicular directions in the crystal, and in an isotropic crystal it is a scalar value, $m^*$.[89] Effective mass has units of electron mass $m_0$, but is often simplified as $m^*/m_0$ such that it is unitless, a convention that is adopted in this review. This value depends on the orbitals present at the band edges and their degree of splitting and hybridization. It usually ranges from 0.1–10 for electrons and holes, though the exact value of $m^*$ differs by application, how it is defined (e.g. number of bands included in the calculation, and Fermi level position), and particular calculation method. Computational codes such as BoltzTraP[90] and EMC[91] are available for calculation of $m^*$ for large data sets. Due to tighter binding of valence band states compared to conduction band states, wide-gap materials tend to have much $m_e^*$ than $m_h^*$, allowing for higher n-type conductivity to be achieved.[92] This is depicted in Figure 4c, where the CBM is far more disperse than the VBM. In most wide-gap oxides, low valence state dispersion is due to highly localized O 2$p$ orbitals lying at or near the VBM and introduces a fundamental limitation to achieving high p-type mobility. This limitation is one reason for exploring chalcogenides, which exhibit more delocalized VBM orbitals and hence may be advantageous over oxides in achieving wide-gap semiconductors with low hole effective mass and high mobility (see Section 1).

Scattering time $\tau$ is much more difficult to calculate from *ab initio* methods, and differs drastically by material. It is typically decomposed into various scattering components:

$$\frac{1}{\tau} = \frac{1}{\tau_i} + \frac{1}{\tau_g} + \frac{1}{\tau_l} + \frac{1}{\tau_n} \tag{4}$$

where $\tau_i$ is ionized impurity scattering (i.e. off of charged dopant ions), $\tau_g$ is grain boundary scattering (i.e. off of crystal grains or dislocation defects), $\tau_l$ is lattice vibration scattering (i.e. off of phonons or polarons), and $\tau_n$ is neutral impurity scattering (i.e. off of uncharged dopants or defects). These carrier scattering mechanisms depend not just on crystal structure but on processing conditions, crystal purity, doping levels, among other conditions, and are covered in detail elsewhere.[93–95] When transparent semiconductors are degenerately doped to concentrations approximately $>10^{20}$ cm$^{-3}$, it is typical for temperature-independent ionized impurity scattering to dominate.[96,97] Most computational screening studies assume ionized impurity scattering dominates in wide-gap chalcogenides such that the constant relaxation time approximation (cRTA) holds. However computational methods and associated software packages, such as the aMoBT code,[98] to consider other scattering effects in the calculation of mobility are advancing, as well as semi-empirical methods such as the $\beta_{SE}$ descriptor.[99] We note that the thresholds for "high" mobility and conductivity are arbitrary, and requirements will differ by application.

### *2.3.2 Mobility and conductivity measurements*

Figure 4g and Figure 4h demonstrate the two most common ways to measure electrical conductivity in thin film materials. In the first configuration in Figure 4g, a four-point probe is used to measure sheet resistance of bulk or thin film semiconductors, and in-plane conductivity can be calculated if film thickness known. The basic concept is to source current through two terminals while generating and measuring voltage from the other two terminals. Vertical, or out-of-plane resistivity can be measured by preparing a sandwich structure with conductive materials as top/bottom electrodes and target material in the middle, and measuring its electrical resistance, also requiring independent thickness measurements to extract electrical conductivity.

Hall effect measurements shown on the left of Figure 4h are often used to obtain Hall mobilities and carrier concentrations (as well as carrier type). When a source current flows through two terminals, carriers deviate under a vertical magnetic field and accumulate at the edges, generating a Hall voltage proportional to carrier concentrations. Mobility can then be back-calculated from conductivity and carrier concentrations, and carrier type is indicated from the sign of the Hall voltage. The measured mobility typically consists of multiple scattering mechanisms, and $\tau$ can be back-calculated using Equation 1 if $m^*$ is known. To reveal the particular scattering mechanism for a thin film, the Hall mobility can be measured as a function of temperature. For example, a temperature-dependent mobility suggests phonon scattering, whereas temperature independent mobility is usually indicative that ionized impurity scattering dominates. This latter case is a typical dominant scattering mechanism in highly conductive n-type transparent conducting oxides, and is also present in several transparent conducting sulfides.

Conductivity can also be measured optically, using light to excite free carriers. This is particularly useful to evaluate intra-grain mobility, and for bulk powder samples which cannot be analyzed with four-point probe or Hall setups due to their granularity. This method relies on the optical absorption at photon energies below the energy gap, which is related to the free electrons or holes. The relationship is summarized by considering carriers as free electrons from the Drude model, from which conductivity, mobility, carrier concentration can be calculated if $m^*$ is known.[100] Mobility can also be backed out from device measurements (e.g. in transistors), as is described for 2D chalcogenides in Section 3.6.

## 2.4 Dopability

In addition to low $m^*$ and high $\tau$, Equation 1 illustrates that high conductivity also requires a sufficiently high carrier concentration $n_n$ for electrons (or $n_p$ for holes). This depends on the material's dopability, which is a function of several parameters. For example, to be a highly conductive wide-gap p-type semiconductor, a material must support the introduction of p-type dopants (i.e. dopant must be soluble and energetically stable), be dopable to a high enough concentration, and not be susceptible to compensation by n-type "hole killer" defects. It has been theorized that n-type dopability is favorable in materials with large electron affinities, while p-type dopability is favorable in materials with small ionization energies.[33] This is one reason why it is more difficult to dope wide-gap materials; in general as the gap increases, the CBM shifts towards vacuum level and the VBM shifts away from it. Additionally, the wider the gap the more likely defect levels are to emerge within the gap and induce carrier compensation. Because of the $3p$, $4p$, or $5p$ character of chalcogen atoms, valence levels of chalcogenides tend to lie closer to vacuum than oxides with $2p$ character of oxygen atoms, suggestive of a higher propensity for p-type doping in chalcogenides compared to oxides. However, while noting this general trend, many chalcogenides are in fact highly n-type dopable as well, and defect calculations, are usually necessary to understand dopability.

### *2.4.1 Computational dopability*

Computationally understanding defect compensation and selecting appropriate dopants requires defect formation energy calculations, which require large DFT supercells and thus are computationally quite expensive. These calculations can estimate the energy level of particular dopants, whether they are shallow or deep, and whether they should lead to n-type or p-type conductivity (see Figures 4d). The defect formation energy, $E^f[X^q]$, is the energy cost to create or remove an isolated defect $X$ with charge state $q$ from a bulk material, and describes how favorable various defects and dopants are to form compared to one another. It is calculated as:

$$E^f[X^q] = E_{tot}[X^q] - E_{tot}[bulk] - \sum_i n_i\mu_i + qE_F + E_{corr} \tag{5}$$

where $E_{tot}[X^q]$ and $E_{tot}[bulk]$ are DFT formation energies of defective and bulk supercells, $-\sum_i n_i\mu_i$ is the summation over chemical potentials of the defect elements, and $qE_F$ is the cost in energy of adding or removing an electron ($E_F$ is the Fermi level). $E_{corr}$ are correction term(s) that account for mirroring of charges, VBM alignment, and other spurious computational effects. The calculated formation energies can be used as inputs to thermodynamic simulations that output equilibrium defect concentration, carrier densities, and Fermi level positions for an assumed set of chemical potentials. Specifics of these defect calculations and thermodynamic simulations are covered in comprehensive reviews.[101,102] Recently, codes such as PyCDT and Pylada have been developed to run defect calculations on a high throughput framework, making them more accessible to the research community.[103,104] Defect calculations can also be used to screen for defect tolerance[105,106,107] and for deep-level defect induced absorption that would decrease optical transparency.[108]

To roughly assess dopability, some high throughput studies have calculated the branch point energy (BPE, also referred to as Fermi level stabilization energy or charge neutrality level), which approximates the position where the Fermi level $E_F$ is pinned when defects are introduced.[109] The BPE is calculated using the following formula:[110]

$$\mathrm{BPE} = \frac{1}{2N_k}\sum_k\left[\frac{1}{N_{CB}}\sum_i^{N_{CB}}\epsilon_{CB_{ik}} + \frac{1}{N_{VB}}\sum_i^{N_{VB}}\epsilon_{VB_{ik}}\right] \tag{6}$$

where $N_k$ is the number of k-points in the DFT calculation, $N_{CB}$ and $N_{VB}$ the number of CBs and VBs averaged over, and $\epsilon_{CB_{ik}}$ and $\epsilon_{VB_{ik}}$ the DFT energy eigenvalues at each k-point. This descriptor is advantageous because it uses only bulk DFT band structure calculations and can be simply incorporated into to high throughput methodologies. BPE has been benchmarked for several binary wurtzite, zincblende and rocksalt systems,[110] and has been used in screenings for p-type transparent conductors.[111,112] However, despite its advantages BPE does not indicate whether a semiconductor is actually dopable in practice, cannot explain deep defect levels, varies depending on how many bands are averaged over, and has not yet been benchmarked for a comprehensive set of structures and chemistries.[113]

### *2.4.2. Experimentally quantifying doping*

Dopants are introduced intrinsically or intentionally during growth, or by a subsequent processing step such as rapid thermal processing or ion implantation. At concentrations greater than approximately 1%, impurity concentration can usually be determined via x-ray fluorescence spectroscopy (XRF), x-ray photoelectron spectroscopy (XPS), energy dispersive spectroscopy (EDS), or Rutherford backscattering spectroscopy (RBS), among other methods. Some highly sensitive techniques such as secondary-ion mass spectrometry (SIMS) can detect impurities down to parts per million or even parts per billion,[114,115] but their quantification requires reference samples. X-ray absorption spectroscopy (XAS) can be useful to determine the location and coordination of defects, i.e. whether they are vacancies $V_A$, interstitials $A_i$, substitutions $B_A$, antisites ($X_A$), or complexes, but also requires standards and detailed modeling. We note that the semiconductor community sometimes uses the term "dopant" for incorporation of chemical substitutions of greater than 1%, even in solid solutions or systems where phase segregation occurs. This appears in multiple studies referenced in this review, so it is important to be clear about diversity of this terminology.

Due to compensation, dopant concentration does not usually equal carrier concentration, so it is essential to measure carrier concentration directly. In the context of this article, the goal of doping a semiconductor is often to increase the majority carrier concentration without significantly reducing the mobility. These two properties can be measured using a Hall effect setup, as described previously and schematically shown in Figure 4h (left). It is highly recommended to ensure that the measured carrier type is the same and the carrier density is similar upon the magnetic field reversal. Even with these precautions, it may be difficult to accurately determine whether a material is doped n-type or p-type from Hall effect measurements of low-mobility materials or magnetic compounds. In such cases the Seebeck coefficient (i.e. thermopower) can be measured using a Seebeck thermocouple probe (Figure 4h, right), which is less sensitive to artifacts. The majority carrier type is indicated by the sign of the Seebeck coefficient, and the carrier concentration is usually inversely proportional to the magnitude of the Seebeck coefficient.

## 2.5 Band positions, alignments, and other properties

Additional criteria essential for the implementation of materials into optoelectronic device applications are the energetic offsets of the band edges and the Fermi level $E_F$ (i.e. ionization energy, electron affinity, and work function). These offsets referenced to those of the other layers in a device, and the band bending of the resulting interface or junction, determine the transport of electrons and holes through the interface of the device. For example, in order to align a top contact n-type wide-gap material to a p-type CdTe solar cell, the position of the CBM of the n-type material must be closely aligned in energy to the CBM of CdTe.[116] As mentioned in 2.1.3, chalcogenide materials tend to have higher VBMs closer to vacuum (lower ionization energies) than their oxide counterparts. This could enable band alignment configurations in devices and junctions that are not possible in oxides. Other properties of interest to wide band gap semiconductor materials and their practical applications include flexibility, lattice matching (i.e. interatomic distance), and microstructure, which will be discussed for materials in Section 3.

### *2.5.1 Computational band alignment*

As depicted in the bottom of the computation funnel of Figure 3, band alignment can be qualitatively assessed using DFT surface calculations of alternating crystal and vacuum slabs to estimate the ionization energy ($IE$, or the VBM with respect to vacuum). These methods are reviewed comprehensively elsewhere.[117,118] Some studies have used these calculations for high throughput screenings, in particular in searches for electrochemical catalysts and electrodes.[52] Band alignments can differ drastically depending on the surface morphology, including crystallographic plane, surface termination and defects. As a result, band alignments in polycrystalline materials can be averaged over the most stable surfaces, but there is not yet a consensus amongst the DFT community about their treatment.[119] In addition, band alignments are often modified in practice when two materials are brought into direct contact (rather than contact with vacuum). Such interfacial band alignment can be accessed from explicit interface slab calculations for any material pair for which the crystallographic relationship at the interface is well-defined. This is difficult to do for most materials without considering interfacial defects or strain, hence such calculations are usually performed for epitaxial lattice-matched interfaces.

### *2.5.2 Experimental band alignment*

Experimentally, work functions ($E_F$ with respect to vacuum level) are typically measured by Kelvin probe or by ultraviolet photoelectron spectroscopy (UPS) from secondary electron cutoff.[120] Another powerful technique is XPS, which can be used to measure valence band offsets and band bending by following shifts in the core levels. Interfacial band alignment experiments can be performed by bottom-up (layer by layer deposition) or top-down (depth profiling) modes, to experimentally measure band bending at interfaces, which is of particular importance to device applications. The latter mode is more common in the literature but less accurate due to chemical modification of the interface by sputtering with Ar ions. When determining band offsets from XPS core level shifts when another layer deposited on the original surface, it is important to keep in mind that the information depth depends on the kinetic energy of the photoelectrons. Complementary to XPS/UPS, Auger electron spectroscopy measurements are another possible way to detect the energy difference between $E_F$ and the VBM. Such experimental assessment of band alignment is important for designing functional optoelectronic devices (see Section 4).

# 3. Materials

Wide band gap chalcogenide semiconductors have unique chemistries and properties compared to their oxide counterparts, which can be understood from the point of view of molecular orbital theory. A simple chemical bonding schematic comparing atomic orbitals (AOs) and approximate molecular orbitals (MOs) is shown in **Figure 6** (see Supporting Information for figure details), and is useful for illustrating bonding-derived distinctions between chalcogenides and oxides. Figure 6a depicts the AO energies for group VI,[121,122] demonstrating that *p* valence energies increase going down group VI (from O 2*p*, S 3*p*, Se 4*p*, Te 5*p*). Figure 6a also demonstrates that *Ch p* (and *s*) orbitals are much closer in energy to each other than to O 2*p* (and O 2*s*, not pictured), which helps explain why chalcogenides often crystallize in stable structures distinct from oxide counterparts but similar to one another (e.g. $CuAlO_2$ vs. $CuAlCh_2$, see Sections 3.3 and 3.4.3). Energy levels of common transition metals (*M* = Cu, Zn, Ag, Cd), which are cations in many compounds discussed in this review, are also plotted in Figure 6a.

Figure 6b depicts molecular orbital schematics for four representative oxide and sulfide compounds (see S.I.). According to molecular orbital theory, orbitals closer in energy to one another are more likely to hybridize. In many common *M*-VI bonds, ionicity tends to decrease going down group VI, in particular in Cu-VI bonds due to Cu's high energy 3*d* orbitals. This promotes a larger *p-d* orbital overlap integral in chalcogenides than oxides, as exemplified in Figure 6b by a larger *p-d* overlap in $Cu_2S$ than $Cu_2O$. Thus higher energy *p* orbitals can hybridize more strongly with metal *d* orbitals. Assuming these orbitals contribute to the VB edge, this can lead to more delocalized VBMs, lower hole effective masses (see Section 2),[6] and higher achievable mobilities in p-type chalcogenides as discussed in Section 1 and subsequently.

sThe trade-off is that the band gap $E_G$ tends to be smaller for chalcogenides than for oxides, e.g. $E_G$ of $Cu_2S$ < $E_G$ of $Cu_2O$ as shown in Figure 6b. However, Figure 6b also demonstrates an exception to this general trend, namely $E_G$ of ZnS > $E_G$ of ZnO. One key difference between these Cu-VI and Zn-VI systems is that metal 3*d* states contribute significantly to the VB in Cu-VI, whereas the VB is primarily of anion *p* character in Zn-VI. We note that many of the compounds in this review are Cu-based (see Tables 1–4), but some compounds still exhibit an exception to this general trend. A postulated design principle in the literature for achieving high p-type doping propensities is a small ionization energy, i.e. higher position of *p*-orbital derived valence bands.[33] In Figure 6b, the VB edge is higher in ZnS than ZnO; ZnS has been shown to be p-type dopable (see Section 3), while ZnO is definitively n-type.[54,55] $Cu_2O$ and $Cu_2S$ have even higher VB edges and higher p-type dopabilities, respectively. The gap of $Cu_2S$ is too low to be considered in this review, but many materials discussed subsequently contain Cu and thus Cu-*Ch* bonds.

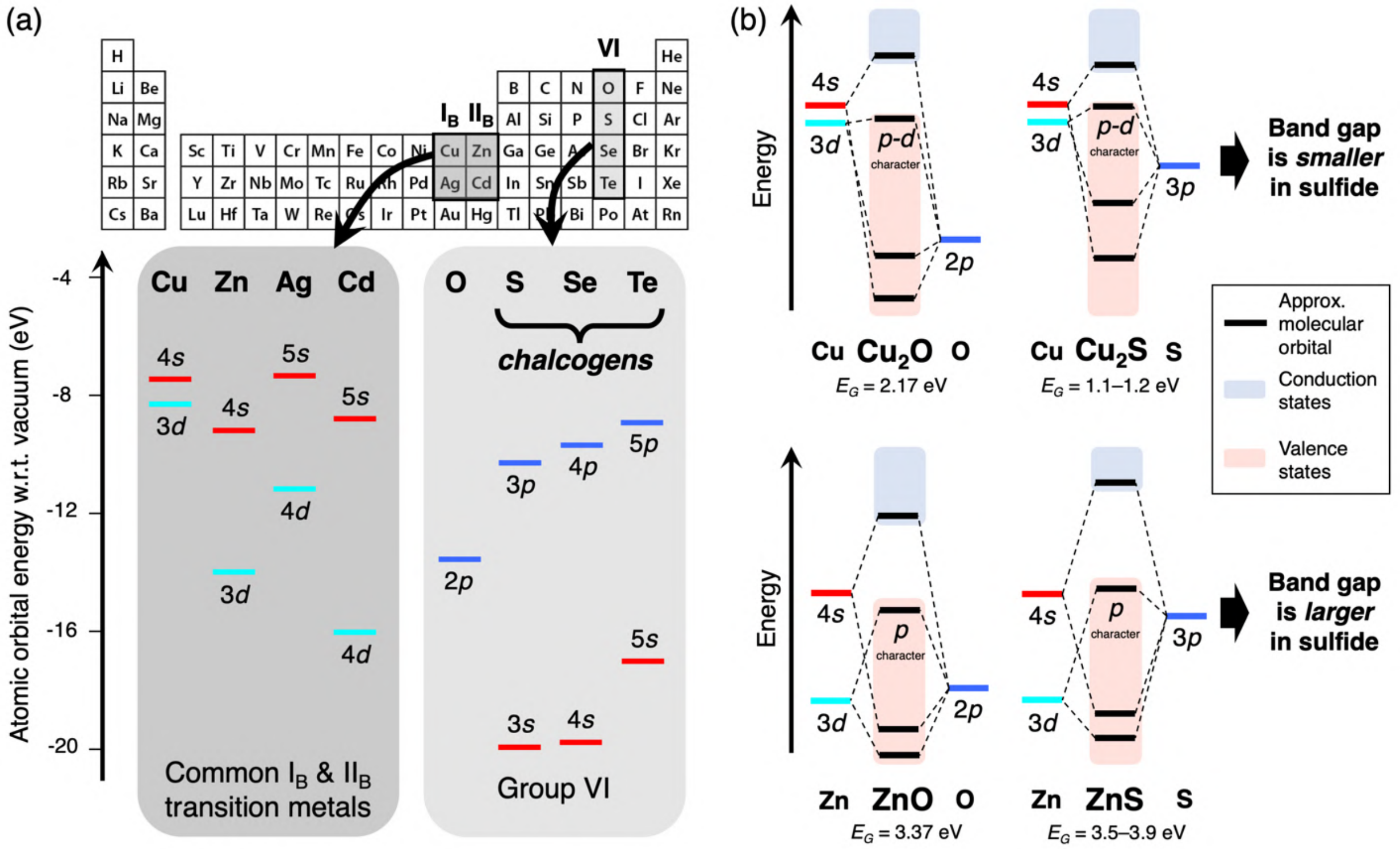

**Figure 6**. (a) Atomic orbital (AO) energies of group VI elements (including chalcogens S, Se, Te) and of common transition metal elements present as cations in compounds throughout this review. Energies are from NIST database LDA calculations,[121] and are aligned to experimental first ionization energies (see Supporting Information for details). (b) Molecular orbital (MO) schematics of representative binary oxides and sulfides: cuprite $Cu_2O$, chalcocite $Cu_2S$, wurtzite ZnO, and wurtzite ZnS (a wide-gap chalcogenide semiconductor, see Section 3). Bonding diagrams and band offsets are approximated from the literature and density of states (DOS) calculations using a schema described in the S.I.[123,124,125,126] We emphasize that these MO diagrams should be seen as schematic illustrations, as they neglect ionic bonding and cannot be directly scaled up to extended solids. For each compound, only the majority orbital contributions to the DOS are included. AO positions (colored bars) are to scale and band edges of each compound (blue and pink shading) are to scale with those of other compounds, but MO positions (black bars) and orbital splitting (dashed lines) are estimates and are not to scale.

In this section, we highlight the major advances in various classes of dopable wide band gap chalcogenide semiconductors ($E_G > 2.0$ eV), and compare their properties and their underlying physics. We will discuss bulk crystal structures, intrinsic and extrinsic dopants, and some alloys and composites, focusing mostly on materials that have been synthesized as polycrystalline thin films. Experimental band gaps and electrical properties discussed are measured at approximately 300 K, unless otherwise denoted. We also highlight some of the binary chalcogenides, ternary chalcogenides, and oxychalcogenides that have been predicted via computational searches. Many of these materials have not yet been explored in-depth for optoelectronic applications, so they open up areas of future experimental research. As stated previously, we will not cover amorphous materials or insulating materials, such as chalcogenide glasses, which are of interest but have been reviewed in depth elsewhere.[43] Materials reported in this section are summarized in Tables 1–4 and a summary figure in the end of this section.

## 3.1 Binary II-*Ch* chalcogenides

Binary, bivalent metal chalcogenides $M^{2+}Ch^{2-}$ (*Ch* = S, Se, Te) are the simplest class of wide band gap chalcogenide semiconductors. The II-VI chalcogenide materials (notated II-*Ch* here) are the most common binaries for electronic applications, and typically contain metal cations from group II and $II_B$ (Zn, Mg, Mn, or Cd, among others). II-*Ch* binaries typically crystallize in wurtzite (WZ), zincblende (ZB, also written "zinc blende" i.e. sphalerite), rocksalt (RS, i.e. halite), or nickeline (NC, i.e. nickel arsenide) structures, as shown in **Figure 7**. These close-packed structures differ in their stacking (hexagonal close-packed in WZ and NC, face-centered cubic in RS) and their cation coordination (tetrahedral in WZ and ZB, octahedral in RS and NC).[127] The WZ, ZB, RS, and NC structures typically crystallize in space groups $P6_3mc$, $F\overline{4}3m$, $Fm\overline{3}m$, and $P6_3/mmc$, respectively.

The structural motifs in these binary chalcogenides are the building blocks for most of the wide band gap chalcogenide materials used today. Alloys and composites can be formed using these binaries as terminal compounds, forming isovalent isostructural alloys e.g. $Zn_xMg_{1-x}S$, heterovalent heterostructural alloys with another lower gap binary e.g. $Cu_xZn_{1-x}S$, or composite materials e.g. $Cu_xS$:ZnS. This strategy has been heavily explored in the space of n-type transparent conducting oxides, and has led to state-of-the-art optoelectronic properties,[2,128] which motivates further exploration for p-type transparent conductors. Since the properties of these alloy and composite systems depend on the terminal compounds, we will focus on the binary compounds and mention only a few of their representative alloys. Next, we review the different chemistries and discuss n-type and p-type dopants in these material systems, both experimentally realized and theoretically predicted.

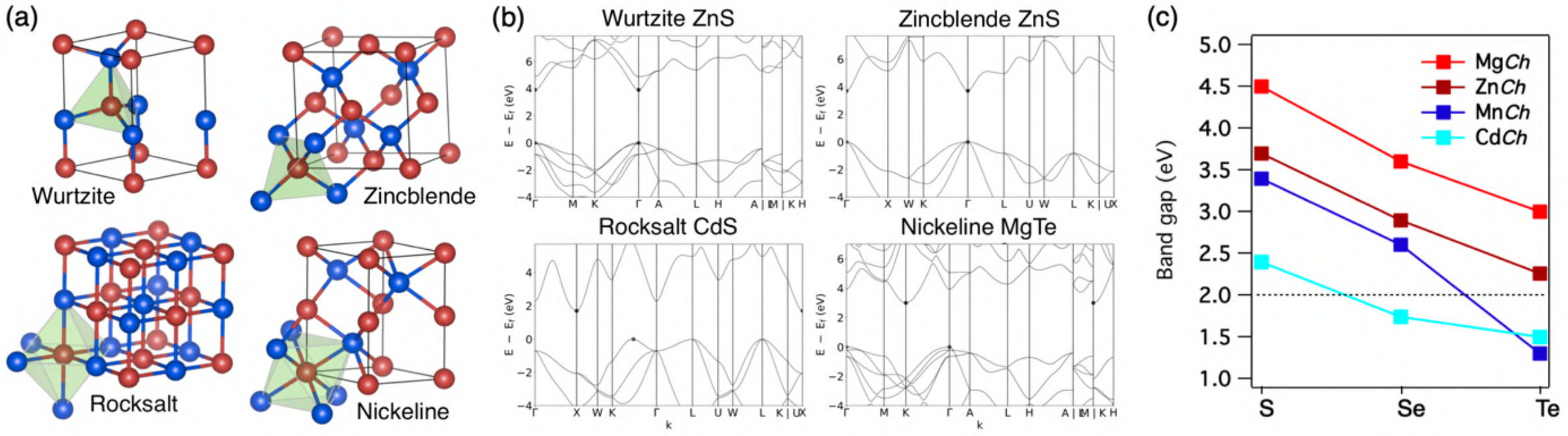

**Figure 7**. (a) The four most prevalent crystal structures of Zn-, Mg-, Mn-, and Cd-based binary, bivalent II-*Ch* metal chalcogenides. (b) Band structures of representative binaries with these crystal structures, calculated from the Materials Project database using a GGA functional and a scissor operation to correct for experimental band gap. (c) Experimental band gaps of the most stable structure of each material in this class, as discussed in this section (see Table 1).

### *3.1.1 ZnCh*

ZnS is one of the most prevalent sulfide minerals on Earth, and it used in numerous electronic device applications. It crystallizes in a low temperature cubic zincblende (ZB) phase ($F\overline{4}3m$), high temperature hexagonal wurtzite (WZ) phase (e.g. $P6_3mc$, most commonly with a 2H stacking polytype), or a mixture of both, depending on growth conditions.[8,129] Both structures have a fairly wide direct band gap of ~3.7–3.8 eV for ZB and ~3.9 eV for WZ.[130,131] These properties make both structures suitable as buffer layers and, if sufficiently dopable, as transparent electrode layers in solar cells. Importantly, ZnS is an non-toxic, earth-abundant material that is cheap and easily synthesizable in a variety of microstructures. It is closely lattice matched with $Cu(InGa)Se_2$, the absorber layer in CIGS solar cells, as well as GaP and other III-V and II-VI alloys.[132] ZnS is also important for its strong and stable luminescence, tunable across the blue to red end of the visible spectrum.[133]

The conductivity of undoped ZnS is usually very low,[134] likely due to low carrier concentration, and at slight off-stoichiometries is reported intrinsically n-type due to interstitial Zn ($Zn_i$) or S vacancies ($V_S$). However, ZnS has been reported to be ambipolar upon extrinsic doping, e.g. n-type by $Al_{Zn}$ (up to ~$10^{-3}$ S cm$^{-1}$)[135] and p-type by $Cu_{Zn}$ (~$10^{-5}$ S cm$^{-1}$ for 0.1% Cu, and even up to ~1 S cm$^{-1}$ for 9%, though this may be influenced by Cu-S impurities; see Section 3.1.7).[136] Other known n-type dopants are $F_S$ (conductivities ~2 ×$10^{-7}$ S cm$^{-1}$),[137] $Cl_S$,[138] $In_{Zn}$,[139] among others, and co-doping strategies have been employed as p-type dopants (e.g. N/Li,[140] and In/Ag/N[139]). Both polymorphs of ZnS have low effective masses of both electrons ($m_e^*$) and holes ($m_h^*$), evident from highly dispersive CBMs and VBMs (see Figure 7). The ambipolar dopability may derive from appropriately aligned band positions with respect to vacuum, with reported electron affinities in zincblende ZnS of approximately 3.8–4.0 eV and ionization energies ~7.5 eV by both computation and experiment (see Section 4.3).[141–143] Additionally, ZnS has minimal compensating defects, as demonstrated computationally in a recent defect screening study of binary chalcogenides.[144] This study found p-type conductivity promising in doping Zn sites with Cu ($Cu_{Zn}$), confirming experimental results, as well as Na ($Na_{Zn}$), K ($K_{Zn}$) and doping S sites with N ($N_S$), as depicted in **Figure 8**a. This is evident by the low defect formation energy when the Fermi level $E_F$ is near the VBM.

ZB ZnSe has a lower band gap than ZB ZnS (~2.7 eV at room temperature, ~2.8 eV at 10 K), but it is still sufficiently transparent for use as a window layer in solar cell applications.[145,146]

It is also intrinsically n-type due to $Zn_i$ or $V_{Se}$, though p-type conductivities up to ~5 S $cm^{-1}$ have been demonstrated in metalorganic vapor phase epitaxy films.[147] Various dopants have been explored experimentally to modify the conductivity of ZnSe, e.g. $N_{Se}$ (p-type, 0.06 S $cm^{-1}$ and mobilities up to 86 S $cm^{-1}$),[148] $Ga_{Zn}$ (n-type, 20 S $cm^{-1}$),[149] $Cl_{Se}$ (n-type, ~333 S $cm^{-1}$),[150] among others. ZnSe has an even lighter effective mass than ZnS. Particularly notable is that the mobility in high-quality, nominally updoped MBE-grown intrinsic n-type ZnSe has been reported up to 550 $cm^2$ $V^{-1}$ $s^{-1}$ at room temperature and is dominated at room temperature by polar optical phonon scattering. This high mobility motivates ZnSe's application in devices such as light emitting diodes (LEDs) and thin film transistors (TFTs).[151]

ZnTe also crystallizes in its ground state in the ZB structure, with a direct gap of ~2.3 eV,[152] and its intrinsic p-type conductivity is reportedly due to $V_{Zn}$. P-type dopants $N_{Te}$ (25 S $cm^{-1}$),[153] $Cu_{Zn}$ (~0.33 S $cm^{-1}$),[154] and $Sb_{Zn}$ (~30 S $cm^{-1}$)[155] improve the p-type conductivity of ZnTe, and several of these dopants have been confirmed computationally with defect studies (see Figure 8b).[144] N-type doping is trickier but has been achieved in epitaxial crystals with $Al_{Zn}$, $Cl_{Te}$, and $Sn_{Zn}$.[156–159] P-type ZnTe is commonly used in CdTe solar cells as a back contact due to its small valence band offset with CdTe of approximately <100 meV (see Section 4),[160] and has been explored for photocatalysis applications.[161]

As expected from bonding trends and valence band positions of the chalcogen anions (see Section 1), ZnS and ZnSe have a wider band gap but lower achievable dopings than ZnTe. A "δ-doping" technique has been used to leverage the wide band gap of ZnS and high dopability of ZnTe.[162] Highly conductive p-type ZnSe and ZnS films were achieved using molecular beam epitaxy (MBE) to insert heavily N-doped ZnTe layers between each layer ZnSe or ZnS layer. The hole concentration in a ZnSe/ZnTe:N δ-doped stack reaches $7\times10^{18}$ $cm^{-3}$, and a $[N_a\text{–}N_d]$ value reaches $5\times10^{17} cm^{-3}$ in a ZnS/ZnTe:N δ-doped stack.[163] This is another technique that could be applied to other material spaces to engineer desired optoelectronic properties.

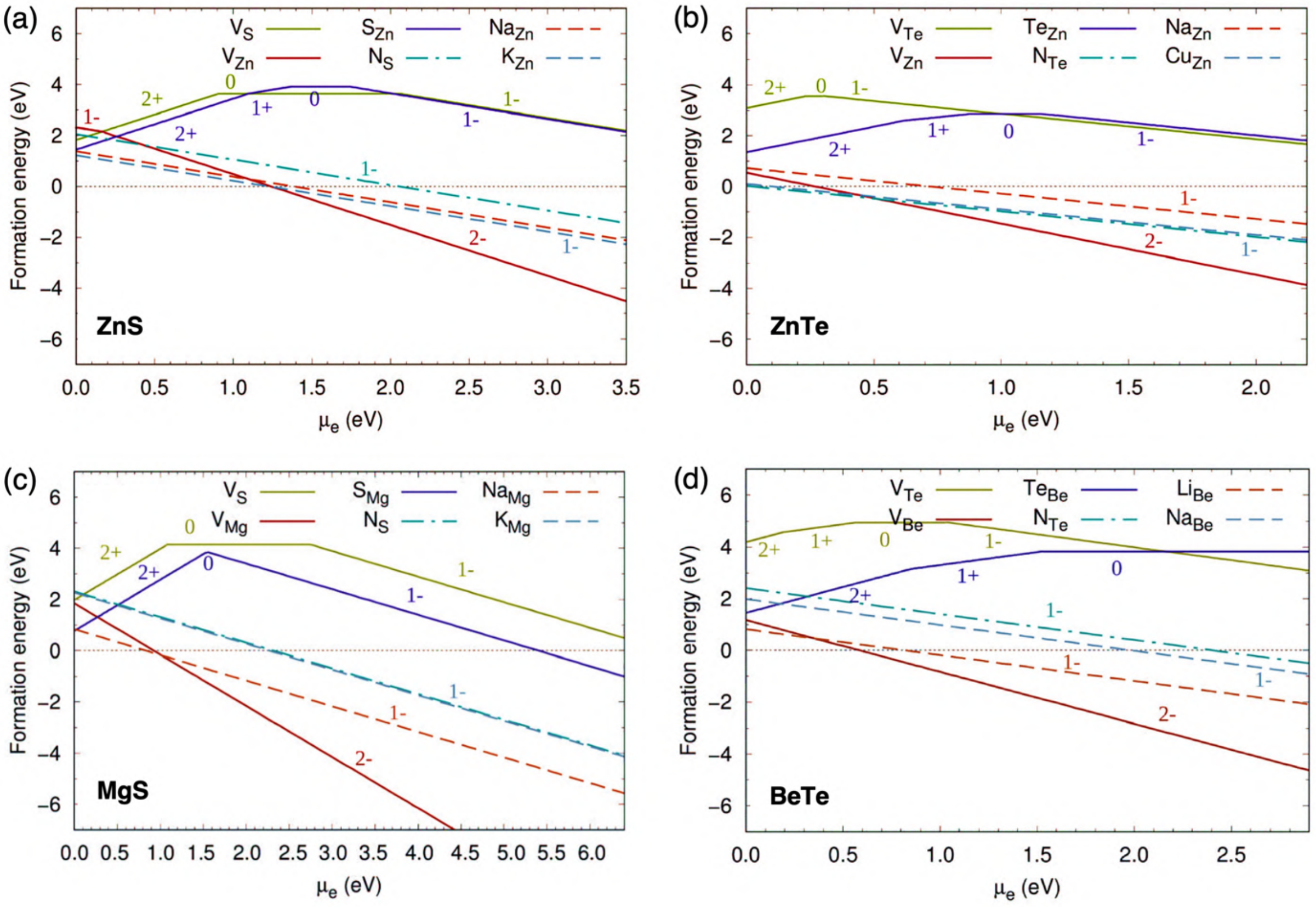

**Figure 8**. Defect formation energy plots for a representative set of binary II-VI chalcogenides: (a) ZnS, (b) ZnTe (c) MgS, and (d) BeTe. The defect formation energy $E^f[X^q]$ of a given dopant $X^q$ is plotted as a function of chemical potential $\mu_e$ at 0 K (i.e. Fermi energy $E_F$). These plots show the lack of compensation for particular dopants, suggesting p-type conductivity. Plots are reproduced from the literature.[144]

### *3.1.2 MgCh*

There are fewer reports on Mg*Ch* compounds compared to Zn*Ch*, likely because they react strongly with moisture and are not stable in air.[67] However, Mg*Ch* have larger band gaps than the other binary chalcogenide semiconductors mentioned here, so development of a stable synthesis pathway could be useful. The experimentally reported band gaps of MgS (ZB), MgS (WZ), MgSe (ZB), MgTe (ZB), and MgTe (WZ) have been estimated from MBE grown $Zn_{1-x}Mg_x(S,Se,Te)$ and $Cd_{1-x}Mg_xTe$ alloy films to be ~4.5 eV (measured at 77 K), 4.87 eV (measured at 77 K), ~3.6–4.05 eV, ~3.5 eV, and ~3 eV, respectively.[164-165,67] Considering their wide band gaps and structural compatibility with other II-VI compounds, Mg*Ch* usually act as a critical component in various alloy structures and enable tuning of properties (see Figure 11). For example, alloying MgS with PbS in $Pb_{1-x}Mg_xS$ can allow for tunable morphology, transparency, band gap, and conductivity.[166] Additionally, the lattice constant and band gaps increase with the increase of Mg content in $Zn_{1-x}Mg_xSe$,[167] and Mg content influences the band gap of $Cd_{1-x}Mg_xTe$ while maintaining a small mismatch with CdTe.[67] Polymorphs of Mg*Ch* can be stabilized when alloyed with other materials. Computational investigation has predicted rocksalt as the stable phase for MgS and MgSe, and the nickeline phase was predicted for MgTe alloys.[168] According to computational defect studies of rocksalt MgS (see Figure 8c) and wurtzite MgTe, alkali substitutional acceptor dopants $Na_{Mg}$ and $K_{Mg}$, and anion substitutional dopant $N_{Ch}$ should lead to p-type dopability. However, the doping ability of other Mg*Ch*-based alloys requires further investigation.[144]

### *3.1.3 MnCh*

MnS is a wide-gap semiconductor that can crystallize in the WZ (γ) structure with a gap of 3.88 eV,[169] in the ZB (β) structure with a gap of 3.8 eV in single crystal samples,[170] rocksalt (α) with a gap of 2.8–3.2 eV and single crystal mobility of 10 $cm^2 V^{-1} s^{-1}$ (the most stable phase at ambient conditions),[164] and in an amorphous phase with a gap of 2.8–3.0 eV.[171] MnS is intrinsically p-type, likely due to doubly-ionized $V_{Mn}$,[172] with a low room temperature reported conductivity of ~$10^{-5}$ S $cm^{-1}$,[173] and MnS has been doped by e.g. $Cd_{Mn}$.[174] Generally, MnSe is stable in its rocksalt (α) structure[175] and has a band gap of approximately 2.5 eV.[176] Its ZB and WZ polymorphs have also been synthesized, with reported gaps of ~3.4 eV and 3.5–3.8 eV,[177,178] respectively, and are structurally compatible with other III-V and II-VI semiconductor systems.[178] MnSe has been synthesized in a high-pressure nickeline (NC) phase as well. Doping has not been explored, but it is likely also p-type due to $V_{Mn}$. MnTe usually crystallizes in a NC phase ground state, while the band gap of the ZB polymorph of MnTe can reach 3.2 eV.[179] NC MnTe has high p-type conductivity compared to the other Mn-based compounds, up to 5–6 S $cm^{-1}$, with Hall mobilities up to 0.5 $cm^2 V^{-1} s^{-1}$, room temperature intrinsic degenerate doping of $5 \times 10^{19}$ $cm^{-3}$, and an estimated hole effective mass of 1.5,[41] but a lower band gap of 1.3 eV. WZ MnTe with a wider band gap of 2.7 eV can be stabilized by alloying with ZnTe,[41] or by growth on $InZnO_x$-coated glass.[180] Investigation into polymorphs with tunable band gaps and conductivities remains to be explored. We note that partially-filled *d* orbitals in elements such as Mn and Co introduce a complication in that there can be additional magnetic degrees of freedom. Any magnetic order present, as well as hybridization between *d* and *p* orbitals, can strongly influence the material's optical and electronic properties and can often reduce the transparency window.

### *3.1.4 CdCh*

CdS is one of the most intensively investigated wide-gap semiconductors, used in a large variety of optoelectronic applications, and has been heavily reviewed.[181–183] This material is the archetype n-type buffer layer used commercially in CdTe and CIGS solar cells, [183,184] and has been used for LEDs, TFTs, photonic/lasing devices, and piezoelectrics. CdS's most common form is hexagonal WZ, ZB phases have also been investigated, and there exists a high-pressure RS phase. Epitaxial growth has produced single phase, single crystal wurtzite material to quantify the room temperature direct gap at 2.5 eV (measured with ellipsometry), and the cubic phase has a similar but slightly lower measured direct gap of approximately ~2.3–2.4 eV.[185] The RS phase has been reported to have an indirect gap of ~1.5–1.7 eV, consistent with band structure calculations.[186] Polycrystalline thin films can be prepared by sputtering, chemical bath deposition (CBD), and thermal evaporation, among other methods, and tend to have a mixed WZ and ZB structure with band gaps of approximately 2.3–2.5 eV. This variation has been explained by the range of substoichiometric sulfur content in reported films due to various deposition temperatures.[185,187] WZ CdS is typically grown n-type, and is ambipolar with nearly intrinsic electron conductivity of $2.8 \times 10^{-2}$ S $cm^{-1}$ and nearly intrinsic hole conductivity of $1.5 \times 10^{-2}$ S $cm^{-1}$.[164]

The temperature dependence of conductivity, electron concentration, and electron mobility in ultrapure CdS samples is shown in **Figure 9**. Hall measurements (see Section 2.3 and 2.4) of carrier concentration and conductivity generally increase as temperature increases (Figure 9a), while mobility decreases with increased temperature due to carrier scattering (Figure 9b). Figure 9c illustrates the sharp absorption edge in CdS due to its direct band gap, the directional dependence of the absorption edge in the crystal, as well as how increasing temperature lowers the absorption edge. Peak Hall n-type mobilities in ultrapure crystals have been measured up to $10^4$ $cm^2 V^{-1} s^{-1}$ at 30–40 K (see Figure 9b), while record room temperature n-type mobilities have been reported up to 160 $cm^2 V^{-1} s^{-1}$ in CdS extrinsically doped with $In_{Cd}$ to $5 \times 10^{19}$ $cm^{-3}$ and conductivities up to ~50 S $cm^{-1}$.[188,189] Hole mobilities are predicted to reach 15 $cm^2 V^{-1} s^{-1}$. Other common n-type dopants include $Ga_{Cd}$,[190] $Al_{Cd}$,[191] and $Cl_S$,[138,192] while p-type dopants include $Cu_{Cd}$ (up to 2 S $cm^{-1}$, likely due to impurity band or hopping conduction)[193,194] and $Bi_S$ (up to $10^{-1}$ S $cm^{-1}$ and mobilities of 3 $cm^2 V^{-1} s^{-1}$).[195] Compared to CdS, CdSe and CdTe have lower band gaps (<2.0 eV)[196,197] and are reviewed extensively elsewhere,[198,199,200] though we note they can also act as substitutional partners with other chalcogenides to form wide-gap semiconductor alloys (see Figure 11).

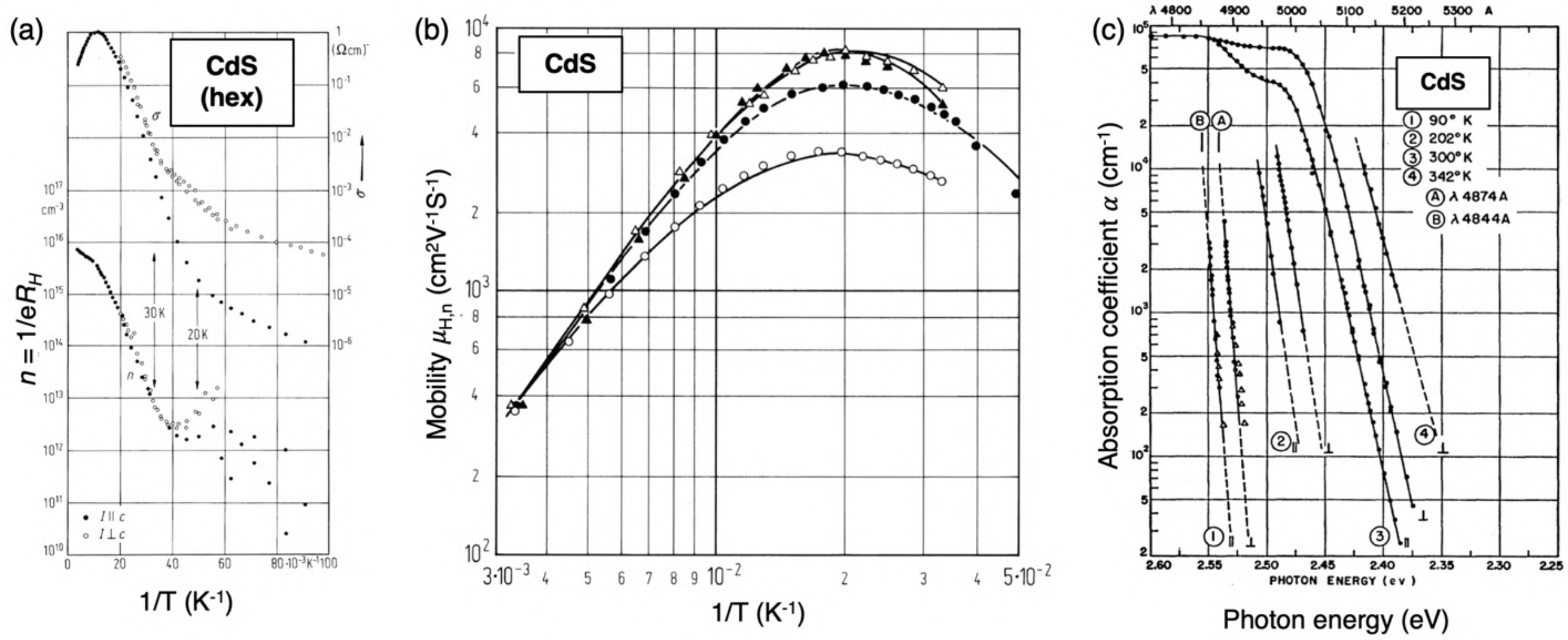

**Figure 9**. (a) N-type conductivity and Hall-derived carrier concentration, (b) electron mobility, and (c) absorption coefficient of wurtzite ("hex") CdS as a function of measurement temperature. Figures are from the "Semiconductors: Data Handbook" (page 808 and 810).[164]

### *3.1.5 BeCh*

BeS, BeSe, and BeTe all crystallize at ambient pressure in the zincblende structure. Alloyed $M_xBe_{1-x}Ch$ compounds ($M$ = V, Cr, C) have been investigated computationally for their ferromagnetism,[201,202] but not yet as intensively for electronic properties. BeS crystallizes in a zincblende phase, with an exceptionally high reported indirect gap of ~5.5 eV,[203] and high pressure wurtzite and nickeline polymorphs have been computationally predicted. BeSe is also stable in the zincblende phase, with a high reported indirect gap ranging in various studies from 4.0 to 5.6 eV.[203,204] $Be_xZn_{1-x}Se$ and (Be,Mg,Zn)Se, which can be lattice matched to Si substrates, are used as transparent buffer layers in ZnSe-based laser diodes. BeS and BeSe have wide band gaps and low effective masses, but neither intrinsic nor extrinsic dopants (e.g. alkali and N) are favorable for p-type conductivity due to compensation.[144]

Zincblende BeTe has an experimental indirect band gap of approximately 2.7 eV (bulk)[203] and 2.8 eV (epitaxial thin film), and a direct gap transition of ~4.1 eV at room temperature. It has been shown to be intrinsically p-type, has been p-doped in the literature by $N_{Te}$ to conductivities up to ~86 S $cm^{-1}$, corresponding to mobilities up to 20 $cm^2 V^{-1} s^{-1}$ and hole concentrations of $2.7 \times 10^{19}$ in MBE films.[205] Due to its high conductivity, BeTe (or pseudograded superlattices of BeTe-ZnSe) is used as a p-type contact in ZnSe-GaAs-based green laser diodes. BeTe is advantageous due to its high-lying valence band, aligned to within 100 meV with GaAs, as well as a near lattice-match with GaAs and ZnSe.[205] BeTe has been explored as an alloy partner with MgTe and MnTe for this application. It has also been theoretically predicted from a high throughput screening to be p-type due to $V_{Be}$ and p-type dopable with $Li_{Be}$, as shown in Figure 8d,[144] but this and other p-type dopants have not yet been confirmed experimentally. No reports of n-type doping of BeTe could be found.

### *3.1.6 CaCh*

CaS crystallizes in the rocksalt structure with a very wide indirect experimental gap of ~4.5 eV, which increases with Bi doping to 4.8 eV (bulk) and 5.2 eV (nanocrystalline).[206] CaS has been studied as a solid electrolyte for its ionic conductivity.[207] (CaS-SrS):$Eu^{2+}$ alloys have been used as phosphors for LEDs, where CaS is the host for luminescence center $Eu^{2+}$.[208] La-doping yields a highly resistive n-type semiconductor.[209] Rocksalt CaSe has a reported direct optical gap of 5 eV,[210] though computations have predicted a wide range of direct and indirect gaps. No investigation of electrical properties could be found. Rocksalt CaTe also exhibits an indirect gap, with an experimental absorption edge around 4.5 eV, and was very recently predicted computationally to be a potential p-type transparent semiconductor with a low hole effective mass.[211] This study acknowledged that $Te_{Ca}$ antisites could lead to hole compensation, but identified a possible extrinsic p-type dopant, $Na_{Ca}$, that should have a lower formation energy than $Te_{Ca}$. Using electron-phonon coupling calculations, the mobility of CaTe was computationally predicted to be ~20 $cm^2\ V^{-1}\ s^{-1}$. Related alkali-earth chalcogenide materials, including SrS and BaS, have been also studied as wide-gap semiconductors, though no conductivity reports could be found.[212,213]

### *3.1.7 II-Ch alloys and composites*

The II-*Ch* materials we just discussed can be alloyed with other isostructural binaries, swapping out cations or anions. Bowing in such binary alloy systems has been studied in depth (see Figure 11). Band gaps in isostructural $A_xB_{1-x}$ alloys can be estimated using Vegard's law with an applied bowing correction:

$$E_G(x) = xE_G(A) + (1-x)E_G(B) - bx(1-x) \quad 6)$$

where $E_G(A)$ and $E_G(B)$ are the band gaps of the end points $A$ and $B$, and $b$ is the bowing parameter. One of the most technologically important alloys in this space is $Cd_{1-x}Zn_xS$, and other representative alloys include $Cd_{1-x}Co_xS$[214] and $Cd_{1-x}Mn_xS$ (note the presence of magnetic cations). Alloys can also be formed between isostructural compounds with different cations and anions, e.g. $Zn_xMg_{1-x}S_ySe_{1-y}$,[165] or with their oxide counterpart, e.g. $Cu_{2-2x}Zn_xO_{1-y}S_y$.[215,216] For example, n-type $ZnO_xS_{1-x}$ has been researched heavily as a front contact to CIGS solar cells.

An example of a II-*Ch* mixed with a non-isostructural and non-isovalent binary is the ternary space of $Cu_xZn_{1-x}S$ heterovalent heterostructural alloys, also denoted as $Cu_xS$:ZnS phase-separated composites past the solubility limit of Cu. This system has recently garnered attention due to its favorable combination of transparency and conductivity, and tunability of these properties, at low processing temperatures (25–100ºC). Cu mixing allows for hybridization with S 3*p* orbitals in the valence band.[217] Recent combinatorial sputtering across the full $Cu_xZn_{1-x}S$ cation alloy space (0<x<1) demonstrated the stabilization of a metastable wurtzite $Cu_xZn_{1-x}S$ alloy between two cubic binary end-points.[218] With 25° C pulsed laser deposition, ZB and WZ $Cu_{0.3}Zn_{0.7}S$ alloy thin films have hole conductivities up to 40 S $cm^{-1}$, hole mobilities up to 1.4 $cm^2$ $V^{-1}$ $s^{-1}$, and a direct band gap of ~3.1 eV, while in composite $Cu_xS$:ZnS zincblende CBD films conductivity increases to 1,000 S $cm^{-1}$ and the "gap" decreases to ~2.4 eV (note: band gaps are not well-defined in composites). This is one of the highest reported p-type conductivities for a wide-gap chalcogenide. Such CBD films have been incorporated as the heterojunction emitter layer in Si solar cells and have been studied as TFTs (see Section 4).[219] These studies demonstrate that exploring combinations of a wide-gap semiconductor (e.g. ZnS) and highly conductive semiconductor (e.g. $Cu_xS$) could be a promising design route for discovering new tunable, low thermal budget, wide band gap p-type semiconductors.

## 3.2 Other binary $M_xCh_y$ chalcogenides

$SnS_2$ is a layered material stacked via Van der Waals forces,[220] and typically crystallizes in a $P\bar{3}m1$ space group in the 2H polytype with a stacking sequence *X-M-XX-M-X*.[164] $SnS_2$ has been reported in single crystals with a forbidden indirect gap of ~2.1–2.2 eV and a direct gap of 2.9 eV, though gaps have been reported in the literature between those values. A direct band gap of 2.2–2.4 eV is observed in $SnS_2$ thin films deposited by successive ionic layer adsorption and reaction,[221] but the values differ in the "Semiconductors: Data Handbook" (2.07 eV indirect, 2.88 eV indirect).[164] The n-type conductivity of $SnS_2$ is reported to range widely from ~1–$10^{-7}$ S $cm^{-1}$, with carrier concentrations $10^{13}$–$10^{18}$ $cm^{-3}$ and electron mobilities in the 15–52 $cm^2$ $V^{-1}$ $s^{-1}$ range, depending on materials quality and deposition method.[222–224] It has been hypothesized using DFT defect calculations that sulfur vacancies $V_S$ act as donors to dominate n-type conductivity,[223] while in CVT grown crystals $Cl_S$ impurities are the dominant donors.[225] Carriers perpendicular to the c axis (in plane) have normal lattice scattering (phonon scattering), while those parallel to c axis (out of plane) conduct via activated hopping. P-type doping has been reported by adding excess sulfur, but conductivity is only <$10^{-7}$ S $cm^{-1}$ and only one report of p-type $SnS_2$ exists, to our knowledge.[225] A high computed ionization potential of 9.54 eV and electron affinity of 7.30 eV suggest application of $SnS_2$ as an n-type buffer layer to thin film solar cells would be difficult (see Figure 24).

$In_2S_3$ is stable at room temperature as β-$In_2S_3$, an ordered defect spinel structure that crystallizes in a $I4_1/amd$ space group, though a higher temperature phase and an amorphous phase have also been reported. The energy gap of β-$In_2S_3$ polycrystalline thin films has been reported within the range of 1.80–2.75 eV.[164,226,227] This discrepancy in the literature spans from whether β-$In_2S_3$ is an indirect semiconductor, and from the possibility of mixed phases or off-stoichiometry. The conductivity mechanism has been studied in single crystal films, with undoped n-type conductivity of $2 \times 10^{-4}$ S $cm^{-1}$ at room temperature. It has been self-doped with S, with the band gap increasing to 2.43 eV in $In_2S_{3.9}$. Extrinsically, it has been substituted with $O_S$ to raise the gap,[228] and doped n-type with $Sn_{In}$ to increase conductivity (~30 S $cm^{-1}$),[229] as well as $Na_{In}$,[230] $V_{In}$,[231] and $Mn_{In}$[232] by a variety of physical and chemical synthesis methods. No reports of p-type doping could be found. $In_2S_3$ has been investigated as a replacement buffer layer to CdS in CIGS, CdTe and CZTS solar cells (see Section 4.1).

Wide-gap III-Se semiconductors have historically been studied for their close lattice match to Si and for Si passivation, but recent work suggests them as possible p-type TCs. $Al_2Se_3$ crystallizes in a tetrahedrally coordinated hexagonal *Cc* structure and has a wide band gap of 3.1 eV.[233] Defect calculations predicted potential p-type conductivity in $Al_2Se_3$ with $Mg_{Al}$ and $N_{Se}$ doping,[144] and this material has a low predicted hole effective mass of 0.56. Layered GaSe (*P*6₃/*mmc*) is also found to have a low planar effective mass of 0.25, and a mixture of Ga and Se *p*-states comprising the VBM states. Defect formation energy calculations suggest $Zn_{Ga}$ and $P_{Se}$ as appropriate dopants, but not alkalis nor $N_{Se}$ doping. Experimental measurements reveal a gap of 2 eV. Ambipolar doping has been reported in single crystals, presumably undoped p-type and Sn-doped n-type with hopping conductivity in both cases,[234] but doping has not been explored in depth.

In Table 1 we tabulate all the experimentally investigated binaries mentioned above and other wide-gap binary chalcogenide semiconductors that merit further investigation into dopability. We list their structure type, gap, doping information, and electronic properties, as reported in the references above and the "Semiconductors: Data Handbook."[164] Computationally predicted wide-gap compounds with p-type conductivity are tabulated in Table 4. Other wide-gap binary chalcogenides with electronic properties that have not been measured and computed materials that have not been confirmed p-type are tabulated in the Supporting Information.

**Table 1**. Experimentally investigated wide-gap binary chalcogenides, and their reported optoelectronic properties.

| Compound | Structure type | Band gap (eV) | Carrier type | Dopant | Conductivity (S cm$^{-1}$) | Refs. |
|---|---|---|---|---|---|---|
| ZnS | ZB | ~3.7–3.8 | n | $Al_{Zn}$ | ~$10^{-3}$ | 135 |
| | | | | $F_S$ | ~$2 \times 10^{-7}$ | 137 |
| ZnS | WZ | ~3.9 | p | $Cu_{Zn}$ | ~$10^{-5}$ | 136 |
| ZnSe | ZB | ~2.7 | n | $Ga_{Zn}$ | 20 | 149 |
| | | | | $Cl_{Se}$ | ~333 | 150 |
| | | | p | $N_{Se}$ | 0.06 | 148 |
| ZnTe | ZB | 2.3 | p | $N_{Te}$ | 25 | 153 |
| | | | | $Cu_{Zn}$ | ~0.33 | 154 |
| | | | | $Sb_{Zn}$ | ~30 | 155 |
| MgS | ZB | ~4.5 | — | — | — | 164 |
| MgS | WZ | ~4.87 | — | — | — | 164 |
| MgSe | ZB | ~3.6–4.05 | — | — | — | 164 |
| MgTe | ZB | ~3.5 | — | — | — | 67 |
| | WZ | ~3 | — | — | — | 67 |
| MnS | WZ | 3.88 | p | $V_{Mn}$[172] | ~$10^{-5}$ | 173 |
| | ZB | 3.8 | — | — | — | 170 |
| | RS | 2.8–3.2 | — | — | — | 164 |
| MnSe | RS | 2.5 | — | — | — | 175, 176 |
| | ZB | 3.4 | — | — | — | 177 |
| | WZ | 3.5–3.8 | — | — | — | 178 |
| MnTe | NC | 1.3 | p | — | ~5–6 | 41 |
| | ZB | 3.2 | — | — | — | 179 |
| | WZ | 2.7 | — | — | — | 41 |
| CdS | WZ | 2.5 | n | $In_{Cd}$ | ~50 | 182 |
| | | | p | $Cu_{Cd}$ | 2 | 193,194 |
| | ZB | ~2.3–2.4 | — | — | — | 185 |
| | RS | ~1.5–1.7 | — | — | — | 186 |
| BeS | ZB | ~5.5 | — | — | — | 203 |
| BeSe | ZB | 4.0–5.6 | — | — | — | 203,204 |
| BeTe | ZB | 2.7–2.8 | p | $N_{Te}$ | 86 | 205 |
| $SnS_2$ | layered | ~2.1–2.2 | n | $V_S$ | ~1 to $10^{-7}$ | 222–224 |
| | $P\bar{3}m1$ | 2.9 | p | S-doped | $10^{-7}$ | 225 |
| $In_2S_3$ | defect spinel $I4_1/amd$ | 1.80–2.75 | n | intrinsic | $2 \times 10^{-4}$ | 228 |
| | | | | $Sn_{In}$ | ~30 | 229 |
| $TaS_2$ | layered $P6_3/mmc$ | 2.3 | n | — | ~100 | 164 |
| $HfS_3$ | monoclinic $P2_1/m$ | 3.1 | p | — | 0.01 | 164 |
| γ-$Gd_2S_3$ | orthorhombic *Pnma* | 3.4 | n | — | 0.004 | 164 |

## 3.3 Ternary chalcopyrite I-III-$Ch_2$ compounds

We now turn our attention to ternary wide-gap chalcogenide semiconductors. In the $I_B^{3+}$-$III_{1+}$-$Ch_2^{2-}$ ternary chalcopyrite category (simplified as I-III-$Ch_2$ herein), the group $I_B$ transition metal cation is $Cu^{1+}$ or $Ag^{1+}$, group III metal cation is $Al^{3+}$, $Ga^{3+}$, or $In^{3+}$, and the chalcogen anion $Ch^{2-}$ is $S^{2-}$, $Se^{2-}$, or $Te^{2-}$. Wide-gap chalcopyrites have been investigated in depth for over fifty years for their intriguing properties and applications.[235–238] Their name stems from the chalcopyrite mineral $CuFeS_2$, and their structure is common for P and As anions as well as chalcogens. **Figure 10**a shows the crystal structure of a representative compound, $CuAlS_2$ with tetragonal $I\bar{4}2d$ space group, consisting of tetrahedral coordination of anions and cations. The chalcopyrite structure is an isoelectronic ternary analog of the *MCh* zincblende structure, described previously, with a unit cell twice as large and with metal species I and III alternating between sites. It is distinct from the *MCh* zincblendes due to its cation ordering and resulting tetragonal distortion, resulting from unequal I-*Ch* and III-*Ch* bond lengths, and quantified by an anion displacement parameter $\mu$.[239] The chalcopyrites have *p-d* and *s-p* hybridization at the VBM and CBM of I-III-$Ch_2$, respectively, compared to a single orbital *p* and *s* character bands in II-*Ch*. This is demonstrated in Figure 10b and Figure 10c for representative structures of zincblende ZnS and chalcopyrite $CuGaS_2$.

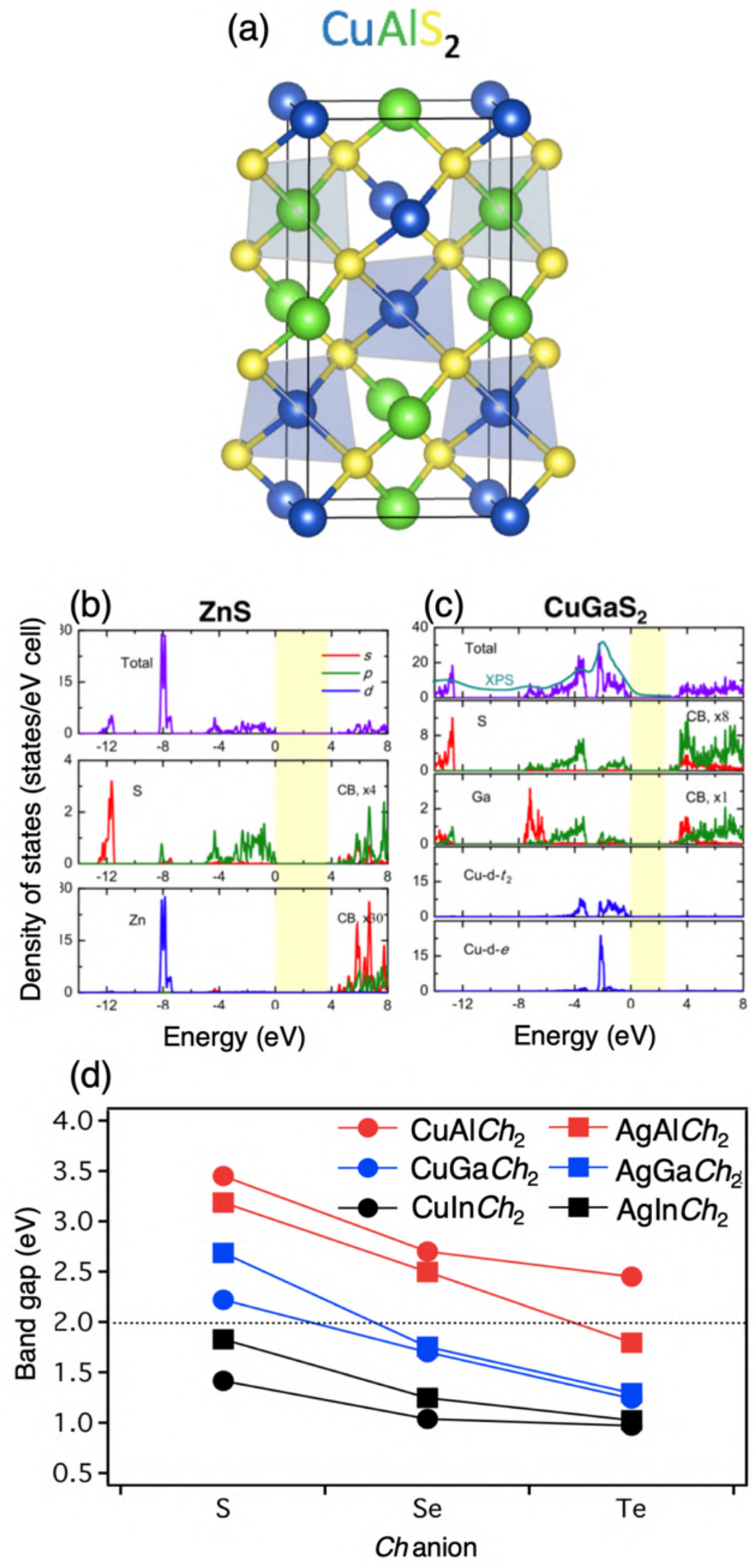

**Figure 10**. Illustration of (a) the chalcopyrite structure, represented by $CuAlS_2$, and the DFT+U+$G_0W_0$ density of states from Zhang et al.[240] of (b) zincblende ZnS compared to (c) chalcopyrite $CuGaS_2$, indicating *p-d* hybridization at the VBM in the chalcopyrite structure. (d) Experimentally measured band gaps of I-III-*Ch*$_2$ semiconductors.

Structural distortions, *p-d* hybridization, and cation electronegativity differences cause gaps of I-III-*Ch*$_2$ to be significantly lower than their binary counterparts.[239] However, these three features lead to a high propensity for p-type doping, and even ambipolar doping in many cases. The VBM *p-d* hybridization and the resulting exceptionally high dispersion at the VBM enables high hole mobilities.[240] For example, the hole effective mass of $CuAlS_2$ is approximately ten times lower than that of delafossite $CuAlO_2$ (the prototypical p-type TCO), and that of $CuAlSe_2$ is even lower. Similarly, the III-*Ch* bond distortion creates *s-p* hybridization at the CBM, and leads to low electron effective masses and high electron conductivity if the material can be doped n-type. The chalcopyrite structure is highly tolerant to off-stoichiometry, facilitating intrinsic doping and the tuning of properties.[241] Intrinsic p-type dopants are typically group I vacancies ($V_I$) and group I on group III antisites ($I_{III}$), while intrinsic n-dopants may be chalcogen vacancies ($V_{Ch}$) and group III on group I antisites ($III_I$), assuming they are sufficiently shallow. Additionally, it has been postulated that cation-terminated grain boundaries in chalcopyrites actually help increase p-type mobility in polycrystalline samples.[242]

Considering all the ion combinations discussed, the I-III-*Ch* system contains 18 possible chemistries. Figure 10d displays the experimentally-derived optical band gap of these materials as a function of chalcogenide anion (representative reports listed in Table 2). Generally, the band gap magnitude is positively correlated with the electronegativity of the *Ch* anion, with the electronegativity difference between the two cations, and with atomic number (e.g. decreases along Al, Ga, In). The various combinations of cations also influence the degree of intrinsic localization and covalent bonding tendencies of the Cu 3*d* (and Ag 4*d*) states, e.g. shorter Cu-S bonds lead to increased *p-d* VB hybridization and decreased hole effective masses.[240] Here, we summarize the eight stable semiconductors in this system whose reported experimental band gaps are above 2 eV: $CuAlS_2$, $CuAlSe_2$, $CuAlTe_2$, $CuGaS_2$, $AgAlS_2$, $AgAlSe_2$, $AgAlTe_2$, and $AgGaS_2$.

### 3.3.1 Cu(Al,Ga)$Ch_2$

$CuAlS_2$ (see Figure 10a) has the widest band gap (~3.4–3.5 eV) among all the I-III-$Ch_2$ chalcopyrite compounds, and hence is the most transparent.[164] It was first investigated as a blue-UV LED material,[243,244] then later as a transparent semiconductor,[30] and has been synthesized using a wide variety of bulk and thin film techniques.[244–248] The p-type conductivity of undoped, near-stoichiometric $CuAlS_2$ only reaches 0.9 S $cm^{-1}$ (bulk)[249] and 0.016 S $cm^{-1}$ (thin film).[250] However, altering the I/III cation ratio allows for drastic changes in electronic properties. Under Cu-rich and Al-deficient conditions ($Cu_{1+x}Al_{1-x}S_2$), the conductivity of bulk $CuAlS_2$ can reach up to 250 S $cm^{-1}$ with high mobilities up to 21 $cm^2$ $V^{-1}$ $s^{-1}$.[30] Similarly, bulk growth under S-rich conditions can increase conductivity to 4.6 S $cm^{-1}$, with lower mobilities of 0.4 $cm^2$ $V^{-1}$ $s^{-1}$.[11] Antisite $Cu_{Al}$ is a doubly ionized acceptor, but at near-stoichiometry its transition level ε(0/−) is deep and should trap holes.[251] To qualitatively explain high off-stoichiometric conductivities, it has been calculated that the increased contribution of Cu 3*d* states to the VBM shifts the VB upward, decreasing the transition level and allowing for more shallow acceptor levels of $Cu_{Al}$.[252] This also decreases Cu-S bond length, further delocalizing the VBM. An alternative possible explanation is the presence of hard-to-measure $Cu_xS$ impurities at Cu-rich synthesis conditions. In addition to intrinsic doping, extrinsic doping has been effective in enhancing conductivity. $As_S$, $Zn_{Al}$, and $Mg_{Al}$ are found to be excellent p-type dopants for $CuAlS_2$, with conductivities reaching 1 S $cm^{-1}$ (thin film),[244] 63.5 S $cm^{-1}$ (thin film),[253] 41.7 S $cm^{-1}$ (bulk),[254] respectively. $CuAlS_2$ has also been reported n-type dopable by $Cd_{Cu}$, $Al_{Cu}$, and $Zn_{Cu}$ at high growth temperatures.[243] It is important to note that the optical band gap drops when p-doping the Al site, likely due to the increased VBM energy and presence of undetected $Cu_xS$ impurities, but does not under Cu-poor and S-rich conditions.[11,249]

$CuAlSe_2$ has a reported experimental band gap at near-stoichiometry of ~2.6–2.7 eV, and has been investigated for multiple applications, e.g. blue LEDs, solar cells, and optical filters.[255] Similar to $CuAlS_2$, $CuAlSe_2$ is ambipolar dopable depending on the synthesis method and dopants. For example, $CuAlSe_2$ thin films grown by co-evaporation have been reported p-type at near-stoichiometry with conductivity $3.3 \times 10^{-3}$ S $cm^{-1}$ and mobilities up to 18 $cm^2$ $V^{-1}$ $s^{-1}$. Cu-rich conditions increased p-type conductivity nearly five orders to 123 S $cm^{-1}$, dropped mobility to 0.76 $cm^2$ $V^{-1}$ $s^{-1}$ at hole concentrations of $10^{21}$ $cm^{-3}$, and lowered the gap to 2.5 eV.[256] This corroborates the expectations of shifting the $\mu$ parameter, increasing the degree of *p-d* hybridization, and pushing the valence levels upwards as discussed previously.[256] Al-rich conditions convert conductivity to n-type (2.3 × $10^{-3}$ S $cm^{-1}$) and raise the gap to 2.87 eV, and a report of epitaxial MBE films also found intrinsically n-type conductivity (50 S $cm^{-1}$) [256,257]. Reported n-type extrinsic dopants are $Zn_{Cu}$ and $Cd_{Cu}$, though $V_{Cu}$, $Zn_{Al}$ and $Cd_{Al}$ p-type compensation causes systematic uncertainties.[258]

$CuAlTe_2$, with reported gaps of ~2.1–2.5 eV, is p-type with typical high resistivities (~$10^{-3}$ S $cm^{-1}$) and mobilities of ~5–6 $cm^2$ $V^{-1}$ $s^{-1}$.[259] Conductivity reportedly can reach up to ~10 S $cm^{-1}$ at room temperature,[260] however its electronic properties have not been as heavily investigated as $CuAlS_2$ and $CuAlSe_2$. The growth of $CuAlTe_2$ thin films by RF sputtering is p-type as-deposited,[261] but convert to n-type after a 140°C anneal presumably due to acceptor-donor compensation.[262] $CuAlTe_2$ was found to form a good Ohmic contact to Mo, and has been investigated computationally and experimentally for thermoelectrics and solar cells, among other applications.[262,263]

$CuGaS_2$ is a promising material for green LEDs, since its band gap is approximately 2.22–2.55 eV.[164,264,265] Reported p-type conductivities of bulk, thin film, and single crystal $CuGaS_2$ samples are 1.7 S $cm^{-1}$ (660°C anneal),[266] 0.7 S $cm^{-1}$,[267] and 0.83 S $cm^{-1}$,[265] respectively. Mobilities have been reported in the single crystal sample up to 15 $cm^2$ $V^{-1}$ $s^{-1}$, with hole concentration 4 × $10^{17}$ $cm^{-3}$. This study revealed activated carrier transport in the temperature range of 5–300 K, and an activated hopping mechanism at temperatures less than 100 K.[268] $CuGaS_2$ has been predicted intrinsically p-type and not ambipolar, and no n-type reports have been found.[269] We note that $CuGaS_2$ has also been recently investigated as an intermediate band solar cell (IBSC) absorber, with $Fe_{Ga}$, $Cr_{Ga}$, Ti, $Ge_{Ga}$, and $Sn_{Ga}$[270,271] deep defects. $CuGaSe_2$ and $CuGaTe_2$ have low gaps of 1.68 eV and 1.24 eV, respectively,[272] but can serve as alloy partners (see Figure 11). We note that $I\bar{4}2d$ chalcopyrite $CuBS_2$ has been computationally predicted as a promising p-type transparent conductor, with a HSE06 direct gap of 3.41 eV, $m_h^*$ of 1, and defect calculations suggesting p-type dopability.[273]

### 3.3.2 Ag(Al,Ga)$Ch_2$

$AgAlS_2$ has a comparable band gap (~3.2 eV) to $CuAlS_2$,[274] though CBD synthesis yields a gap of 2.3 eV. However, it has been reported as unstable in air, which limits its practical applications.[275] There are few studied of its electrical properties, but it has potential to be alloyed with other chalcopyrite semiconductors. We also note that mixed ionic-electronic conductivity in this system (and in $AgGaS_2$) has been reported.[276] The experimental gap of $AgAlSe_2$ ranges from 2.5 to 2.7 eV,[277] but there are no studies of its n- or p-dopability to our knowledge. Computations predict deep extrinsic dopants $Ge_{Al}$ and $Sn_{Al}$ for use as an IBSC absorber.[278] $AgAlTe_2$ has an band gap energy of ~2.3 eV, in agreement with computational predictions, and has also garnered interest as an IBSC absorber.[279]

$AgGaS_2$, with a gap of ~2.7 eV,[280] is considered a semi-insulating material due to its extremely low reported p-type conductivity of <$10^{-5}$ S $cm^{-1}$,[281,282] though amorphous films have conductivities ~1 S $cm^{-1}$.[283] $AgGaS_2$ has been reported with both p-type and n-type conductivity, with low dopings and mobilities up to 30 $cm^2$ $V^{-1}$ $s^{-1}$ at room temperature.[284] Under S-rich conditions, $V_{Ag}$ or $V_{Ga}$ lead to p-type doping, while in S-poor conditions (e.g. annealing in vacuum) $V_S$ defects lead to n-type doping. The proposed dominant scattering mechanism is neutral impurity or acoustic phonon scattering.[282] We note that this compound has the highest effective mass out of all of the I-III-$Ch_2$ chalcopyrites. $AgGaSe_2$ and $AgGaTe_2$ have gaps below 2 eV. Replacing Ag with Li to form LiAlTe$_2$ results in a compound with an HSE03 gap of 3.11 eV and $m_h^*$ of 0.73, though dopability has not been studied.[211]

### 3.3.3. I-III-*Ch*$_2$ alloys

The I-III-*Ch*$_2$ semiconductors can form various multinary alloys to stabilize polymorphs, and to tailor the band gap, conductivity, and band edges towards a particular application. To assist in visualization of this alloy space, **Figure 11** plots the band gap vs. lattice constants for all of the I-III-*Ch*$_2$ chalcopyrite structures discussed previously, as well as a few others with gaps below 2 eV. Plotted alongside are the II-*Ch* zincblende compounds discussed in Section 3.1, and binary zincblende III-V semiconductors GaP, GaAs, and InP for reference. Note that the effects of band gap bowing in chalcopyrites are not plotted, but many of these band gap dependencies have been shown computationally and experimentally to be close to linear with bowing parameters usually less than 0.5 eV (see below).[239,285,286] Also, alloy band gaps are only drawn here between isostructural systems with one substitution, though non-isostructural and multicomponent alloying are regularly utilized across this space, e.g. zincblende (Be,Mg,Zn)Se used for UV lasers[287] and quinternary alloys $CuAl_xGa_{1-x}(S_{1-y}Se_y)_2$.[288] We briefly discuss three strategies – alloying I, III, or *Ch* ions – and mention representative chalcopyrite alloys.

**(Cu,Ag)-III-*Ch*$_2$.** By alloying Cu-III-*Ch*$_2$ with Ag-III-*Ch*$_2$, one can obtain $Cu_xAg_{1-x}$III-*Ch*$_2$. For example, $Cu_xAg_{1-x}AlS_2$ polycrystalline thin films prepared by chemical spray pyrolysis at 360 °C have been reported, albeit with a tendency to oxidize due to residual oxidant in the precursor. It was observed that the optical band gap changes nonlinearly with respect to x, though bowing parameter was not reported.[289] Structural and optical properties of bulk $Cu_xAg_{1-x}GaS_2$ materials have been investigated, with a large bowing parameter of 0.8 eV.[290]

**I-(Ga,Al)-*Ch*$_2$.** Alloys of $CuGa_{1-x}Al_xS_2$ were found experimentally to fit Vegard's law, with a near-linear gap vs. lattice constant relation and a bowing parameter of 0.34 eV.[291,292] $CuGa_{1-x}Al_xSe_2$ has been more intensively investigated for its absorption, and has a reported bowing parameter of 0.28 eV.[247,286] In particular, $Co_{Cu}$ doping has been explored to alter absorption.[293,294] This composition has also been deposited as a layered heterostructure, which is another strategy to achieve a mixture of properties.

**I-III-(S,Se,Te)$_2$.** $CuAl(S_{1-x}Se_x)_2$ single crystals have a reported bowing parameter of 0.34 eV, while that of $CuGa(S_{1-x}Se_x)_2$ is approximately zero.[286] Both systems have been studied as solar absorbers. Transport properties of $CuAl(S_{1-x}Se_x)_2$ have been computed for use in thermoelectrics, finding p-type conductivity and a low theoretical power factor.[295]

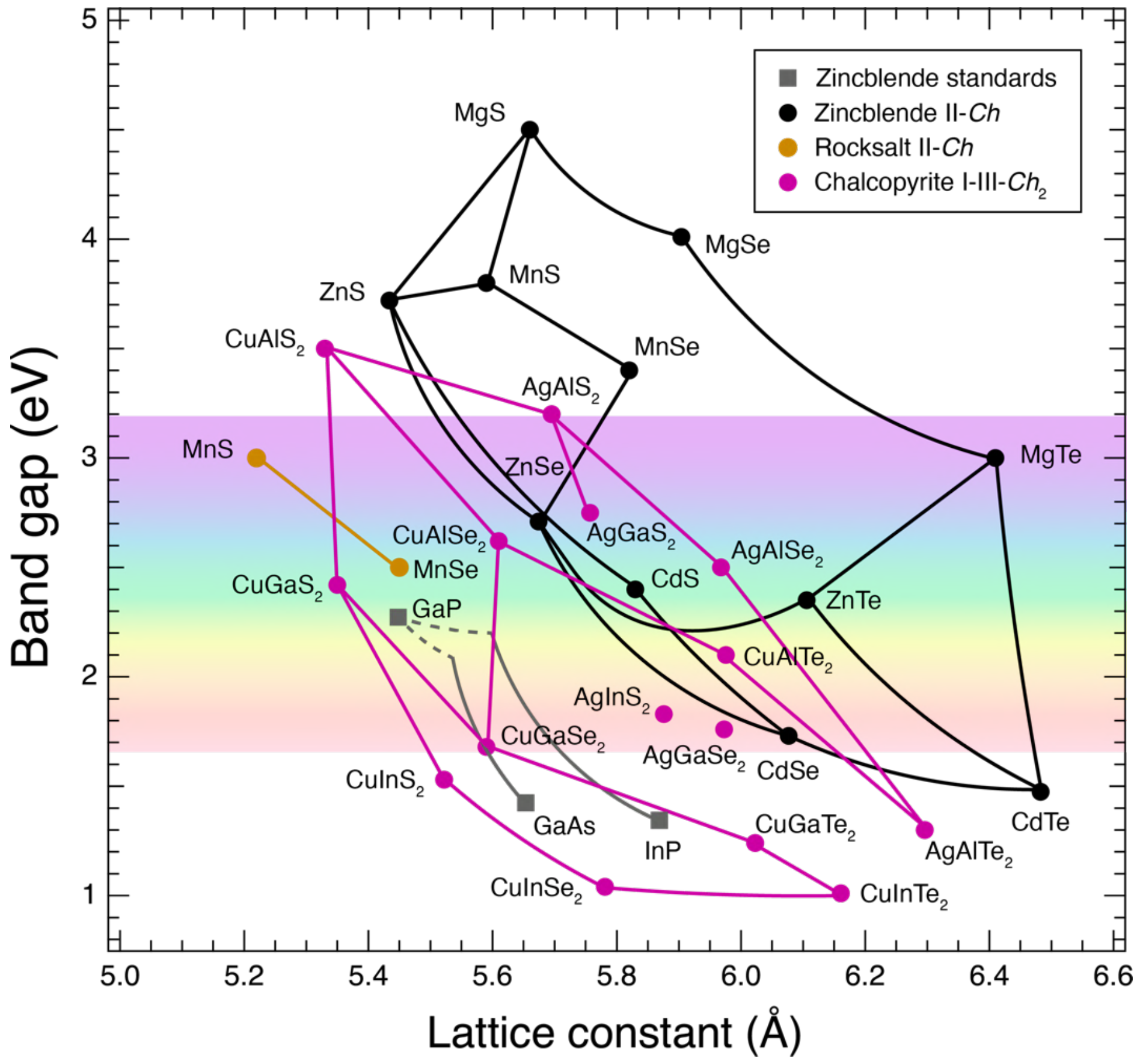

**Figure 11**. Room temperature band gap vs. lattice constant for the zincblende II-VI chalcogenides, rocksalt II-VI chalcogenides, and ternary I-III-$VI_2$ chalcopyrite chalcogenides discussed in this review. A few binary standards are plotted for reference (dashed lines indicate structural phase transformations), and we include some structures below 2 eV to help conceptualize the alloy space. Bowing is plotted for the binary systems to align with literature reports.[296] Bowing is not included for ternary systems, though it has been studied in-depth for many ternary chalcopyrites.[286] We note that the zincblende structure is not the most thermodynamically stable polymorph for several of these systems (e.g. MnS, MnSe, CdS), but we plot zincblende for ease of visualization. Lattice constant data used to make this plot is from the Inorganic Crystal Structure Database (ICSD),[35,36] and is discussed and tabulated in the Supporting Information.

## 3.4 Other Ternary Chalcogenides

### *3.4.1 α-$BaM_2Ch_2$, Pnma structure*

$BaCu_2S_2$ exists in two structures, a low temperature orthorhombic structure (*Pnma*, α-$BaCu_2S_2$)[297] and high temperature layered tetragonal structure (*I*4/*mmm*, β-$BaCu_2S_2$).[298] The $BaCu_2S_2$ reviewed here as a wide band gap semiconductor is the α phase, and we note that β-$BaCu_2S_2$ is primarily investigated for its thermoelectric properties.[299] As shown in **Figure 12**a, Ba occupies 7-fold S-coordinated sites while Cu atoms are tetrahedrally coordinated by S. In contrast with the layered delafossite structure, Cu-centered tetrahedrons are uniformly distributed in the 3D crystal. $BaCu_2S_2$ has a direct gap (see Figure 12b) reported experimentally between 2.1-2.5 eV in thin films.[12,300,301] $BaCu_2S_2$ has been synthesized by RF sputtering, reaching hole mobilities of 3.5 $cm^2\ V^{-1}\ s^{-1}$ at dopings of ~$10^{19}$ $cm^{-3}$, and a conductivity of 17 S $cm^{-1}$,[301] as well as by spin coating, reaching higher dopings of ~$4 \times 10^{20}$ $cm^{-3}$ and conductivities up to 33.6 S $cm^{-1}$. This high p-type conductivity has been explained due to (1) short (~2.71 Å) Cu-Cu distances along a one dimensional chain, allowing for a broad band along which hole conductivity takes place (see Figure 12c),[300] (2) hybridization of S 3*p* and Cu 3*d* states at the upper VB, resulting in low hole effective mass of ~0.8,[302] (3) holes from $V_{Cu}$ acceptors, and (4) a notably high ionization energy compared to other p-type chalcogenides (4.84 eV).[303] The p-type conductivity is limited by $V_S$ donor defects.[304] Combinatorial RF sputtering was used to explore the influence of off-stoichiometries and processing temperature on $BaCu_2S_2$ properties, as shown in Figure 12d,[12] finding conductivity to increase from 0.1 to 53 S $cm^{-1}$ as the film transforms from amorphous to crystalline at a processing temperature of 250 °C. Higher processing temperatures lead to a drop of conductivity and increase in absorption. Both Ba and Cu off-stoichiometry can increase the conductivity at high processing temperatures, though degenerate doping limits this effect at lower temperatures. The combination of reasonably high transparency, high p-type conductivity, and moderate processing temperatures makes α-$BaCu_2S_2$ a promising wide band gap chalcogenide semiconductor.

The semiconducting properties of $BaCu_2Se_2$ and $BaCu_2Te_2$ have also been characterized, but they are not discussed here due to their gaps below 2 eV.[305] $BaAg_2S_2$ (*I*4/*mmm*, not to be confused with its $P\bar{3}m1$ polymorph) has been calculated with HSE to have a gap just greater than 2 eV, but its mobility and conductivity have not been explored experimentally.[303] Other wide-gap compounds have been computationally predicted in this structural family including $SrCu_2S_2$ (HSE gap 2.27 eV) and $SrCu_2Se_2$ (HSE gap 2.03 eV), with formation energies just above the convex hull ($E_{hull}$ of 0.054 eV/atom and 0.027 eV/atom, respectively) according to theoretical calculations,[303] but are difficult to synthesize.[12] It is possible that the Cu-Cu or Ag-Ag chains and hybridized VBMs could lead to high hole conductivities in these alternative structures as well.

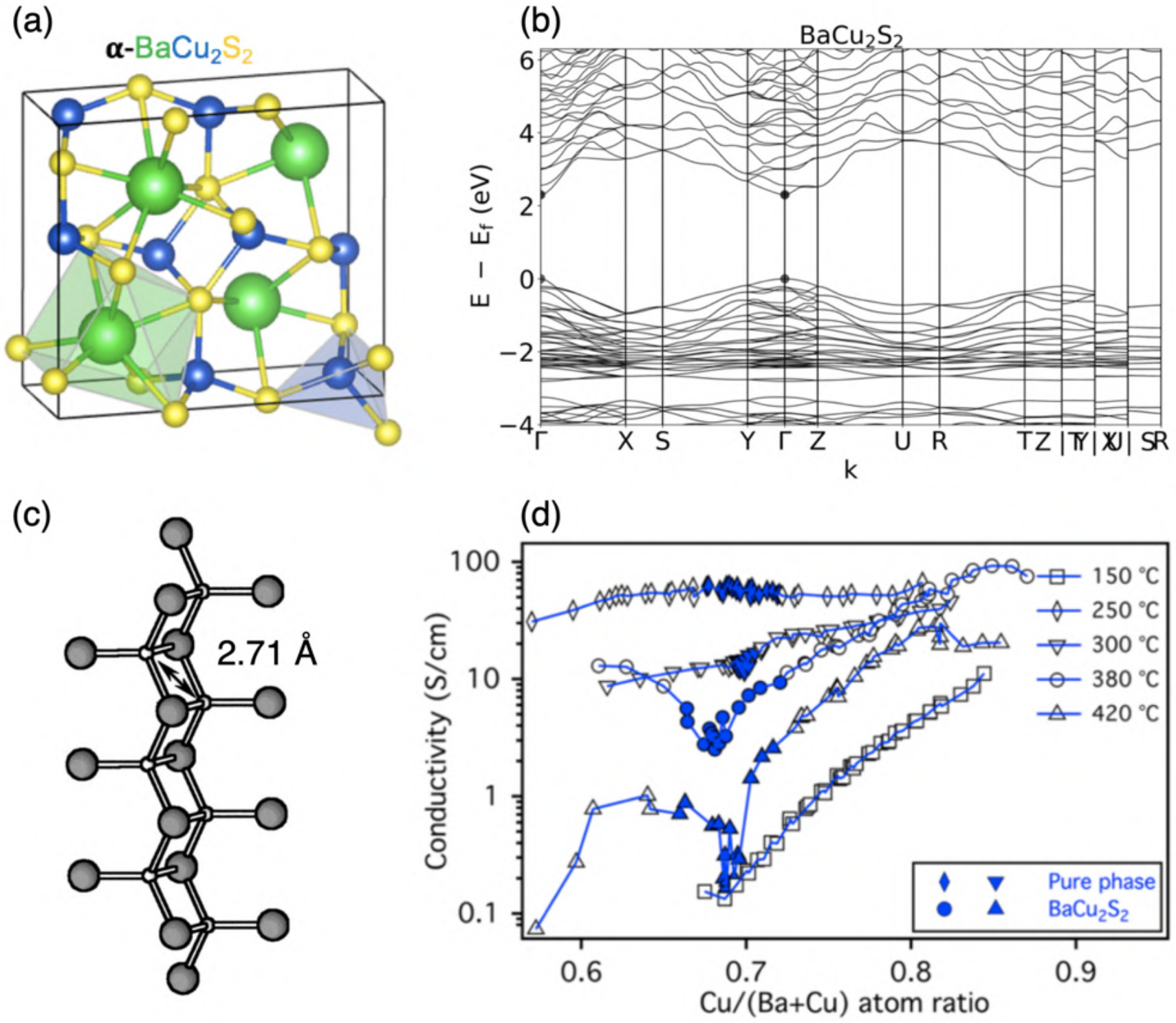

**Figure 12**. (a) The α-$BaCu_2S_2$ unit cell, (b) electronic band structure of α- $BaCu_2S_2$ from the Materials Project database, with a scissor operation applied to correct for experimental band gap, (c) 1D chains of Cu-S within the structure from Park and Keszler,[300] where small circles are Cu atoms and large circles are S atoms, and (d) conductivity of α-$BaCu_2S_2$ as a function of processing temperature and Cu/(Ba+Cu) ratio.[12] Open circles indicate the presence of phase segregation within the film.

### *3.4.2 $Cu_3MCh_4$ sulvanite and sulvanite-like materials*

Compounds of the form $Cu_3MCh_4$ represent a group of copper-based chalcogenide semiconductors, where *M* is group $V_B$ transition metal V, Nb, Ta, and *Ch* is S, Se, Te. They share the cubic sulvanite structure (space group $P\overline{4}3m$), which leads to isotropic optical and electrical properties. **Figure 13**a illustrates the structure for representative compound $Cu_3TaS_4$. *M* cations are located at the corner of the unit cell, while Cu cations are edge centered. All *M* and Cu cations are tetrahedrally coordinated by *Ch* anions. This isotropic cubic structure avoids the need for special crystal substrates for epitaxial growth of common anisotropic Cu based wide band gap p-type semiconductors, which also require high growth temperatures and are difficult to make Ohmic contact to.[109] Kehoe et al. calculated indirect fundamental band gaps for this system, which decrease down the *Ch* group and up the *M* group. The band structure of $Cu_3TaS_4$ is shown in Figure 13c, and corresponds to an average electron effective mass of 1.36 and a hole effective mass of 1.01, which could lead to high hole mobilities.[306] Similar to the other Cu-based chalcogenides, Cu 3*d* states mix with *Ch np* states at the VBM, helping delocalize the hole transport. The vacant center of the cubic unit cell creates a "channel" along the (100) crystallographic plane, opening a pathway for ionic conductivity.[307] Considering the scope of this review, we focus on the wide-gap sulvanites $Cu_3TaS_4$, $Cu_3TaSe_4$, $Cu_3NbS_4$, and $Cu_3NbSe_4$.

$Cu_3TaS_4$ thin films prepared by ALD are reported to have an indirect band gap of 2.7 eV. In contrast, computed PBEsol+U gaps are 2.1 eV (indirect) and 2.6 eV (direct).[132] Polycrystalline PLD films have an absorption onset at 3 eV, hole conductivities of 1.6 S $cm^{-1}$, mobilities of 0.2–0.4 $cm^2$ $V^{-1}$ $s^{-1}$ limited by sample morphology, and hole concentrations $5 \times 10^{19}$ $cm^{-3}$. The gap of $Cu_3TaSe_4$ drops to 2.35 eV (computed indirect 1.71 eV and direct 2.22 eV).[13] Synthesized pellets confirm p-type conductivity according to Seebeck measurements, but has only been reported at ~$3 \times 10^{-3}$ S $cm^{-1}$.[308] $Cu_3NbS_4$ has a computed indirect gap of 1.82 eV and direct gap of 2.3 eV, with a reported experimental optical gap of ~2.6 eV.[309] Its effective mass is similar to $Cu_3TaS_4$, and a hole conductivity of 0.1–0.2 S $cm^{-1}$ was reported for sintered pellets.[309] Similarly, bulk $Cu_3NbSe_4$ has been synthesized from stoichiometry elements.[308,309] Its predicted gap is below 2 eV, but a direct optical gap has been reported in the literature as ~2.2 eV.[310] P-type conductivity was measured up to 1.9 S $cm^{-1}$ in pellets. Synthesis of a p-n junction has been reported by thermal deposition of $Cu_3NbSe_4$ on an n-type silicon substrate.[311] Computational alloy studies found direct and indirect band gaps to vary almost linearly with x in solid solutions of $Cu_3Ta(S_{1-x}Se_x)_4$, $Cu_3Nb(S_{1-x}Se_x)_4$, and $Cu_3Ta_{1-x}Nb_xS_4$.[308] This was confirmed in experimental studies of $Cu_3Nb(S_{1-x}Se_x)_4$, finding x = 0.4 to have the highest alloy conductivity of ~3 S/cm at room temperature.[309] Single crystal $Cu_3VS_4$ has reported p-type conductivity ranging from $10^{-3}$–10 S $cm^{-1}$,[312] with a high Hall mobility of approximately 4 $cm^2$ $V^{-1}$ $s^{-1}$. A database screening of ternary compounds predicts an HSE06 gap of 3.68,[273] but the experimentally reported gap is ~1.3 eV.[313] $Cu_3VSe_4$ and $Cu_3VTe_4$ are p-dopable but have even lower gaps.

$Cu_3PCh_4$ has the same stoichiometry as $Cu_3MCh_4$, but its crystal structure is slightly different (i.e. "sulvanite-like") and P belongs to group $V_A$. Single crystal x-ray diffraction of $Cu_3PS_4$ shows that it crystallizes in either an enargite structure or orthorhombic wurtzite-derived structure with the space group *Pmn*2 or $Pmn2_1$.[314] As in $Cu_3TaS_4$, all cations are tetrahedrally coordinated by *Ch* anions. Single crystals of $Cu_3PS_4$ (indirect optical band gap of 2.38 eV) and $Cu_3P(S_3Se)$ (indirect optical band gap of 2.06 eV) have also been prepared by chemical vapor transport, resulting in p-type conductivity of 0.2–1.0 S $cm^{-1}$, and they have been implemented as cathodes for photoelectrolysis of water.[315] In addition, by replacing Cu with Ag, it has been calculated that $Ag_3PS_4$ (band gap of 2.88 eV) and $Ag_3PSe_4$ (band gap of 2.09 eV) are potential photocatalyst candidates.[316] A compound with a similar structure, $KAg_2PS_4$, has been predicted from first principles to have a gap of 2.53 eV and $m_h^*$ of 0.77, but dopability is not known.[211]

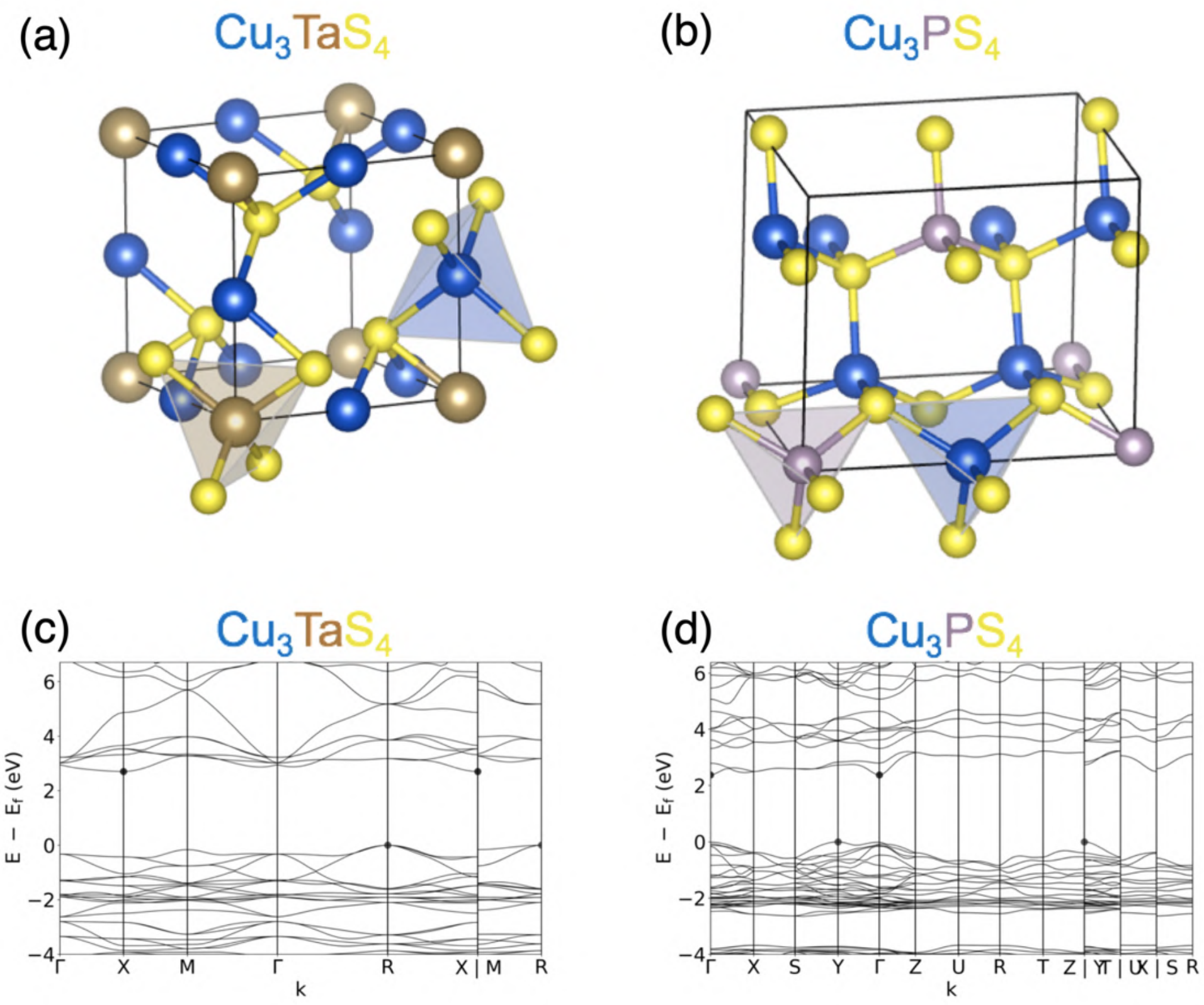

**Figure 13**. (a) Crystal structure of sulvanite $Cu_3TaS_4$, (b) sulvanite-like $Cu_3PS_4$, and (c, d) their corresponding GGA electronic band structures from the Materials Project database, with a scissor operation applied to correct for experimental band gap values.

### *3.4.3 $ABCh_2$ delafossites*

The first predicted p-type wide-gap conducting oxide was $CuAlO_2$, which crystallizes in the delafossite structure (*R*3*m* space group) and was selected based on chemical modulation of the valence band.[6] Subsequently, other delafossite oxides were synthesized as p-TCOs, including the highly conductive (albeit rather absorbing) Mg-doped $CuCrO_2$[317] and ambipolar $CuInO_2$.[318] Cu-based delafossites of the $ABO_2$ structure have been the prototypical TCO candidates due to their structural *A*-O-*B*-O-*A*-O channel for hole conductivity, hybridized delocalized VBM from Cu 3*d* and O 2*p* orbitals, and closed *d*-shells to minimize visible absorption. It is reasonable that these design parameters could translate to chalcogenides, yet to date no wide-gap delafossite chalcogenides have been investigated experimentally as transparent conductors. This may be due to polymorphism; note that $CuAlS_2$, though similar in stoichiometry, crystallizes as a chalcopyrite not a delafossite.

Anticipating promise in this class of ternary chalcogenides, assuming they can be stabilized, computationally predicted wide-gap oxide and chalcogenide delafossite compounds $ABCh_2$ ($Ch$ = O, S, Se, Te) for p-type dopability have been recently investigated.[112] This study looked not only at materials in the Materials Project, but used a global structure prediction algorithm to propose 79 new compounds. After screening for wide HSE06 gap (>2 eV), low $m_h^*$ (>1.5), and low $E_{hull}$, the following class of Y- and Sc-based delafossites emerged from the selection criteria: $A$YS$_2$ ($A$=Ag$^†$, K, Au$^*$), $A$ScS$_2$ ($A$=K, Rb, Au$^*$), $A$YSe$_2$ ($A$=K$^*$, Cs), KScSe$_2^*$, and BaCaTe$_2^*$, with "*" indicating a new material not present in the ICSD or Materials Project database at the time of publication and "†" indicating a structure which, during the structure relaxation calculation, transformed from delafossite to a more stable structure (see Supporting Information). In particular, $AgYS_2$ was recommended for synthesis, with an HSE gap of 3.16 eV and a hole effective mass of 0.71. The crystal structure of another promising compound, $KYS_2$, is shown in **Figure 14** for reference. Branch point energy calculations suggested mid-gap Fermi stabilization energies for nearly all of these compounds, which is, as discussed in Section 2.4.1, not conclusive of doping type and thus defect calculations are still needed. This study is aimed towards finding p-type transparent conductors, but these predicted structures could also be used for other applications, such as those taking advantage of their magnetic properties.

### *3.4.4 ABCh₃ perovskites*

Perovskites are another class of ternary materials common as both n-type and p-type transparent conducting oxides, e.g. n-type $CdSnO_3$ (CTO),[319] n-type La:$BaSnO_3$,[320] p-type In:$SrTiO_3$ (ISTO),[321] and p-type $LaCrO_3$.[322] A computational study screening for new sulfide perovskite materials as photoelectrochemical (PEC) absorbers discovered several compounds with wide band gaps ($E_G$ > 2 eV) and low hole effective masses ($m^*$ < 1) that could be investigated as p-type transparent conductors, including $BaZrS_3$ (see Figure 14), $BiGaS_3$, $BiScS_3$, $CaZrS_3$ and $ZrCdS_3$ (see Supporting Information for other compounds and more information).[323] These structures were screened for defect tolerance to ensure no defect states were present in the middle of the gap, though p-type dopability was not confirmed and remains to be investigated.

### *3.4.5 A₂B₃Ch₄ and dimensional reduction*

With the exception of the materials discussed in Section 3.1, many binary chalcogenides exhibit gaps lower than 2 eV. Dimensional reduction is an interesting strategy to expand the space of wide band gap ternary compounds by starting with small band gap binary constituents and widening their gaps.[324] For example, small gap binary chalcogenides $MCh$ ($Ch$ = S, Se, Te) can achieve a wider gap by admixing with $Cs_2Ch$ to effectively reduce dimensionality (i.e. introduce layers) and reduce orbital overlap. A tradeoff of this procedure, however, is a decreased dispersion due to weakened covalent bonding and thus larger hole effective masses, which is the opposite trend as going from II-$Ch$ to I-III-$Ch_2$ material systems. Based on this method, a dimensional reduction on Zn$Ch$ and $Cs_2Ch$ was performed, and $Cs_2Zn_3Se_4$ (see Figure 14) and $Cs_2Zn_3Te_4$ ternaries were proposed as promising p-type TCs, with HSE06 gaps of 3.61 eV and 2.82 eV and hole effective masses of 1.23 and 1.25, respectively. Undoped $Cs_2Zn_3Se_4$ and $Cs_2Zn_3Te_4$ were found to be intrinsically p-type materials. However, the free hole concentration may be limited by low energy of native donor defects, e.g., Zn interstitials $Zn_i$.[325] Another dimensional reduction study with Hg$Ch$ proposed $Tl_2Hg_3S_4$ and $K_2Hg_3S_4$ with gaps 2.22 eV and 2.2–2.6 eV, respectively.[326] The dimensional reduction strategy has only been applied to a few classes of binary chalcogenides, which have yet to be explored experimentally, so this strategy could benefit from further exploration.

| $KYS_2$ | $BaZrS_3$* | $Cs_2Zn_3Se_4$* | $La_2O_2Se$ | $ZrOS$ |
|---|---|---|---|---|
| $R\bar{3}2/m$ delafossite | *Pnma* perovskite | *Ibam* | $P\bar{3}m1$ oxyselenide | $P2_13$ oxysulfide |

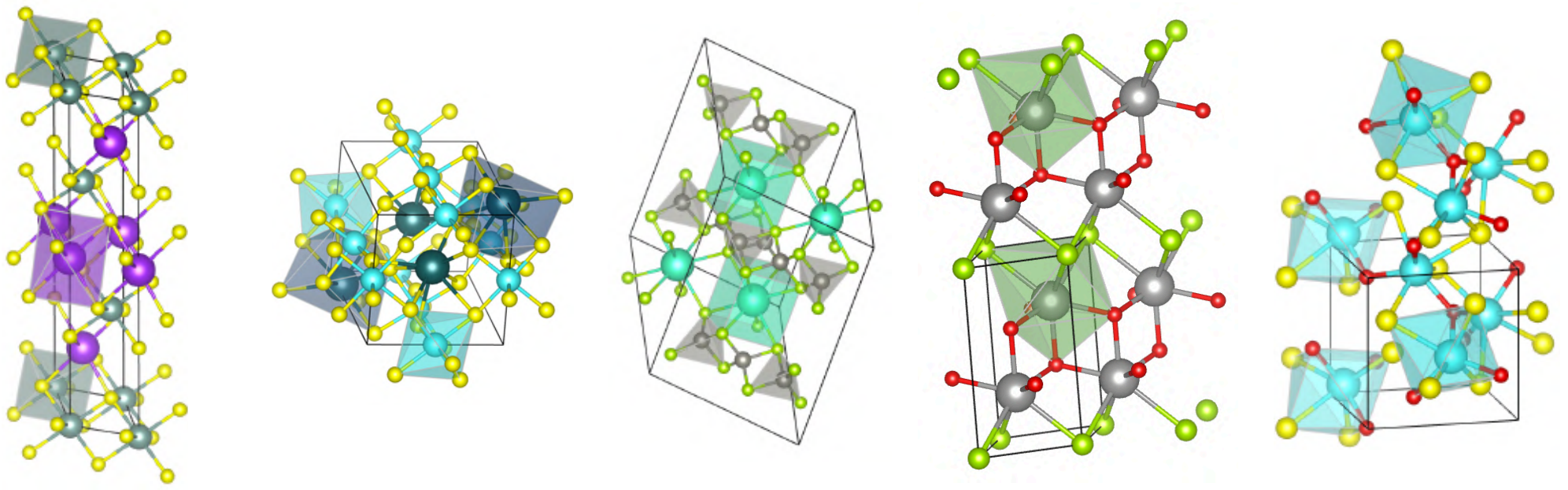

**Figure 14.** Chemical compositions and crystal structures of a representative set of computationally predicted wide-gap p-type dopable chalcogenides discussed in the text, with polyhedra to emphasize coordination. "*" indicates a new material not present in the ICSD or Materials Project database when the prediction study was performed.

### *3.4.6 $A_3BCh_3$ and other ternaries*

$Cu_3SbS_3$ with the wittichenite structure ($P2_12_12_1$ space group), has a computed HSE gap of 2.02 (indirect) and 2.14 (direct), and nanowires have been reported with an experimental optical gap of 2.95 eV.[327] This material has been doped with $O_S$ to raise the gap. A screening of antimony based thermoelectric sulfides computed $Li_3SbS_3$, $Na_3SbS_3$, and $Ca_2Sb_2S_5$ to have gaps of 2.96 eV, 3.14 eV and 2.11 eV, respectively, and low hole effective masses (not reported).[328] To our knowledge these materials have not yet been synthesized experimentally. This suggests incorporating information learned through screenings of thermoelectric materials to find new wide-gap chalcogenides. Other wide-gap ternary chalcogenides predicted computationally include spinels and spinel-like $AB_2Ch_4$ compounds e.g. $Ba_2GeSe_4$, $Ba_2SiSe_4$, (both predicted p-type dopable), $SrAl_2Se_4$ $Al_2ZnS_4$ (unlikely p-type dopable), and $Al_2CdS_4$, as well as $BaB_2Se_6$ and IrSbS, among others (see Table 4 and Supporting Information). In **Table 2**, we tabulate the experimentally verified ternary chalcogenides mentioned above. At the end of the table and in the Supporting Information, we include some other experimentally achieved ternary chalcogenides with wide band gaps that have not to our knowledge been explored in-depth for their electronic properties.

**Table 2.** Experimentally realized wide-gap ternary chalcogenides, and their reported optoelectronic properties.

| Compound | Structure type | Band gap (eV) | Carrier type | Dopant | Conductivity $S\ cm^{-1}$ | Refs. |
|---|---|---|---|---|---|---|
| $CuAlS_2$ | chalcopyrite | ~3.4–3.5 | p | intrinsic | bulk (0.9) | 249 |
| | | | | | thin film (0.016) | 250 |
| | | | | Cu-rich | 250 | 30 |
| | | | | S-rich | 4.6 | 11 |
| | | | | $As_S$ | 1 | 244 |
| | | | | $Zn_{Al}$ | 63.5 | 253 |
| | | | | $Mg_{Al}$ | 41.7 | 254 |
| $CuAlSe_2$ | chalcopyrite | ~2.6–2.7 | p | near-stoichiometry | $3.3 \times 10^{-3}$ | 256 |
| | | | | Cu-rich | 123 | 256 |
| | | | n | Al-rich | $2.3 \times 10^{-3}$ | 256 |
| | | | | intrinsic | 50 | 257 |
| $CuAlTe_2$ | chalcopyrite | ~2.1–2.5 | p | — | $\sim 10^{-3}$ | 259 |
| $CuGaS_2$ | chalcopyrite | 2.22–2.55 | p | — | bulk (1.7) | 266 |
| | | | | | thin film (0.7) | 267 |
| | | | | | single crystal (0.83) | 265 |
| $AgAlS_2$ | chalcopyrite | ~3.2 | — | — | — | 274 |
| $AgAlSe_2$ | chalcopyrite | 2.5–2.7 | — | — | — | 277 |
| $AgAlTe_2$ | chalcopyrite | ~2.3 | — | — | — | 279 |
| $AgGaS_2$ | chalcopyrite | ~2.7 | p | — | $<10^{-5}$ | 280 |
| $BaCu_2S_2$ | α-orthorhombic | 2.1–2.5 | p | — | 0.1–53 | 12 |
| $BaAg_2S_2$ | $CaAl_2Si_2$-type | 2 (calculated) | — | — | — | 303 |
| $Cu_3TaS_4$ | sulvanite | 3 | p | — | 1.6 | 132 |
| $Cu_3TaSe_4$ | sulvanite | 2.35 | p | — | $3 \times 10^{-3}$ | 308 |
| $Cu_3NbS_4$ | sulvanite | 2.6 | p | — | 0.1–0.2 | 309 |
| $Cu_3NbSe_4$ | sulvanite | ~2.2 | p | — | 1.9 | 310, 311 |
| $Cu_3PS_4$ | enargite (sulvanite-like) | 2.38 | p | — | 0.2–1.0 | 315 |
| $Cu_3P(S_3Se)$ | enargite (sulvanite-like) | 2.06 | p | — | 0.2–1.0 | 315 |
| $Ag_3PS_4$ | enargite (sulvanite-like) | 2.88 (calculated) | — | — | — | 316 |
| $Ag_3PSe_4$ | enargite (sulvanite-like) | 2.09 (calculated) | — | — | — | 316 |

## 3.5 Quaternary and mixed-anion chalcogenides

Mixing chalcogenides with oxygen or anions from other groups (e.g. halides), typically in layered motifs, results in wide band gap mixed-anion chalcogenides such as *Ln*CuO*Ch* (*Ln* = La, Pr, Nd) and *M*Cu*Ch*F (*M* = Ba, Sr).

### *3.5.1 LnCuOCh*

We limit our discussion to LaCuO*Ch* as a representative of the *Ln*CuO*Ch* oxychalcogenides family, since it has been most extensively investigated. (Pr,Nd)CuO*Ch* have similar optoelectrical properties including band gap and p-type doping, but chalcogen type significantly influences carrier concentration as shown in **Figure 15**c.[329] Reported conductivities are in the planar direction. We note that these compounds are difficult to make Ohmic contact with due to their layered, anisotropic structure, and that no reports of n-type conductivity could be found in any of these systems.

LaCuOS thin films have been prepared by RF-sputtering with a direct band gap of 3.1 eV, and p-type conductivity confirmed by positive Seebeck coefficients.[330] Its undoped conductivity of $1.2 \times 10^{-2}$ S cm$^{-1}$ can be increased to $2.6 \times 10^{-1}$ S cm$^{-1}$ by partial substitution of La with divalent cations like $Sr^{2+}$.[331] Figure 15a shows the alternative stacking of La-O and Cu-S layers.

By substituting S with Se to form LaCuOSe, the band gap decreases from ~3.1 to ~2.8 eV but remains direct.[332] Figure 15b illustrates the HSE06 electronic band structure of LaCuOSe and its increased VBM dispersion due to Cu 3*d* and Se 4*p* mixing, with a VBM effective mass in the *Γ*–*M* direction of 0.30.[333] Epitaxial thin films of $LaCuOS_{1-x}Se_x$ solid solutions have been prepared to investigate hole transport properties, as shown in Figure 15c. The Hall mobility increases with x between ~0.5 cm$^2$ V$^{-1}$ s$^{-1}$ for LaCuOS and 8 cm$^2$ V$^{-1}$ s$^{-1}$ for LaCuOSe. This mobility is high for typical p-type transparent conductors, and the increase with x is expected due to increased hybridization with Cu 3*d*. The hole concentrations in undoped $LaCuOS_{1-x}Se_x$ are around $10^{19}$ cm$^{-3}$ to $10^{20}$ cm$^{-3}$, so the doping is close to degenerate. $Mg_{La}$ has been shown to degenerately dope LaCuOSe, increasing carrier concentration by an order (up to $2 \times 10^{20}$ cm$^{-3}$) while only decreasing mobility by half, and raising hole conductivity up to the high reported value of 140 S cm$^{-1}$.[334] The layered structure induces a distinctive conduction mechanism in LaCuOSe. It was initially proposed that hole generation occurs in the $(La_2O_2)^{2+}$ layer, then holes are transported through the $(Cu_2Ch_2)^{2-}$ layer (hole conduction layer), successfully avoiding the influence of ionized impurity scattering.[335] It was later suggested by computation that the heavy hole doping from Mg may be due to the off-stoichiometry of Cu and Se rather than Mg doping at the La sites, and that Sr should be a more optimal dopant than Mg.[333] Unlike LaCuOS and LaCuOSe, LaCuOTe has an indirect band gap due to the stronger La-Te interactions compared with other La-*Ch*.[336] A. comparison of experimental and theoretical XRD patterns for LaCuOTe is shown in Figure 15d to demonstrate a single phase sample without other impurities.[337] Synthesis of a ceramic pellet of LaCuOTe yielded an optical band gap of 2.31 eV, conductivity of 1.65 S cm$^{-1}$, and exceptionally high Hall mobility of 80 cm$^2$ V$^{-1}$ s$^{-1}$.[337] We note that other quaternary lanthanum copper oxysulfides such as $La_3CuO_2S_3$ (gap of 2 eV)[338] have been investigated, suggesting further exploration of this chemistry, but electronic properties have not been reported. Substitutions of *Ln h* have also been tried in *Ln*CuO*Ch*, as shown in the left section of Figure 15c.[329]

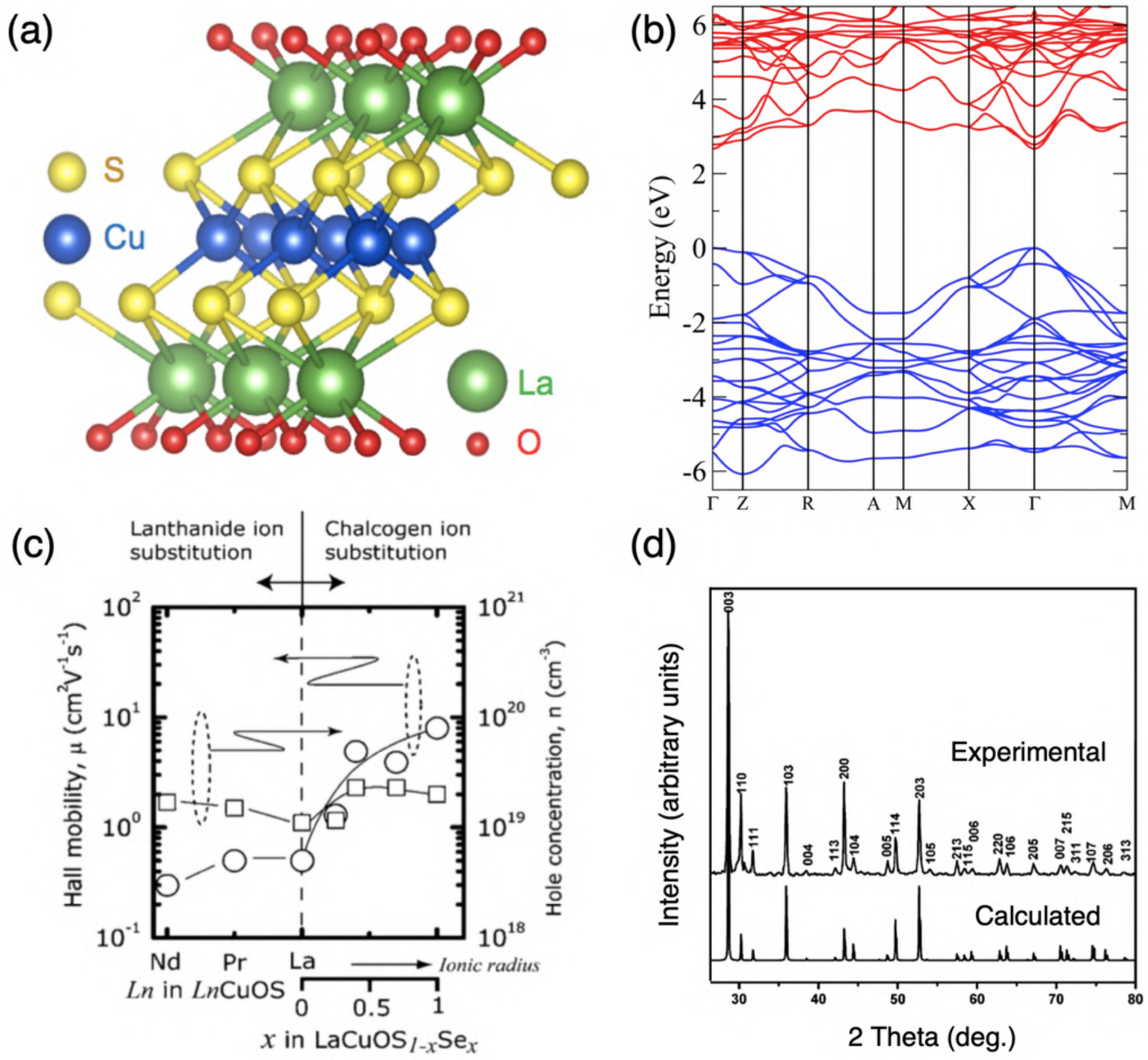

**Figure 15**. (a) Layered structure of LaCuOS. (b) HSE06 electronic band structure of LaCuOSe, adapted from Scanlon et al.[333] (c) Hole transport properties of *Ln*CuO*Ch* (*Ln* = La, Pr, Nd; *Ch* = $S_{1-x}Se_x$) at 300 K as a function of chemical composition.[329] (d) Experimental and theoretical x-ray diffraction powder patterns of LaCuOTe.[337]

### *3.5.2 MCuChF*

Replacing the La-O layer of LaCuO*Ch* with a Ba-F layer results in another layered transparent semiconductor, BaCu*Ch*F. BaCuSF and BaCuSeF were first investigated in bulk form by solid state reaction and their p-type conductivity was confirmed by positive Seebeck coefficients.[339] The conductivity of undoped BaCuSF and BaCuSeF was reported as $8.8 \times 10^{-2}$ and $6.1 \times 10^{-2}$ S $cm^{-1}$, respectively, but could be enhanced significantly in both materials by doping with $K_{Ba}$ to achieve conductivities of 82 S $cm^{-1}$ and 43 S $cm^{-1}$. The band gaps of BaCuSF and BaCuSeF were reported as 3.2 eV and 3.0 eV, respectively, and did not shift with doping.[339] Epitaxial, undoped BaCuTeF thin films have been synthesized by PLD, with reported conductivity of 167 S $cm^{-1}$ and optical band gap close to 3 eV.[340] Elevated conductivity was reported in c-axis oriented BaCuTeF, likely correlating with a lower in-plane effective mass and increased mobility due to decreased scattering with fewer oxidized grain boundaries.[132] Investigations into alloys $BaCuS_{1-x}Se_xF$ and $BaCuSe_{1-x}Te_xF$ found the band gap to decrease linearly (negligible bowing) and conductivity to increase with increasing x, as shown in **Figure 16**.[163]

SrCuSF, a layered material of the same family, has a reported band gap of 3.1 eV and p-type dopability, with Na substituting for Sr and O for S.[341] Recently, wide band gaps have been predicted computationally in the BaAg*Ch*F system of BaAgSF, BaAgSeF, and BaAgTeF, with mBJ-SR gaps of 3.13 eV, 2.85 eV and 2.71 eV, respectively, and in-plane hole effective masses of 0.98, 0.94 and 0.42, respectively.[342]

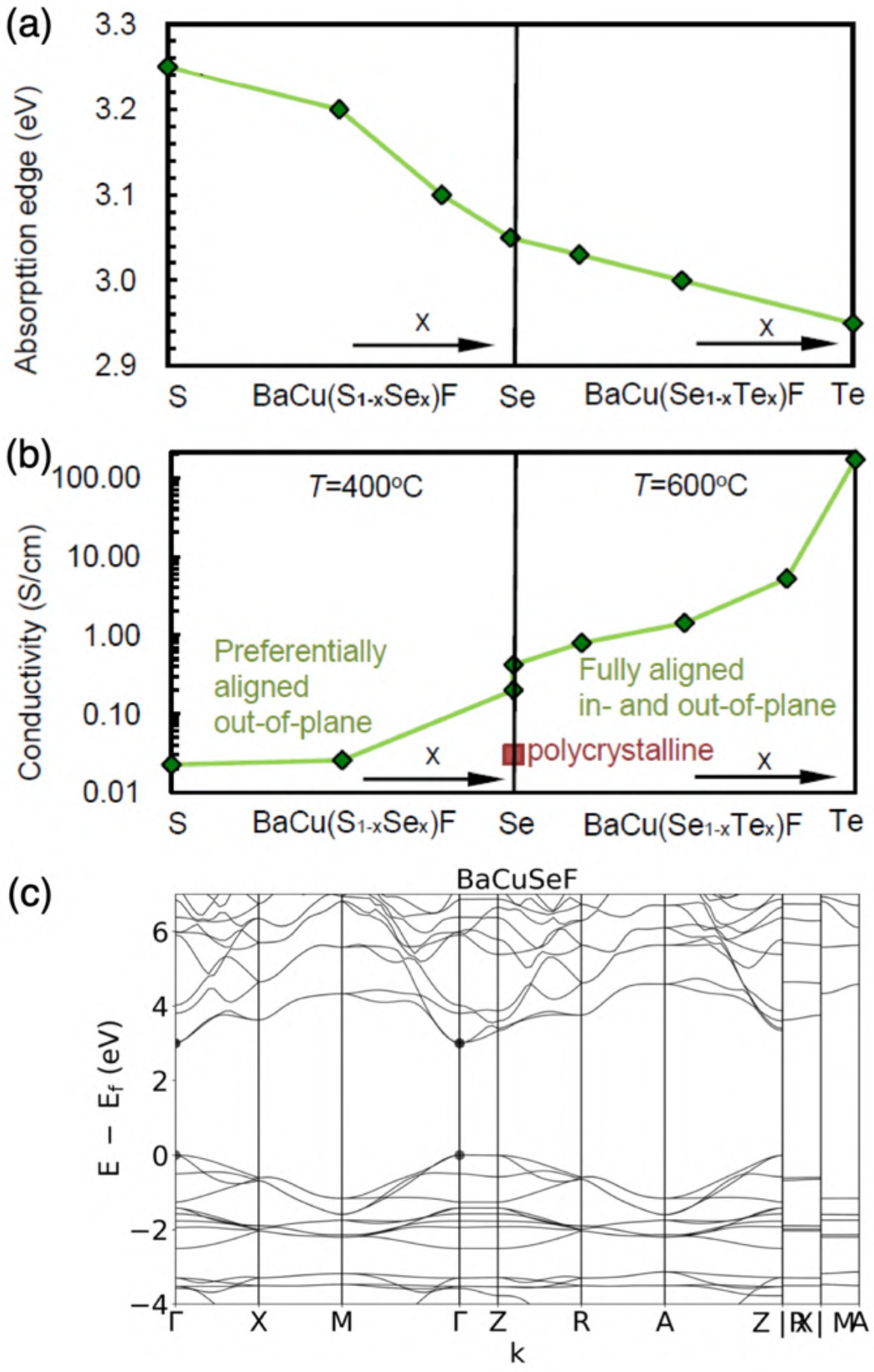

**Figure 16.** (a) Absorption edge and (b) conductivity of $BaCu(Ch_{1-x}Ch_x)F$ thin film solid solutions.[163] (c) GGA electronic band structure of BaCuSeF from the Materials Project database, with a scissor operation applied to correct for experimental band gap values.

### *3.5.3 MSCN*

Copper thiocyanate, CuSCN, is a coordination polymer that typically crystallizes in a hexagonal, layered $P6_3mc$ structure (β-CuSCN). Cu atoms are approximately tetrahedrally coordinated with three S and one N in a mixed ionic-covalent bond, while all other bonds are colinear, including a C≡N triple bond. The wide band gap of β-CuSCN (~3.6–3.9 eV), its facile solution synthesis, and unusual quasi-molecular bonding character enable its device applications and its interest as a p-type transparent conductor.[15] Like other Cu-based p-type chalcogenides discussed previously, the upper valence bands have hybridized Cu 3*d* and S 3*p* character. However, the CBM at the K symmetry point has mostly a cyanide antibonding character, which leads to a more disperse VBM than CBM.[343] Dip coating synthesis on glass substrates resulted hole conductivities of $8 \times 10^{-3}$ S $cm^{-1}$ in undoped films and up to 2 S $cm^{-1}$ with Cl doping (lowering the gap to 3.4 eV).[344] The field-effect mobility of CuSCN is reported in the range of 0.001–0.1 $cm^2V^-$

$^{1}s^{-1}$.[345] P-type conductivity has been explained by Cu vacancies, and coupled C-N vacancies $V_{(CN)}$ have been proposed as an alternate method to increase conductivity.[343] Layered β-CuSCN has been intensively investigated recently as a solid electrolyte window layer in dye sensitized solar cells, as a hole transport layer in perovskite solar cells, and as a heterojunction partner with n-type semiconductors in TFTs and other devices.[346,347] We mention that monoclinic α-AgSCN ($C2/c$) has also reported to be a wide band gap material (3.4 eV).[348] It is studied as a battery electrode,[349] but has not been as heavily investigated for optoelectronic properties.[350] A related compound, mixed-anion $Pb(NCS)_2$, was predicted as a scintillator material with a wide band gap and low $m_h^*$.[351]

### *3.5.4 Quintenary layered compounds*

The approach of consecutively stacking O-based and S-based layers to create layered oxychalcogenide crystals has been expanded to quintenary systems. For example, thin films of $Sr_2Cu_2ZnO_2S_2$ have a reported optical band gap of 2.7 eV, with a $E_F$ near the VBM indicating p-type conductivity. As with most Cu-*Ch* based chalcogenides, the VBM character is primarily hybridized Cu 3*d* and S 3*p*, while the CBM consists of a dispersive Zn 4*s* band.[352] Na doping leads to a p-type conductivity of 0.12 S cm$^{-1}$. This material system has been expanded by replacing Zn with Ga, In,[353] and Sc,[354] and adjusting O and S stoichiometries accordingly. The conductivity of $Sr_2CuGaO_3S$ was improved to $2.4 \times 10^{-2}$ S cm$^{-1}$ by Na doping.[353] Additionally, $Sr_3Cu_2Sc_2O_5S_2$ combines a wide band gap of 3.1 eV and a high conductivity of 2.8 S cm$^{-1}$, which originates from a high hole mobility >150 cm$^2$ V$^{-1}$ s$^{-1}$.[354]

### *3.5.5 Computationally predicted oxychalcogenides*

The concept of introducing a chalcogen anion in addition to an oxygen anion has been a fruitful design route for p-type transparent semiconductors, yet only layered mixed-anion materials from the spacegroup *P*4/*nmm* (and closely related I4/*mmm* for $Sr_2Cu_2ZnO_2S_2$) have been explored in-depth experimentally.[352] There is a tradeoff between high conductivity and transparency in these materials, and they are often grown epitaxially which can be costly and challenging to produce good device contacts. Additionally, each of these materials includes Cu, which may be detrimental to some device applications. This success of layered Cu-based oxychalcogenides supports the continued investigation of new oxychalcogenides with different structures, compositions, and properties.

Previous screenings on p-type transparent conducting oxides have included mixed oxychalcogenides in their findings. In the set of potential p-type TCOs calculated by Hautier and coworkers using the Materials Project, the sulfate $Hg_2SO_4$ and oxysulfides ZrOS (see Figure 14), and HfOS emerged as promising candidates.[92] The calculated $m_h^*$ of these materials are 1.06, 0.96, and 0.9, and GW band gaps are 3.8 eV, 4.3 eV, and 4.5 eV, respectively. Defect calculations suggested all three to be intrinsically p-type. The p-type nature of $Hg_2SO_4$ is described by an overlap between the O/S *p* orbitals and Hg 6*s* orbitals (Hg has a filled 5*d* shell), while hole conductivities of ZrOS and HfOS are due to their mixed-anion nature. It is notable that oxysulfide ZrOS was found to have both a lower m* and higher $E_G$ than its analog sulfide $ZrS_2$.

Sarmadian et al. screened all oxides in the AFLOWLIB library (12,211 total compounds) for low effective mass and wide band gap, and also included branch point energy (BPE) calculations as a proxy for p-type dopability.[11192] Interestingly, out of all the oxides screened, oxychalcogenide materials proved to be the most promising. In particular, this study identified the class of $P\bar{3}m1$ lanthanide oxyselenides, $Ln_2SeO_2$ (*Ln* = La, Pr, Nd, Gd) to be the most viable candidates. Their disperse VBMs ($m_h^*$ of 0.69–0.92), wide band gaps (HSE06 gaps of 2.76–3.49 eV), and BPE levels at or near the VBM are all promising for wide band gap p-type semiconductor applications. $La_2SeO_2$ (see Figure 14) was studied in greater depth with defect formation energy calculations, suggesting that in anion-rich conditions $Na_{La}$ should produce p-type conductivity without compensation from anion vacancies. If realized experimentally, this material could pose as an isotropic, Cu-free alternative to LaCuOSe. Additionally, this study revealed the following materials to consider for future investigation: $Tb_2Ti_2S_2O_5$, $NaVS_2O_8$, $Hg_2SO_4$, $Hg_2SeO_3$, $Ba_3Bi_2TeO_9$, and $CaTe_3O_8$ (though Tb- and Hg-containing compounds are impractical; see Table 4 and Supporting Information for computed properties). Additionally, oxychalcogenides $Gd_2O_2Se$ and $Y_2OS_2$, among others, have been identified as potential scintillator materials, each with a low $m_h^*$ and a relatively wide band gap.[351] Overall, prediction, synthesis, and characterization of oxychalcogenides pose a promising direction for future research. However, they also pose unique challenges in terms of synthesis and stability that must be addressed to advance the field.

### 3.5.6 Quaternary single-anion compounds

The most common quaternary chalcogenide structure is the kesterite structure $I_2$-II-IV-$Ch_4$, the quaternary extension of chalcopyrite. The prototype p-type kesterite $Cu_2ZnSnS_4$ (CZTS) has a gap ~1.4–1.5 eV (ideal for solar absorption), but there may be wide-gap materials within this ternary space or in related structural families (e.g. wurtzite-derived stannite). For example, kesterite and stannite $Cu_2ZnGeS_4$ have reported experimental gaps of 2.27 eV and 2.07, respectively, and hole conductivity in the stannite phase has been reported up to ~1 S $cm^{-1}$.[355] α- and β-$Cu_2ZnSiS_4$ have higher band gaps of ~3.0 eV and ~3.2 eV, respectively,[356] but is effectively insulating.[355] Quaternary *Ln*-based $ALn_xM_yCh_z$ chalcogenides, where *A* is an alkali or alkali-metal (K, Rb, Cs, Ba, Sr, Cd, Mg) and *M* a transition metal (*e.g.* Cu, Ag, Zn), have been synthesized and solved for crystal structure. Band gaps tend to lie within a window of 2–2.6 eV, but there are not many published investigations of optoelectronic properties.[357–359] One such investigation reported the (010) band gaps of $CsYZnSe_3$, $CsSmZnSe_3$, and $CsErZnSe_3$ (*Cmcm* space group) as 2.41 eV, 2.63 eV, and 2.63 eV, respectively. $BaYCuS_3$, $BaNdCuS_3$, and $BaNdAgS_3$ have reported gaps of 2.61, 2.39, and 2.31, respectively.[360] In a table in the Supporting Information we include $Li_2GePbS_4$, a computationally predicted quaternary wide-gap p-type chalcogenide derived from the sulvanite structure, but one can imagine other such compounds derived from binary and ternary structures. To compare properties of quaternary and mixed-anion chalcogenide semiconductors, we summarize the materials above in **Table 3**, and list others in the Supporting Information.

**Table 3**. Quaternary & mixed-anion chalcogenide semiconductors

| Compound | Structure type | Band gap (eV) | Carrier type | Dopant | Conductivity (S $cm^{-1}$) | Refs. |
|---|---|---|---|---|---|---|
| $Cu_2ZnGeS_4$ | kesterite | 2.27 | p | VS | 1 | 355 |
| LaCuOS | tetragonal layered | 3.1 | p | $Sr_{La}$ | 0.26 | 330, 331 |
| LaCuOSe | tetragonal layered | ~2.8 | p | $Mg_{La}$ | 140 | 332, 334 |
| LaCuOTe | tetragonal layered | 2.31 | p | — | 1.65 | 337 |
| BaCuSF | tetragonal layered | 3.2 | p | $K_{Ba}$ | 82 | 339 |
| BaCuSeF | tetragonal layered | 3 | p | $K_{Ba}$ | 43 | 339 |
| BaCuTeF | tetragonal layered | 3 | p | — | 167 | 340 |
| CuSCN | layered | ~3.6–3.9 | p | — | 2 | 15,344 |
| AgSCN | layered | 3.4 | — | — | — | [348] |
| $Sr_2Cu_2ZnO_2S_2$ | tetragonal | 2.7 | p | Na | 0.12 | [352] |
| $Sr_2CuGaO_3S_2$ | tetragonal | — | p | Na | $2.4\times10^{-2}$ | [353] |
| $Sr_3Cu_2Sc_2O_5S_2$ | tetragonal | 3.1 | p | — | 2.8 | [354] |

## 3.6 2D chalcogenides

Two dimensional (2D) chalcogenide semiconductors have been heavily investigated in recent years due to their unique properties resulting from 2D confinement. We will briefly discuss the optical and electronic properties of wide-gap 2D chalcogenides here, though they have been extensively reviewed elsewhere.[361,362,363] 2D materials typically possess weak interlayer van der Waals (vdW) forces with strong in-plane covalent bonds, but are not limited to this configuration. For example, several oxide-based bulk materials can be exfoliated into atomically thin crystals, e.g. $MoO_3$ and $WO_3$, while in other cases the stoichiometry can differ from bulk form, e.g., $TiO_2$.[364] 2D chalcogenides appear to be more favorable than 2D oxides for electronic applications due to their tunable band gaps and higher mobilities. Monolayer oxides tend to have larger band gaps and lower dielectric constants compared to their bulk forms,[364,365] which suggests that thinning a 2D chalcogenide could be advantageous to attaining larger band gaps or higher transparency under visible light. In this section, we focus on wide band gap (>2 eV) 2D compounds in the following categories: transition-metal chalcogenides (TMC), post-transition-metal chalcogenides (PTMC) and transition-metal phosphorus trichalcogenides (MPT). As shown in **Figure 17**a, 2D chalcogenides share various crystal structures, such as the trigonal prismatic structure for $WS_2$ (2H-phase) from the TMC category,[366] β-phase GaSe from PTMC,[367] and $MnPS_3$ from MPT.[368]

Figure 17b summarizes predicted and experimental mobilities from the reviewed wide-gap 2D chalcogenide materials. We report mobility because it is generally more important than conductivity for optoelectronic carrier transport applications. The band gaps in 2D materials are typically a function thickness and tend to differ substantially from bulk counterpart band gaps, as demonstrated in Figure 17c. It has been demonstrated that transition-metal dichalcogenides (TMDC), such as $WS_2$, exhibit indirect to direct band transition upon exfoliation from bulk crystals to monolayers.[369] Additionally, the carrier mobilities in monolayers may decrease by orders of magnitude, potentially due to charge traps introduced by substrates or surface defects, while the bulk crystal mobilities are typically limited by phonon scattering.[370] Future advances in this field will require synthesis of large and uniform atomically thin layers, comprehensive understanding of the 2D band structures and interfacial effects, and charge transfer processes across interfaces, not only to restore the mobilities to the bulk values but also improve their technological potentials. Phase engineering, heterostructure and superlattice construction, and band alignment tuning of 2D chalcogenides open up many prospects in optoelectronics.

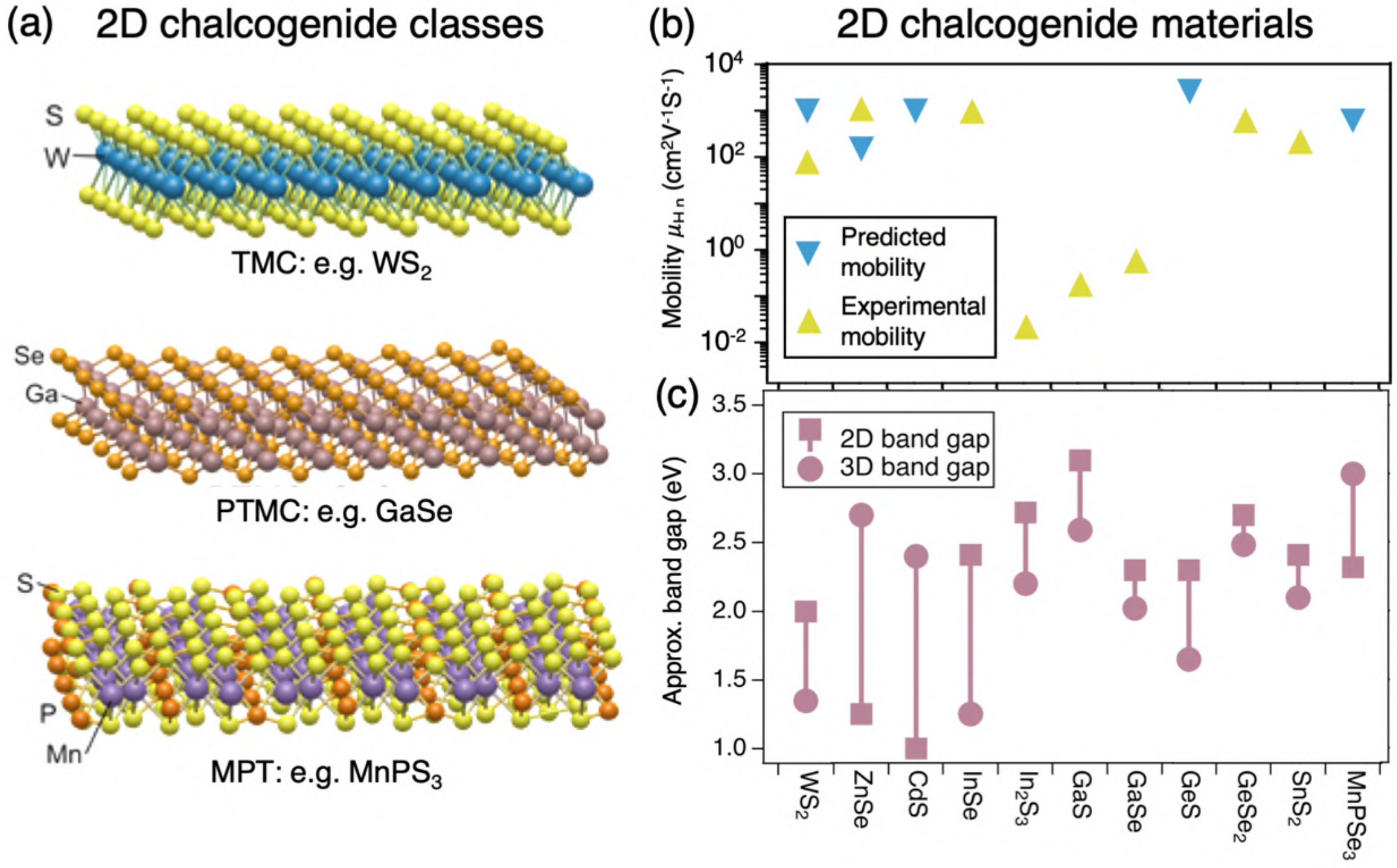

**Figure 17**. (a) Crystal structures of representative wide band gap 2D chalcogenide materials from three classes, namely $WS_2$ from transition-metal chalcogenides (TMC), GaSe from post-transition-metal chalcogenides (PMT), and $MnPS_3$ from transition-metal phosphorus trichalcogenides (MPT). (b) Predicted (blue marker) and experimental (yellow marker) mobilities from a representative set of 2D materials. (c) A comparison of 2D (square marker) and bulk (i.e. 3D, circle marker) band gaps for the same set of 2D chalcogenide materials.

### *3.6.1 Transition-metal chalcogenides (TMC)*

$WS_2$ in the 2H-polytype structure is an n-type indirect gap semiconductor with a band gap of ~1.35 eV.[371] Strong photoluminescence has been demonstrated from monolayer $WS_2$ due to the transition from its bulk indirect band gap to its monolayer direct band gap[366,373] of ~2 eV at room temperature.[374] Monolayer $WS_2$ provides the widest band gap and highest predicted phonon-limited mobility (over 1,000 $cm^2\ V^{-1}\ s^{-1}$) among TMDCs ($MCh_2$, $M$ = W, Mo; $Ch$ = S, Se, Te) at room temperature.[375] Due to defects, charge traps, and Coulomb impurities, reported experimental mobility of monolayer $WS_2$ is only 83 $cm^2\ V^{-1}\ s^{-1}$, even after improvements via a combination of a thin layer of $Al_2O_3$ and thiol chemical functionalization.[376] It has been shown that in gated charge transport measurements, mechanically exfoliated monolayer $WS_2$ with a nitrogen plasma treatment has higher electron mobility (184.2 $cm^2\ V^{-1}\ s^{-1}$) than the scalable metal-organic CVD monolayer $WS_2$ (30 $cm^2\ V^{-1}\ s^{-1}$) at room temperature,[377,378] which reflects that there is still room to improve the synthesis of monolayer $WS_2$.[379] One recent strategy to increase monolayer mobility is doping with other transition metals, such as V and Zr.[380] The thin-layer nature of monolayer $WS_2$ opens up the possibility of manipulating its band structure by forming heterostructures. It is observed that conductivity enhancement and band bending occurs at the interface of $WS_2$-based heterojunctions.[381] 2D $WS_2$ has been explored as a hole-transport layer in perovskite solar cells, demonstrating higher performance and better stability than transitional conductive polymers.[382,383]

ZnS has a gap of ~3.7–3.8 eV (zincblende) in bulk form, as discussed in Section 3.1.1. In monolayer ZnS, DFT GGA calculations predict a wide band gap of 2.07–2.77 eV.[384] Hydrothermal synthesized ZnS nanoparticles (15–25 nm) suggest a wider gap in the range of ~3.9–4.8 eV.[385]

Monolayer ZnSe, a direct band gap semiconductor with a pseudohexagonal (ph) lattice structure,[386,387] demonstrates a high carrier mobility of 1250 $cm^2V^{-1}s^{-1}$ and generates a much higher photocurrent than in its bulk form.[388] The DFT calculated band gap of ph-ZnSe is reported to be 1.25 eV, using input atomic structural parameters from x-ray absorption fine structure (XAFS).[388] A tetragonal structure ZnSe (t-ZnSe) has been predicted under lateral pressure, with a wide band gap of 2–3 eV and retention of its direct character up to 7% biaxial strain.[389] Computed n-type mobility of t-ZnSe is ~138.8 $cm^2\ V^{-1}\ s^{-1}$, making it a possible n-type channel material for transparent TFTs.[389]

CdS is also discussed previously as a wide band gap bulk semiconductor (~2.4 eV).[390] Several studies have investigated the synthesis and characterization of colloidal 2D nanosheets of CdS and $CdS_{1-x}Se_x$.[391,392] CdS nanosheets and nanoparticles have been intensively studied for photocatalysis by assembling into heterojunctions with other 2D materials, e.g., graphene/graphene oxide,[393,394,395] $MoS_2$,[396,397] ZnSe,[398] etc., and monolayers or nanosheets can be stacked sequentially to form heterostructures. CdS/ZnSe in-plane heterostructures are predicted by HSE06 to have type-II band alignment and 2.31 eV direct band gap with the "armchair configuration", with predicted electron and hole mobilities of ~884 $cm^2V^{-1}s^{-1}$ and ~1920 $cm^2V^{-1}s^{-1}$, respectively.[398]

### 3.6.2 Post transition-metal chalcogenides (PTMC)

The band gap of monolayer InSe is calculated to be 2.41 eV by DFT with the screened exchange (sX) hybrid density functional, in contrast to its calculated bulk indirect gap of 1.2–1.3 eV.[399] Predicted electron mobilities in atomically thin InSe are above $10^3$ $cm^2\ V^{-1}\ s^{-1}$.[400,401] 2D InSe inherits the advantageous properties of bulk InSe, such as a low $m_e^*$ and shallow dopant sites.[399,401] Several studies using InSe in optoelectronic applications have shown promising characteristics.[402,403] GaS in bulk form has a PBE predicted indirect band gap of 1.59 eV and a direct band gap of 2.59 eV.[404] Another first-principles study predicts a direct band gap of 1.785 eV in monolayer GaS but a direct gap of 2.035 eV in bilayer GaS (less than 0.2 eV above its indirect band gap).[405] Bulk GaS has a hole mobility of ~80 $cm^2\ V^{-1}\ s^{-1}$,[406] while the reported electron mobility for 1.5 nm thick GaS is 0.2 $cm^2\ V^{-1}s^{-1}$.[407] As a transistor, thin GaS demonstrates typical n-type conduction behavior with an on/off ratio of $10^4$.[408] Band gap and mobility values enable GaS nanosheets as possible photodetectors.[405]

Monolayer GeS has predicted electron mobilities of 2,950–3,680 $cm^2\ V^{-1}\ s^{-1}$, depending on the axis direction, and an indirect band gap of 2.34 eV,[409] in contrast to a calculated electron mobility of 2,430 $cm^2\ V^{-1}\ s^{-1}$ and an HSE06 predicted indirect band gap as 2.29 eV.[409,410] Both external electric field and strain can tune the band gap.[411] Monolayer GeS has been investigated for applications such as gas sensors,[412] catalysis,[413] and photocatalysis.[414]

Bulk $SnS_2$ is a p-type semiconductor with a band gap of ~2.1 eV, as discussed in Section 3.2.[415,416] Monolayer $SnS_2$ has a computed indirect band gap of 2.41 eV (by HSE06).[417] A monolayer $SnS_2$ FET demonstrates n-type conduction,[418] with an electron mobility of ~5 $cm^2\ V^{-1}\ s^{-1}$ reported at room temperature for exfoliated few-layer $SnS_2$. The FET devices in solution provide enhanced carrier mobilities up to 230 $cm^2\ V^{-1}\ s^{-1}$.[419] This result emphasizes the sensitivity of the electronic properties of an atomic thin material on surface interactions.

The n-type semiconductor β-$In_2S_3$ has a reported band gap of 2.3 eV in thin film form and is the stable phase at room temperature (see Section 3.2).[420] Synthesis of thin β-$In_2S_3$ nanosheets based on aqueous solution routes and CVD methods have been reported.[421–423] A photodetector based on β-$In_2S_3$ nanosheets has demonstrated a high responsivity and external quantum efficiency, while the calculated carrier mobility is 0.023 $cm^2V^{-1}s^{-1}$ with an on/off ratio of $1.5\times10^3$.[423] Bulk $As_2S_3$ is known as the mineral orpiment (layered $P2_1/n$ structure), and is highly resistive p-type semiconductor with a band gap of 2.5 eV.[424] An indirect band gap of 2.12–3.18 eV has been predicted for monolayer $As_2S_3$, influenced by layer thickness and external strains as is observed in other 2D chalcogenides. [424]

Bulk $TlInS_2$, $TlGaSe_2$, $TlGaS_2$ from the ternary thallium chalcogenide family $TlMCh_2$ (*M* = Ga or In, *Ch* = S, Se) have wide band gaps of >2 eV.[425,426,427] These compounds share a quasi-2D layered structure in bulk form, and potentially can be exfoliated to atomic-thick layers.[426] The field-effect mobility of a thin $TlGaSe_2$ sheet (~220 nm) was reported at 2.1 × $10^{-3}$ $cm^2V^{-1}s^{-1}$ with a DFT calculated band gap of 1.95 eV, though this is not technically 2D.[428] Recently, a high-throughput screening suggested that $MgAl_2S_4$, a layer-structured ternary chalcogenide material with a direct band gap of 2.015 eV (PBE), is stable in its monolayer form. [429]

### 3.6.3 Metal phosphorus trichalcogenides (MPT)

$MPS_3$ (*M* = Fe, Mn, Ni, Cd, Zn) and $MPSe_3$ (*M* = Fe, Mn) constitute another family of layered 2D chalcogenides, some of which have been synthesized or exfoliated into monolayers.[368] The optical band gaps of monolayer $MPS_3$ and $MPSe_3$ materials vary from 1.3 to 3.5 eV, potentially suitable for photoelectrochemical applications.[368] For example, monolayer $MnPSe_3$ is predicted as a potential material for spintronic devices with electrically controllable spin polarization, while monolayer $Ag_{0.5}Sc_{0.5}PSe_3$ can serve as an efficient photocatalyst for water splitting.[430,431] The HSE calculated single layer band gaps of $CdPS_3$ and $CdPSe_3$ are 3.30 eV and 2.25 eV, respectively. Computed monolayer gaps are 1.93 eV and 1.29 eV (PBE), respectively, while the band gaps of bulk $CdPS_3$ and $CdPSe_3$ are 2.96 and 2.18 eV, respectively.[430] Monolayer $MnPSe_3$ has a high predicted electron mobility up to 625.9 $cm^2V^{-1}s^{-1}$ and a predicted direct band gap of 2.32 eV.[432] To date, not many optoelectronic properties of 2D MPT materials have been experimentally demonstrated.

## 3.7 Summary of optoelectronic properties

With this breadth of wide-gap chalcogenide materials discussed above, it is important to summarize all of the materials and their most important properties. To provide an overview of the wide-gap chalcogenide semiconductors covered in this section, we compare the maximum achieved conductivity in the literature and the reported experimental band gaps of each material in **Figure 18**a (see Tables 1–3 for tabulated data). Error bars indicate the range of reported band gaps in the literature. This figure demonstrates that there are over twenty wide-gap chalcogenides with reported conductivities greater than 1 S $cm^{-1}$, and several with reported conductivities greater than 100 S $cm^{-1}$. Figure 18b includes computationally predicted wide band gap p-type semiconducting chalcogenides that have not yet been explored in-depth experimentally, plotting the average computed hole effective masses ($m_h^*$) as a function of computed band gap. We have compiled relevant computed data and computational methods in **Table 4** for compounds confirmed p-type by experiment or defect calculations, and include a table with compounds which have not yet been confirmed p-type in the Supporting Information. We note that band gaps are computed with a variety of functionals, as denoted in Table 4, so values from one study may not be directly comparable to those from another. Figure 18b indicates both compounds with confirmed p-type doping and those that have yet to be explored, with the marker size corresponding to the likelihood of p-type dopability (note that "BPE" indicates the material is *not* definitively n-type, see Section 2.4.1). Though this is just a small representative set, we observe a slight positive correlation between gap and effective mass which reflects the very broad trend across materials of decreased band dispersion with widening gap. However, the key finding from across these computational studies is that many of these wide band gap chalcogenide compounds demonstrate exceptions to this trend: they have $m_h^*$ of approximately 1 or less, which could indicate high p-type conductivities. Additionally, these materials contain cations from across the periodic table. We note that the calculation methodology differs by study, and that some studies look more in-depth into p-type dopability than others (indicated by size of marker). As indicated by Figures 3.12a and b, we have covered materials with a wide range of conductivity and band gaps, paving the way for applications in a variety of optoelectronic devices which are discussed subsequently.

**Table 4**. Wide band gap chalcogenides computationally explored in the literature for p-type optoelectronic applications. We note that some of these have been explored experimentally (see Section 3), while some warrant experimental confirmation. GW is expected to be the most accurate, followed by HSE06 and GLLB-SC, then by PBE+U (see Section 2.2.1). A "d" indicates direct band gap, and hole effective mass $m_h{}^*$ is averaged over three crystallographic directions. Under the column "p-type dopable?", "yes (exp)" indicates experimentally confirmed p-type dopability and "yes (defects)" indicates predicted p-type via defect formation energy calculations. More computed compounds that have not been confirmed p-type are listed in the Supporting Information.

| Material | Space group | Predicted band gap (eV) | DFT functional | Predicted $m_h{}^*$(avg) | $m_h^*$ calc. method | $E_{hull}$ (eV/atom) | p-dopable? | Ref. |
|---|---|---|---|---|---|---|---|---|
| ZnS | $F\bar{4}3m$ | 3.5 | HSE06 | 0.7 | EMC code | — | yes (exp) | 144 |
| ZnSe | $F\bar{4}3m$ | 2.7 | HSE06 | 0.63 | EMC code | — | yes (exp) | 144 |
| ZnTe | $F\bar{4}3m$ | 2.6 | HSE06 | 0.4 | EMC code | — | yes (exp) | 144 |
| MgS | $Fm\bar{3}m$ | 5.5 | HSE06 | 0.96 | EMC code | — | yes (defects) | 144 |
| MgTe | $F\bar{4}3m$ | 3.24 | HSE03 | 0.95 | BoltzTraP | 0.9 | yes (defects) | 211 |
| CaTe | $Fm\bar{3}m$ | 2.18 (d=3.50) | HSE03 | 0.60 | BoltzTraP | 0 | yes (defects) | 211 |
| SrSe | $Fm\bar{3}m$ | 3.03 (d=3.68) | HSE03 | 0.83 | BoltzTraP | 0 | yes (exp) | 211 |
| YbSe | $Fm\bar{3}m$ | 2.43 (d=3.48) | HSE03 | 0.67 | BoltzTraP | 0 | yes (exp) | 211 |
| YbTe | $Fm\bar{3}m$ | 1.76 (d=3.09) | HSE03 | 0.54 | BoltzTraP | 0 | yes (exp) | 211 |
| GaSe | $P6_3/mmc$ | 2.6 | HSE06 | 0.25 | EMC code | — | yes (defects) [144] | 144 |
| BeTe | $F\bar{4}3m$ | 2.6 | HSE06 | 0.56 | EMC code | — | yes (defects) | 144 |
| BeTe | $F\bar{4}3m$ | 2.45 (d=4.04) | HSE03 | 0.42 | BoltzTraP | 0 | yes (defects) [144] | 211 |
| GeS | $Pnma$ | 1.85 (2.5 for monolayer) | HSE06 | 0.35 | EMC code | — | yes (defects) | 144 |
| $Al_2Se_3$ | $Cc$ | 3.1 | HSE06 | 0.14 | EMC code | — | yes (defects) [144] | 144 |
| IrSbS | $P2_13$ | 3.08 | HSE06 | 0.39 | BoltzTraP (Dryad DB) | 0 | yes (defects) | 273 |
| RbAuSe | $Cmcm$ | 3.40 | HSE06 | 1.0 | BoltzTraP (Dryad DB) | 0 | yes (defects) | 273 |
| KAuSe | *Cmcm* | 3.42 | HSE06 | 1.0 | BoltzTraP (Dryad DB) | 0 | yes (defects) | 273 |
| $CuBS_2$ | $I\bar{4}2d$ | 3.41 (d) | HSE06 | 1.0 | BoltzTraP (Dryad DB) | 0 | yes (defects) | 273 |
| $Ba_2GeSe_4$ | $P2_1/m$ | 3.01 (d) | HSE06 | 0.60 | BoltzTraP (Dryad DB) | 0 | yes (defects) | 273 |
| $Ba_2SiSe_4$ | $P2_1/m$ | 3.96 (d) | HSE06 | 0.75 | BoltzTraP (Dryad DB) | 0 | yes (defects) | 273 |
| $BaB_2Se_6$ | *Cmce* | 3.53 | HSE06 | 0.78 | BoltzTraP (Dryad DB) | 0 | yes (defects) | 273 |
| $Cu_3VS_4$ | $P\bar{4}3m$ | 3.68 | HSE06 | 0.93 | BoltzTraP (Dryad DB) | 0 | yes (exp) | 273 |
| $Cu_3NbSe_4$ | $P\bar{4}3m$ | 3.13 | HSE06 | 0.82 | BoltzTraP (Dryad DB) | 0 | yes (exp) [309] | 273 |
| $Cs_2Zn_3Se_4$ | *Ibam* | 3.61 | HSE06 | 1.23 | PBE, top VB | — | yes (defects) | 325 |
| $Cs_2Zn_3Te_4$ | *Ibam* | 2.82 | HSE06 | 1.25 | PBE, top VB | — | yes (defects) | 325 |
| $Hg_2SO_4$ | $P2/c$ | 3.8 | GW | 1.06 | BoltzTraP | 0 | yes (defects) | 273 |
| ZrOS | $P2_13$ | 4.3 | GW | 0.96 | BoltzTraP | 0 | yes (defects) | 211 |
| HfOS | $P2_13$ | 4.5 | GW | 0.9 | BoltzTraP | 0.007 | yes (defects) | 211 |
| $Sb_2S_2O$ | $P\bar{1}$ | 1.24 | DFT-GGA | 1.38 | BoltzTraP | 0.002 | yes (defects) | 211 |
| $La_2SeO_2$ | $P\bar{3}m1$ | 3.49 | HSE06 | 0.92 | Avg. w/in 26 meV | 0 | yes (defects) | 111 |

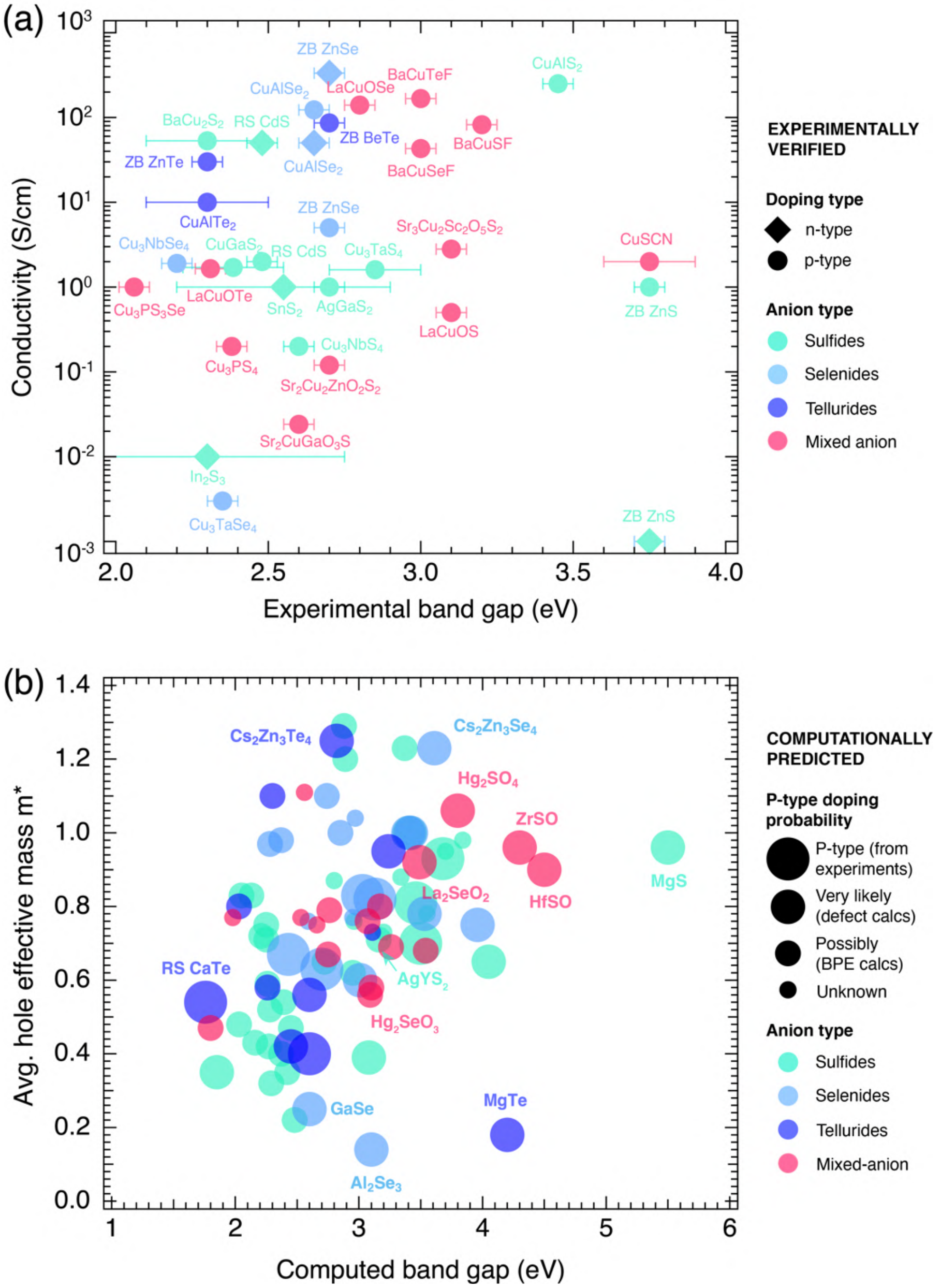

**Figure 18**. (a) Summary of the experimentally realized wide-gap semiconducting TCs discussed in this section, with the highest reported conductivity plotted as a function of experimental band gap. Error bars indicate the range of reported band gaps. Both n-type (diamonds) and p-type (circles) materials are plotted here. This plot only includes representative semiconductors where both conductivity and band gap are reported, and reports references in the text (Tables 1–3). (b) Computed properties of the predicted wide-gap p-type conductive materials discussed in this section. The size of the marker distinguishes the degree of certainty of p-type dopability, based on branch point energy (BPE) calculations (medium-sized) or defect calculations (large-sized). Data and references for each material are given in the text and Table 4 and the Supporting Information (note varying levels of theory for DFT calculations).

# 4. Applications

In this section, we summarize the various roles that wide band gap chalcogenide semiconductors play in different electronic devices, focusing when appropriate on their unique advantages compared to other materials. Devices discussed here are photovoltaic (PV) solar cells, thin film transistors (TFTs), photoelectrochemical (PEC) water splitting devices, and light emitting diodes (LEDs), and a schematic of their stacking configurations and corresponding band diagrams is given in **Figure 19**. Other areas of interest that we will not focus on here include photodetectors, transparent diodes, laser diodes, upconverters, scintillators, gas sensors, thermoelectrics, and electrochromic windows, and are reviewed elsewhere.[1,434–442] Some traditional applications of wide band gap semiconductors, such as power electronics and radio-frequency transistors,[443,444] have not been explored for chalcogenides to our knowledge, and hence are not reviewed here. For each device discussed in this section, we address the following considerations: (1) how wide band gap chalcogenide semiconductors contribute to each device, (2) which specific wide band gap chalcogenides have been implemented in each device so far, (3) what the performances are of such devices containing wide band gap chalcogenide semiconductors, and (4) what some prospective directions are for future research incorporating new wide-gap chalcogenide materials.

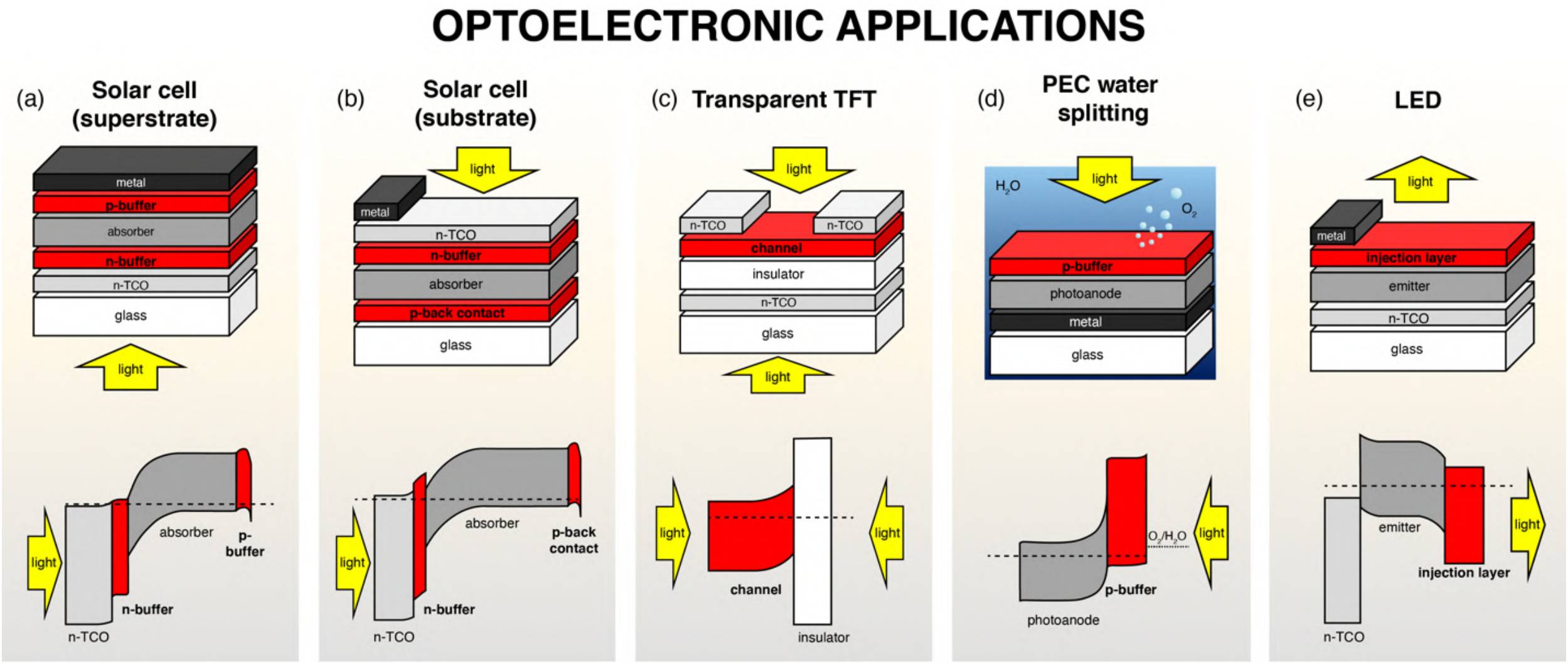

**Figure 19.** Comparative stack schematic of the devices discussed in this section, with corresponding representative electronic band diagrams under zero bias: (a) superstrate solar cell, e.g. CdTe,[445] (b) substrate solar cell, e.g. CIGS,[446] (c) transparent thin film transistor (TFT),[447] (d) photoelectrochemical (PEC) water splitting device (e.g. a one-terminal devicephotoanode shown only), and (e) light emitting diode (LED). Red coloring highlights the layer in which a wide-gap semiconductor can be implemented, labeling the functionality of the layer in each device. and the dotted line indicates the Fermi level $E_F$. Note that the y axis in the device stack corresponds to the x axis in the band diagrams (i.e. device thickness), and the thicknesses are not to scale.

## 4.1 Solar cells

Solar cells, also known as photovoltaic (PV) cells, are one of the most prominent applications that incorporate wide-gap chalcogenide semiconductors. Here, we will focus primarily on CdTe[198] and Cu(In,Ga)(Se,S)[448] i.e. CIGS (or CIGSe) devices, since they are the most commonly studied inorganic thin film solar cells that use wide-gap chalcogenide layers, but we will mention other PV applications (e.g. perovskite contacts, tandem window layers, IBSC absorbers). In this past decade (2010 – 2020) outstanding progress has been made in CdTe, CIGS(Se), and perovskite technologies, achieving certified power conversion efficiencies (PCE) exceeding 22% (22.1%, 23.4%, and 25.2%, respectively).[449]

### *4.1.1 CdTe solar cells*

Homojunction devices are not attractive for CdTe solar cells due to CdTe's limited n-type dopability,[450] so a p-n heterojunction is typically formed between p-type CdTe and another n-type semiconductor. Figure 19a describes the layer stack of a typical CdTe solar cells in a "superstrate" configuration. A typical CdTe/CdS solar cell structure (from bottom up) consists of (1) an n-type transparent conductive oxide (TCO) on a glass substrate, usually fluorine-doped tin oxide ($SnO_2$:F or FTO), indium tin oxide (ITO), or cadmium stannate ($Cd_2SnO_4$),[451,452] (2) n-type buffer layer (also called emitter layer or window layer) on top of the TCO, such as CdS,[453,454,455,456] (3) the p-type CdTe absorber layer, with a post-deposition $CdCl_2$ treatment, (4) a p-type Ohmic back contact buffer layer to CdTe, such as $Cu_xTe$,[457] Te,[458] or ZnTe:Cu, and (5) a metal contact. Wide-gap chalcogenides are essential in layer (2), as a window layer between CdTe and the n-type TCO, and in (4) as the buffer layer between CdTe and the metal back contact. Actual device stacks can be more complicated, with multiple buffer layers, but this is a basic schematic.

**Front buffer layer to *p*-CdTe**. One of the major purposes of this layer is to form a proper band alignment to the CdTe absorber to collect electrons and block holes. A flat or small positive conduction band offset between the CdTe and the window layer, as shown in the band diagram in Figure 19a and under bias in **Figure 20**a, has favorable band alignment for high cell performance. Since the light absorbed in the window layer does not contribute to the photocurrent, the band gap of the window layer should be as wide as possible to transmit maximum light to the junction and absorber. Other requirements are the ability for this layer to be deposited uniformly as a very thin film (5–50 nm) and to form thermochemically stable junctions. The band gap of CdS is ~2.4 eV and photoemission experiments have shown a valence band offset of $\Delta E_{VB}$ = ~1 eV at the CdS/CdTe interface,[453] which corresponds to a conduction band offset of ~ -0.1 eV.

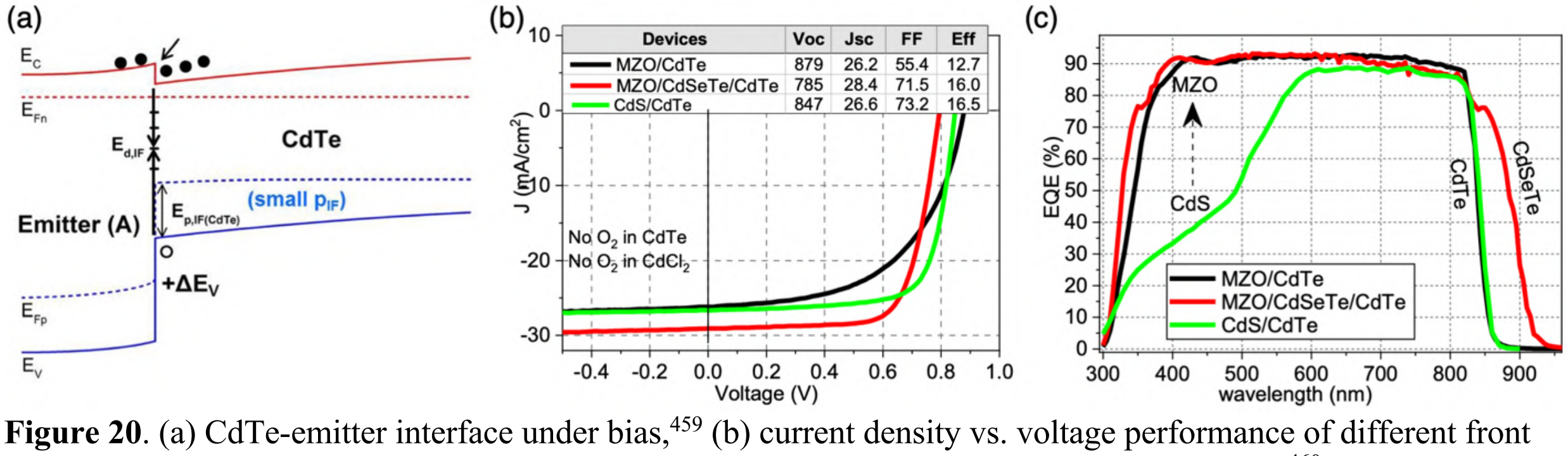

**Figure 20**. (a) CdTe-emitter interface under bias,[459] (b) current density vs. voltage performance of different front interfaces in CdTe solar cells, and (c) a corresponding external quantum efficiency (EQE) plot.[460] In (b) and (c), CdS and $Mg_xZn_{1-x}O$ (MZO) are compared as emitter layers.

Historically CdS ($E_G$ ~ 2.4 eV) was adopted as the standard "window layer" in PV, and it still is often used in research.[453,454,455,456] Although CdS provides favorable band alignment with CdTe, a spectrum loss is evident for photon energies larger than 2.4 eV (< 550 nm). Such losses can be avoided by using buffer layers with higher band gaps and similar alignment. An increased transmission can be achieved by replacing the CdS with n-type ZnS ($E_G$ of ~3.7 eV), but the band alignment at ZnS/CdTe interface is not optimal. Calculations show that the valence band offset at the interface is $\Delta E_{VB}$ = -0.2 eV, while there is a very large positive conduction band offset of $\Delta E_{CB}$ = ~1.1 eV.[461] Modifying the band alignment by alloying the ZnS with CdS has been proposed, with demonstrated cell efficiencies greater than 10%. Alloying of ZnS with ZnO has been also attempted.[462] More recently, CdS has been replaced with a $Mg_xZn_{1-x}O$ (MZO) alloy. Figure 20b and c demonstrate current density vs. voltage (JV) and external quantum efficiency (EQE) data from devices with different front interfaces showing the loss of current collection in CdS and how it improved when replaced with MZO. However, a new wide-gap n-type chalcogenide with good alignment could also be explored.

**Back contact to p-CdTe**. Forming a low barrier Ohmic back contact to CdTe to extract photogenerated holes is challenging and is an ongoing subject of research. The Ohmic back contact does not necessarily need to be transparent, but wide band gap p-type back contacts are helpful to reflect photo-generated electrons. They would also allow for use as window layers in tandem devices or bifacial PV. CdTe has a very high ionization energy of ~5.9 eV, which would require a metal with a work function greater than or equal to 5.9 eV. However, such a metal does not exist. Direct deposition of metals with high work functions such as Au onto CdTe generally leads to an interfacial reaction, which causes Fermi level pinning at the interface and a drop-in device performance due to series resistance.

In order to decrease the barrier between CdTe and the metal contact, intermediate layers (buffers) are typically used. One approach, which has yielded some of the highest CdTe solar cell efficiencies, is using a very thin (~1–3 nm) Cu or Cu-containing layer between the CdTe/metal interface. The diffusion of Cu into CdTe is hypothesized as enhancing the p-type conductivity of CdTe, while it may also help reduce the back barrier between CdTe and the metal. Successful demonstration of such contacts includes Cu/Au bi-layers,[463] Cu-containing graphite paste,[454] $Cu_2Te$,[464] $CuCl_2$ and Cu-doped ZnTe.[465] However, Cu may also be responsible for degradation,[466] so prospective area of research is finding a p-type non-Cu buffer layer that makes an Ohmic contact with CdTe.

An alternative strategy to mitigate this challenge is to add a thicker (~100–300 nm) electron reflector between the metal and CdTe. In addition to acting as an Ohmic contact for holes in the valence band, this layer raises the conduction band upward to create an energy barrier for electrons. Approaches to implement such an electron reflector include (1) creating positive conduction band bending in the rear interface while maintaining a near-flat valance band offset, by heavy p-type doping near the back CdTe interface, and (2) implementing a contact with sufficiently low electron affinity and high work function to reverse the normal downward band bending. Wide-gap p-type chalcogenide materials with proper alignment are optimal for this latter strategy. P-type ZnTe, with an band gap of ~2.3 eV, is currently is a prototypical electron reflector.[467] The valence band offset at the CdTe/ZnTe interface has been reported to be around 100 meV,[468,469] which is ideal for hole conduction through the interface. In addition, it has been reported that ZnTe can be highly p-type doped with Cu or N (>$10^{20}$ $cm^{-3}$), [470,471] which can create an effective tunnel junction to the metal layer. First Solar Inc. has used ZnTe as a buffer layer and credited it with recent improvements in its world record efficiency solar cells.[472] However, it is possible that another wide-gap p-doped chalcogenide may enable even higher efficiencies, so the materials previously discussed in this review may be interesting to explore for this application.

**Top contact to n-CdTe**. Achieving a high performing p-type transparent semiconductor also opens up the possibility of a device architecture using n-doped CdTe rather than p-doped CdTe, and creating a p-n junction with n-CdTe and a p-type wide band gap chalcogenide. A transparent p-type contact would need to let light through to the absorber, have its VB closely aligned to that of CdTe, and be of high enough quality to not induce interfacial recombination or device degradation. For example, high quality monocrystalline n-type CdTe devices have been grown with amorphous Si:H as a p-type contact, but the VB alignment is not ideal.[473] Chalcogenides tend to have ionization potentials closer to vacuum, so they may have closer VB alignment to CdTe. Recently, a wide-gap p-type $Cu_xS$:ZnS nanocomposite (see Section 3.1.7) was used as a transparent top contact to a monocrystalline n-type CdTe device, resulting in a high Voc of nearly 1 V, but such devices have yet to be optimized.[219] Implementation of other p-doped wide band gap semiconductors into n-type CdTe devices could be an interesting area for future research, and may enable exploration of new solar cell architectures.

### *4.1.2 CIGS solar cells*

High efficiency thin film solar cells based on alloys of $CuInSe_2$ (CIS) have been an ongoing research effort for decades. $CuInSe_2$ and its alloys crystallize in the chalcopyrite structure, as discussed previously (see Figures 10 and 11). The electronic and optical properties of CIS can be tuned by substituting In with Ga or Al and substituting Se with S. This leads to absorber material collectively called "CIGS" such as $CuIn(S,Se)_2$, $Cu(In,Ga)Se_2$, $Cu(In,Ga)(S,Se)_2$, etc., with a band gap range of ~1.1 eV to 1.7 eV for $Cu(In,Ga)Se_2$.[474] CIGS has emerged as one of the most promising PV materials for low-cost and high-quality thin film solar cells because of its tunable band gap, high absorption coefficient due to direct character of this gap ($\alpha > 10^5$ $cm^{-1}$), and high tolerance to off-stoichiometry, defects, and impurities.[475] CIGS absorbers are all intrinsically p-doped due to the presence of Cu vacancies ($V_{Cu}$), which are shallow acceptors.

A simplified structure of a typical CIGS cell is illustrated in Figure 19b. In contrast to the "superstrate" device configuration of CdTe with illumination from the glass side (Figure 19a), CIGS cells typically use a "substrate" configuration with illumination from the non-glass side. The front contact is a degenerately doped n-type TCO, usually Al-doped ZnO (AZO). A resistive ZnO layer is generally combined with the highly conductive TCO to form AZO/ZnO bi-layers to minimize shunting. Cells require both the front contact and the buffer layer to have a wide enough band gap to minimize spectral losses. The standard back contact is Mo, which sometimes selenizes to form a thin interfacial $MoSe_2$ layer. Wide band gap p-type oxides such as $MoO_x$ have also been researched at the back interface,[476] but no work on wide-gap chalcogenides for this layer could be found.

Various buffer layers have been used between the front contact and the CIGS absorber. The buffer layer is an n-type wide band gap material which forms a p-n junction with the p-type CIGS absorber. Similar to CdTe, the major selection criteria for the buffer layers are (1) a wide band gap to avoid undesirable absorption, (2) a suitable conduction band-offset, critical for minority carrier transport, and (3) a low defect density and controllable reactions at the interfaces. The second requirement is often the trickiest; although tuning the band gap can be achieved by varying the In/Ga or S/Se ratios in $Cu(In,Ga)(S,Se)_2$, this also alters CB alignment. For a typical, high-efficiency CIGS material composition of Cu/(In+Ga) = ~0.9 and Ga/(In+Ga) = ~0.3, the electron affinity is reported to be 4.5 eV.[477] Therefore, a wide band gap material with electron affinity of ~4.2–4.5 eV is preferred, which will produce flat band offsets, or a 0.3 eV spike in the conduction band going from CIGS to a buffer.

CdS has been widely adopted as a buffer layer, with a spike-type CB offset. The experimental determination of the VB offset at the CIGS/CdS interface is $\Delta E_{VB}$ = 0.8–1.1 eV,[478] leading to an estimated 0.0–0.3 eV CB offset. Theoretical calculations suggest the lack of a strong dependence of $\Delta E_{VB}$ on the Ga content, since the introduction of Ga into $CuInSe_2$ primarily shifts the CB to higher values,[479] and experimental observations corroborate the theory.[479] Thus, for CIGS solar cells with a higher Ga content, finding an alternative n-type buffer layer is especially important.

For this reason, and due to the small band gap of CdS (2.4 eV) and spurious optical absorption, alternative buffer layers have been developed. Binary chalcogenides such as n-type CdSe, ZnS, ZnSe, and ZnTe have been investigated as potential buffer layers.[480] Currently, Zn(S,O) is extensively used. Adjusting x in $ZnO_{1-x}S_x$ can tune the gap from that of wurtzite ZnO (3.2 eV) to that of zincblende ZnO (3.6 eV), and conduction band offsets, $\Delta E_C$, from -0.1 to 0.3 eV at the Zn(O,S)/CIGS interface, albeit with large band bowing at intermediate compositions. This allows for largely unimpeded electron transfer, forming a foundation for high-efficiency Zn(O,S) buffer based CIGS solar cells.[481] N-type $In_2S_3$ has also been investigated as a buffer layer, resulting in ~16% efficiency cells.[482] As discussed in Section 3, the optical band gap of the $In_2S_3$ are 1.9 – 2.8 eV, where the range is due to different phases, deposition techniques and the In/S ratio. CB offsets of the $In_2S_3$/CIGSe interface have been reported in a wide range, from -0.45 to 0.5 eV.[483] However, the interfacial chemistry is complex, and several reports have observed chemical reactions occurring at the interface.[484,485,486] The highest efficiencies in CIGS solar cells with CdS and Zn(O,S) buffer layers are 22.3% and 22.8%, respectively, [487] and the search for the next generation of buffer layers is still underway.

Other chalcogenide materials mentioned in Section 2 have been also researched for CIGS solar cells. A thin p-type $CuGaS_2$ layer, sandwiched between a p-type $CuInS_2$ absorber layer and a Mo back contact, has a beneficial effect on the performance of thin-film CIGS solar cells fabricated on glass substrates.[488] A thin interlayer of $CuAlS_2$ has been also considered promising for its high thermal stability and wide band gap. Besides CIGS, n-type and p-type wide-gap chalcogenide materials have been researched for $Cu_2ZnSn(S,Se)_4$ solar cells (i.e. CZTS, $CZTS_{1-x}Se_x$), with quaternary kesterite crystal structures (see Section 3.5.6) and similar material parameters as CIGS. Similar to CIGS, CdS and ZnS buffer layers have been used for CZTS solar cells.[489–491] However, since CZTS devices have similar interfacial properties but lower efficiency than CIGS, this review does not go into detail and instead we refer the reader elsewhere.[492]

### *4.1.3 Additional PV applications*

**Silicon PV.** An ongoing field of research in silicon heterojunction (SHJ) photovoltaics is the development of hole-selective contacts to Si to replace traditional homojunctions formed by diffusion doping.[493] An example of an SHJ device is a HIT cell ("heterojunction with intrinsic thin layer"), with a heterojunction between amorphous Si and crystalline Si. Wide-gap p-type transition metal oxides such as $MoO_x$ and $WO_3$ have been reported as hole-selective contacts with high $V_{OC}$, and transparent graphene and carbon nanotubes have been investigated as well.[494,495] This is another space where implementation of a wide-gap chalcogenide-based semiconductor with sufficient alignment to the Si VBM could be fruitful, so long as passivation of the absorber is maintained. For example, it was recently demonstrated that a nanocomposite of $Cu_xS$:ZnS could be used as a hole-selective contact to n-type Si, and proof-of-concept devices were fabricated with $V_{OC}$ up to 535 mV and $J_{SC}$ of 21 mA/cm$^2$, using 1 sun illumination.[496] There also may be applications for wide-gap chalcogenide contacts in bifacial Si solar cells.

**Hybrid organic-inorganic halide perovskite solar cells**, based on $CH_3NH_3PbI_3$ (i.e. "MAPI") and related absorbers, have gained huge prominence in the last decade, but there are still many major engineering challenges to be solved. The challenge of relevance to this review article is finding a high performing hole transport layer (HTL or HTM).[497] Typically, polymer materials such as Spiro-OMeTAD have been used for this layer, but their conductivities are very low, and finding optimal replacements is an active area of research. Wide-gap chalcogenides could serve as this layer due to their high mobilities and adequate band alignment. As depicted in **Figure 21**a and b, CuSCN is a common material used as an HTM in both perovskite solar cells and dye sensitized solar cells (DSSC).[344,498,499] Kesterite $Cu_2ZnSnS_4$ has been also applied as an HTM to perovskite solar cells.[500] An additional criteria would be to function dually as a capping layer to protect the perovskite absorber from degradation under ambient atmosphere and heating, so materials with high stability are preferable for this application. CuSCN has been shown to be more thermally stable than the classic Spiro HTM (see Figure 21c).[501]

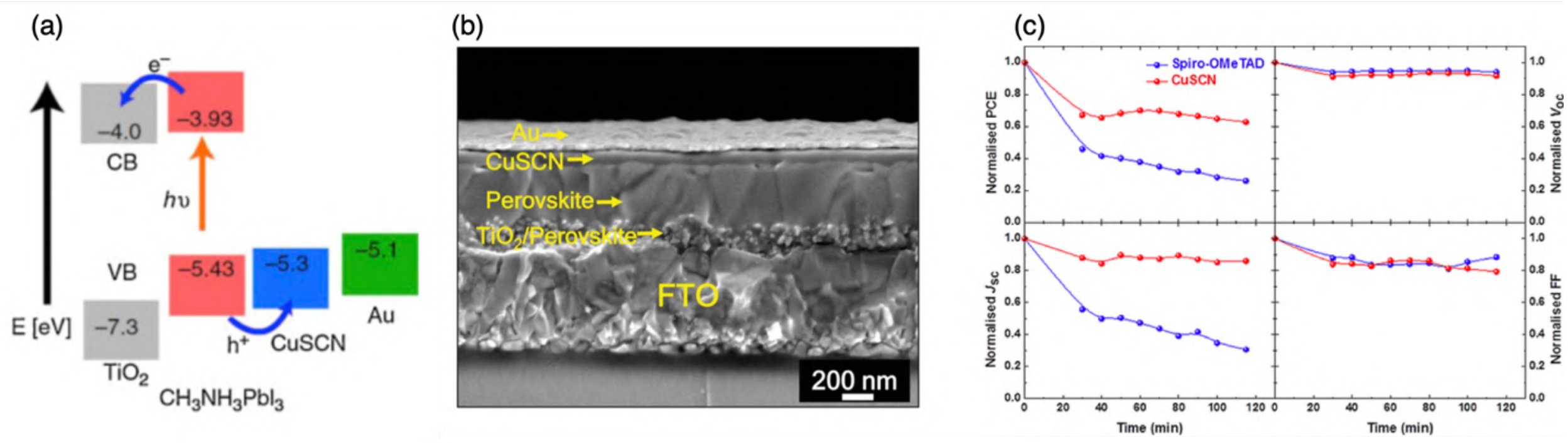

**Figure 21**. CuSCN as a wide-gap mixed-anion chalcogenide in a perovskite solar cell. (a) Band alignment in a perovskite solar cell, demonstrating the adequate VBM position of CuSCN with respect to $CH_3NH_3PBI_3$,[502] (b) TEM interface image from Arora et al.,[498] and (c) reduced thermal degradation of cells using CuSCN, compared to Spiro-OMeTAD, from Jung et al.[501]

**Tandem or multijunction photovoltaics** consist of two stacked solar cells that selectively absorb different sections of the solar spectrum.[503] In order to make contact between the two layers, these devices require a tunnel p-n junction in between the two cells that can selectively transport current but let enough light through to be absorbed by the bottom cell. N-type TCOs are currently used in this application, and development of an adequate p-type transparent conductor would have important implications. This is another area where a wide-gap p-type chalcogenide could be useful, in particular for tandems with chalcogenide absorbers.

**Intermediate band solar cells (IBSCs)** are a creative approach to achieve efficiencies beyond the Shockley-Queissar limit.[504] Rather than optimizing the fundamental band gap for solar absorption, IBSCs intentionally introduce a deep level, mid-gap defect such that absorption occurs at two different wavelengths, from the CB to the deep level and from the deep level to the VB.[270] Wide-gap chalcopyrite chalcogenides have been explored as the absorbers for IBSC due to their wide, direct gap, and their ease of dopability. There have been investigations on possible IBSC dopants in chalcopyrites $CuAlS_2$, $CuGaS_2$, $AgGaSe_2$, and $AgGaTe_2$,[271,278,505] but very few device related publications could be found in literature.

## 4.2 Other devices

### *4.2.1 Transparent thin film transistors*

Transparent thin film transistors (TFTs) require wide band gap semiconductors as the active/channel layer to ensure visible range transparency. An advantage of wide band gap semiconductors for TFT applications is that their intrinsic carrier concentrations tend to be lower than for narrow band gap materials (since the former are more difficult to dope), thus achieving low "off" current. Another advantage is that the wider band gap generally results in a larger band offset, leading to better control of the carrier concentrations by the gate voltage. Simultaneously, high electronic quality of the channel's interfaces with the insulator is generally required to maintain high carrier mobilities without scattering on interfacial defects. Thus, wide band gap semiconductors with high mobility but low carrier concentrations are suitable as the active layer in transparent TFTs. N-type oxides such as amorphous In-Ga-Zn-O (IGZO) meet these requirements and are the conventional choice of channel layer material in unipolar switching devices.[511] On the other hand, chalcogenide wide band gap semiconductors may enable bipolar devices, complementary metal oxide semiconductor (CMOS) technology and transparent circuits, since they have better propensity for p-type doping the oxides.

Figure 19c depicts a typical schematic of a transparent TFT and its band alignment under no bias. The source electrode is grounded, a fixed voltage bias is applied to the drain electrode, and the gate voltage $V_g$ is used to control the carrier concentration in the channel, e.g. the current between source and drain. For example, in an n-type TFT when $V_g$ is above the threshold voltage $V_{th}$, the channel layer CBM bends down and is closer to $E_F$, leading to a higher concentration near the insulator/channel layer interface, called the "on" state. In contrast, if $V_g$ is lower than $V_{th}$, the CBM bends upwards and causes a lower carrier concentration, i.e. the "off" state. Thus, the current between source and drain, $I_{DS}$, shows a clear on-off switching as a function of $V_g$. $I_{DS}$ from p-type TFTs changes inversely compared to n-type TFTs (**Figure 22**). There are several p-type oxides used as TFT channels, mainly based on CuO or SnO.[7] But their performance is not as good as for their n-type counterparts, likely due the low hole mobility originating from localized oxygen $2p$ orbitals. By introducing a chalcogen, more covalent bonds are formed between cations and anions and a higher mobility is expected, which may lead to better performance.

There have only been a few reported attempts to implement wide-gap chalcogenides into transistors, and their high achieved mobilities offer inspiration to continue to explore this material space. For example, the fabrication of an n-type $SnS_{2-x}Se_x$ TFT by spin coating resulted in an n-type mobility greater than 10 $cm^2\ V^{-1}\ s^{-1}$, comparable to the mobility of IGZO based TFTs (~20 $cm^2\ V^{-1}\ s^{-1}$).[506,507] TFTs using p-type ZnTe as the channel layer were first reported as far back as 1967, with mobilities of 1–10 $cm^2\ V^{-1}\ s^{-1}$.[508] Since then, p-type ZnTe TFTs were integrated with n-type ZnO TFTs to form a complementary logic circuit, but the low on-current limited the on/off ratio to only ~100.[509]

N-type CdS TFTs have also been reported to function in both accumulation and depletion mode, with a mobility of 1.25 $cm^2\ V^{-1}\ s^{-1}$.[510] CdS is promising due to its high mobility >100 $cm^2\ V^{-1}\ s^{-1}$ and its low temperature processing for flexible electronics.[511] Additionally, n-type CdSe was reported decades ago with a mobility of 4–30 $cm^2\ V^{-1}\ s^{-1}$.[512] P-type PbS TFTs have been prepared chemically, with low mobility in as-deposited devices but improvement after thermal annealing.[513,514] We note that channel layers are not limited to wide band gap semiconductors. For example, TFTs using narrower band gap (<2 eV) semiconductors as channel layers have been reported, e.g. $CuInTe_2$.[515]

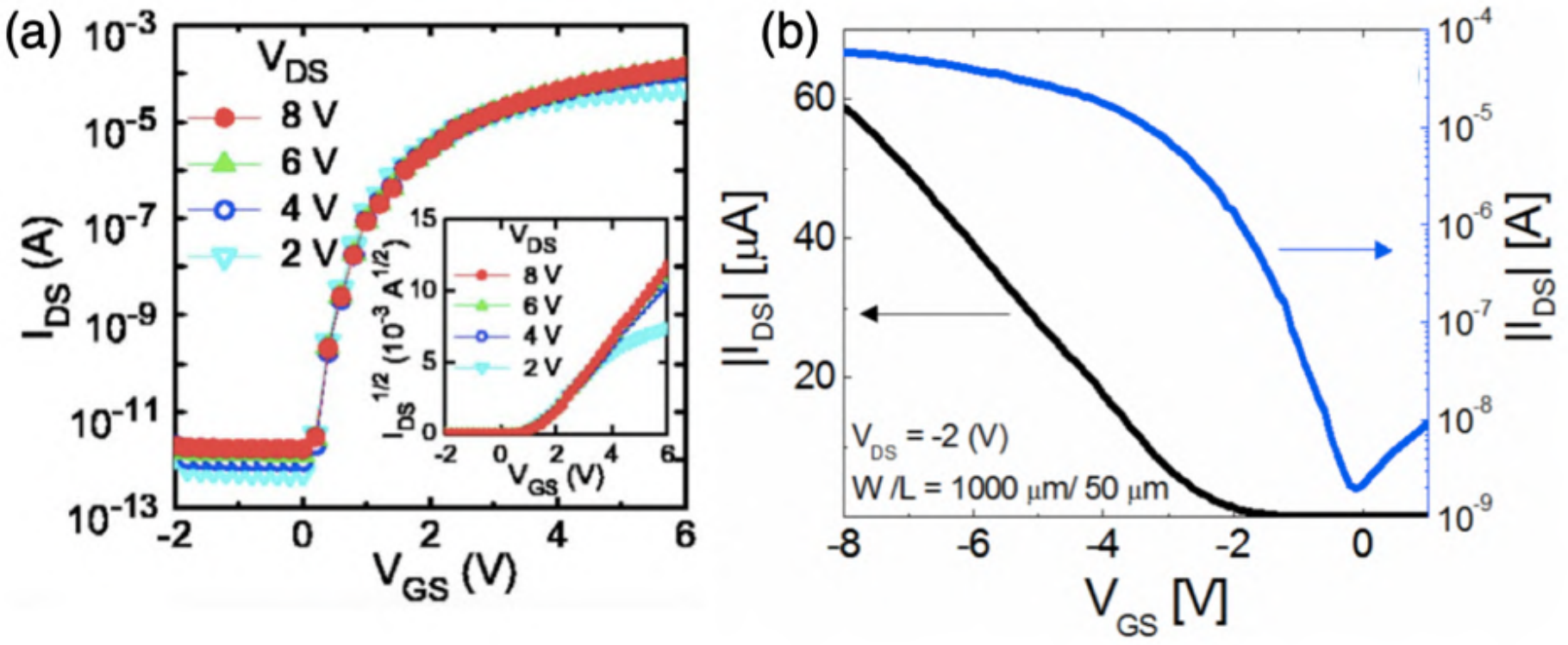

**Figure 22**. Example transfer curves of (a) an n-type TFT (IGZO, reproduced from the literature[516]), and (b) a p-type transparent TFT (SnO, reproduced from the literature[517]).

### 4.2.2 Water splitting photocatalysts and photoelectrocatalysts

Wide band gap chalcogenides have suitable properties for applications in both two types of water splitting mechanisms — as photocatalysts for photocatalytic (PC) devices and as photoelectrocatalysts for photoelectrochemical (PEC) devices — and have been extensively investigated.

**Photocatalysts.** Due to relatively high CB positions compared to oxides, wide band gap chalcogenides are attractive choices as photocatalysts for "overall water splitting." The three fundamental criteria of a photocatalyst material for $H_2$ generation by water splitting are (1) suitable band alignment, (2) optimal band gap, and (3) resistance to corrosion in electrolyte solution. In particular, the CBM potential must be lower than 0 V vs. the normal hydrogen electrode (NHE) to reduce water, and the VBM potential must be higher than 1.23 V vs. NHE to oxidize water. Although the difference between water reduction and oxidation potentials is only 1.23 eV, for a single semiconductor photoelectrode the band gap must lie within the range of $1.6 < E_G < 2.4$ eV.[518] This is due to the required overpotential to drive the hydrogen evolution reaction (HER) and oxygen evolution reaction (OER) at the same time, as well as to account for different loss mechanisms in practical devices. Other desirable properties are high carrier mobility to enhance charge transfer, and good electron and hole mobilities to increase charge separation and to reduce recombination. A trade-off between crystallinity and active surface area is also of interest, as a higher degree of crystallinity improves charge transport while a nanoporous or amorphous structure provides more active sites for catalysis.

Multinary metal chalcogenides such as ZnS:$CuInS_2$, ZnS:$CuGaS_2$, $ZnIn_2S_4$, and $NaInS_2$ have been explored in search of improved photocatalytic properties in the visible spectrum.[519–522] For a visible light driven hydrogen evolution reaction, the band gaps of ZnS, $CuGaS_2$, and $AgGaS_2$ are too wide to absorb enough light. For $CuInS_2$ and $AgInS_2$, the conduction band potential is too low. Alloyed wide-gap chalcogenides with optimized compositions (e.g. $CuIn_{0.3}Ga_{0.7}S_2$ and $AgIn_{0.1}Ga_{0.9}S_2$) can balance the optimal properties of band gap and CB alignment and tune them to desirable levels.[523,524] Computational screenings have recently provided a promising avenue for searching for photocatalysts. Based on first principles analysis, $Ba_2ZnSe_3$, with a band gap of 2.75 eV, was predicted to be a potential candidate for visible light responsive photocatalytic water splitting.[525] Lower optical absorption of wide band gap chalcogenides in the solar spectrum can be overcome by careful tuning of morphology. CdS in nanoparticle, nano-rod, and nanosheet structures exhibit high photocatalytic efficiency for hydrogen generation.[526–528] ZnS has been investigated with various metallic dopants (Ni, Cu, Pb).[529] Additionally, expensive noble metals such as Pt, Ru, Ag, and Au are required as co-catalysts for HER. Wide-gap chalcogenides such as $MoS_2$ and $WS_2$ have been investigated as inexpensive substitutes.[530,531]

Other wide band gap chalcogenides remain to be investigated. Figure 24 labels the HER and OER potentials ("$H^+/H_2$" and "$O_2/H_2O$", respectively), indicating the wide-gap chalcogenides from this review that can have application as water splitting photocatalyst. Materials with CBMs above the hydrogen reduction potential can be suitable HER catalysts, those with VBMs below the oxygen reduction potential are prospective OER catalysts, and those which meet both criteria can be applicable for overall water splitting, given that the stability and required overpotential criteria are also met.

**Photoelectrocatalysts.** In photoelectrochemical (PEC) water splitting, photogenerated electrons and holes split the water into hydrogen and oxygen. An external bias may be necessary based on the band alignment of the semiconductor being used. An alternative kind of water splitting device involves combining a photoelectrode with photovoltaic (PV) cells in a multi-junction hybrid structure, also known as a hybrid photoelectrode, and is essentially a solar cell submerged in electrolyte solution. Tandem device architecture with a wide band gap top absorber is employed to generate the required potential for driving the water splitting reactions and to efficiently utilize the solar spectrum.[532]

Wide band gap chalcogenides are particularly attractive in single material systems, or as top electrodes in tandem systems, due to their high open circuit potential. With proven PV device performance, low cost, and tunable band gap, wide band gap varieties of different known chalcogenides are actively being investigated for PEC device applications. Band gaps of $Cu(In,Ga)S_2$ can span 2.05–2.45 eV based on In and Ga content, and current generation capacity up to -5.25 $mA/cm^2$ at saturation has been demonstrated from such photocathodes.[533] Different $CuGaS_2$ based photocathodes are also investigated (see **Figure 23**), which in conjunction with $Co:BiVO_4$ photocathode, showed solar water splitting without external bias.[534,535] CdS can enhance the performance of p-type chalcogenide photocathodes by creating a p-n junction, resulting in enhanced photocurrent and onset potential. CdS-modified $CuGaS_2$ and $(Ag,Cu)GaSe_2$ electrodes have exhibited stability beyond 10 days in electrolytes and $H_2$ evolution has been demonstrated.[536,537] $In_2S_3$ also has similar applications and has been investigated with $CuInS_2$ photocathodes due to suitable band alignment.[538,539]

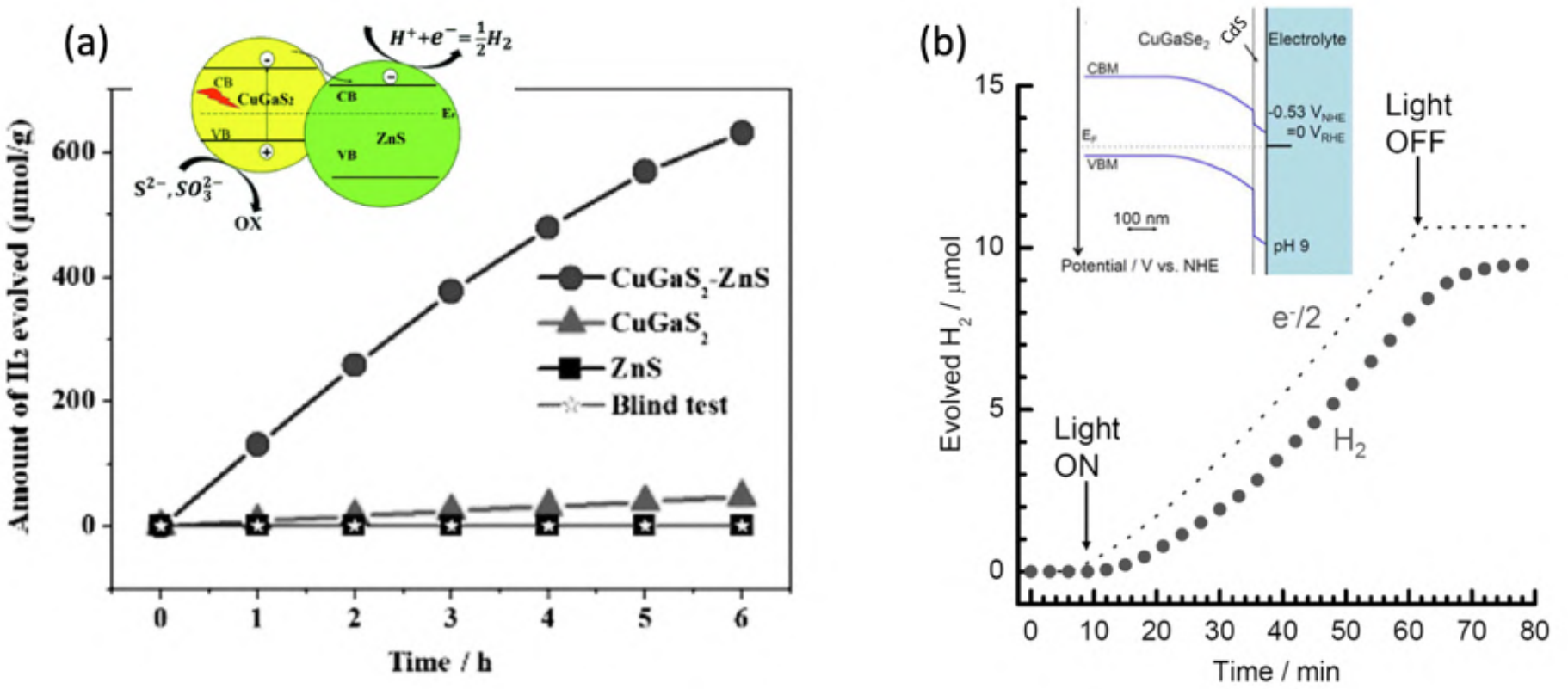

**Figure 23**. (a) Photocatalytic $H_2$ production from $CuGaS_2$:ZnS nano-heterostructure under visible light irradiation; reproduced from the literature.[520] (b) Photoelectrochemical $H_2$ evolution from a $Pt/CdS/CuGaSe_2$ electrode (0.09 $cm^2$) under 300W Xe lamp with an electrode potential of -0.1V RHE; reproduced from reference.[536]

#### *4.2.3 Diodes*

**Light emitting diodes** (LEDs) allow for direct conversion of electrical energy to light. A simple device structure and band diagram schematic is depicted in Figure 19e. Holes and electrons are injected via the hole transport and electron injection layer, respectively, and combine in the active layer to emit light. Some wide band gap chalcogenides are used in LEDs as transport and injection layers. Their wide band gaps enable light to escape the device, maximizing the luminescence quantum efficiency. Band alignments of the transport and injection layers with the emitting layer, as well as their high conductivities, are important interface and materials properties for this application. Here, we mention some examples of wide band gap chalcogenides used in LEDs. In addition to acting as transport layers, wide-gap chalogenides have been used as active layers too (e.g. ZnS,[540] CdS,[541] etc.).

N-type CdS has been studied as an electron transport layer in LEDs by forming a heterojunction with poly p-phenylene vinylene (PPV).[542] The close band alignment between CdS and PPV, compared to that between Al and PPV, lowers the energy barrier for electron injection. Concurrently, CdS blocks holes from flowing to the negative electrode, and helps electron injection. 2D materials such as $MoS_2$, $WS_2$, and $TaS_2$ have also been investigated as hole injection layers to replace contact materials in organic LED (OLED) applications, since these chalcogenides won't cause degradation of the organic components like ITO electrodes do.[543]

A few p-type chalcogenides have also been deployed in LEDs as hole transport or injection layers. Hole-injection barrier formed between oxygen plasma treated LaCuOSe:Mg and *N*,*N*′-diphenyl-*N*,*N*′-bis (1,1′-biphenyl)-4,4′-diamine (NPB) can be as low as 0.3 eV, approximately half that of a conventional ITO and NPB interface, indicating its potential as an efficient transparent anode for OLEDs and other organic electronic devices.[544] A chemically inert interface with a hole injection barrier of 0.11 eV was formed at the interface of BaCuSeF and ZnPc (zinc phthalocyanine), probably providing improved device performance for OLEDs and OPVs compared with ITO electrodes.[545] A p-type CuSCN layer can help balance electron and hole injection rates in n-ZnO nanorod/p-CuSCN heterojunction LEDs, enhancing efficiency of radiative recombination.[546] CuSCN is advantageous due to its wide band gap, hole-transport characteristics, and low-temperature solution-processability.[547]

In the past, ZnS has been heavily studied as a candidate active material for blue LEDs,[540,548] but these research efforts decreased with the advent of GaN-based LEDs. More recently, core/shell nanocrystal quantum dots of CdSe-ZnS[549] and CdSe-CdS,[541] among other systems, have been studied as efficient electroluminescent materials to be incorporated into light emitting diodes. The good alignment between CdSe nanocrystals and the metal electrode contributes to a low operating voltage. The color of the emission can be tuned by voltage due to quantum size effects.[550] Single crystal p-type ZnTe can also be made into green LED using an Al-diffusion technique at the pn junction.[551]

## 4.3 Band offsets of wide-gap chalcogenides for devices

In Section 3 we have discussed a large set of wide-gap chalcogenide semiconductors that could be used for device applications, and assessed their optoelectronic properties. However, in Section 4 we have seen that the materials used in solar cells, transparent transistors, etc., are limited to only a few candidates, primarily CdS, ZnS, ZnTe, $In_2S_3$, CuSCN and $SnS_2$. What is preventing researchers from incorporating less conventional chalcogenides into these devices as contacts or active layers? First may be simply the lack of recognition of which materials are available and what their corresponding properties are. We hope to have alleviated this barrier with the discussion of potential materials and their properties in Section 3, including a summary in Figure 18 and tabulated values in Tables 1–4. Second, in order to select an appropriate material for a device application, it is essential to understand the alignment of the valence and conduction bands at interfaces and the types of junctions that will form. This can be a somewhat trickier constraint, as alignment changes drastically depending on environment, grain orientation, interface configuration, etc.

To help assess band alignment for current and future device applications, in **Figure 24** we compile and tabulate the band offset positions reported in the literature for many of the wide-gap chalcogenide semiconductors discussed previously. When available, we have reported measurements from x-ray and ultraviolet electron spectroscopy (XPS/UPS), referenced to vacuum levels. If these have been not experimentally measured, we have reported calculated offsets and denoted these cases with a * suffix (see S.I. for level of theory and DFT functional used). Band offsets are arranged in increasing order of ionization potential, and color represents reported doping type in the literature. The grey materials on the right side of the graph are known absorbers in solar cell applications (CdTe, CIGS, Si, GaAs). References for all reported values are tabulated in the Supporting Information. The dotted lines correspond to the HER and OER potentials. We also note that uncertainties in XPS/UPS data are typically >0.1 eV, and that it is strictly a surface measurement. DFT alignments can vary by crystallographic facet orientations considered in the model and averaging scheme, so they may not robustly represent experimental alignments.

This figure illustrates that the band edges of the wide-gap chalcogenides reviewed here vary drastically, from VBMs of $SnS_2$ at ~9.5 eV (CBM: ~ 7.3 eV) to BaCuSeF at ~3.6 eV (CBM: ~ 0.9 eV). Considering the wide array of possible alloys between these compounds, as well as new chalcogenides with band edges not yet known, this gives an immense amount of possibilities for tuning band alignment to a given application of interest. For example, from comparing BeTe and ZnTe we can hypothesize that a $Be_xZn_{1-x}Te$ alloy may result in a VBM aligned with that of CdTe. In addition to the transport properties we've discussed so far, these band offsets are key first steps to understanding band bending and charge transport at the interfaces of the heterojunctions we have discussed in this section. The next steps are explicit interface simulations via slab supercell approaches and junction modeling, and are covered in depth elsewhere (see Section 2.5.1). Ultimately experimental band alignment measurements of real interfaces is needed to assess band positions and bending. This has been performed for many of the chalcogenide interfaces discussed in Section 4,[120,286,483,332,552] but not for the less extensively studied materials.

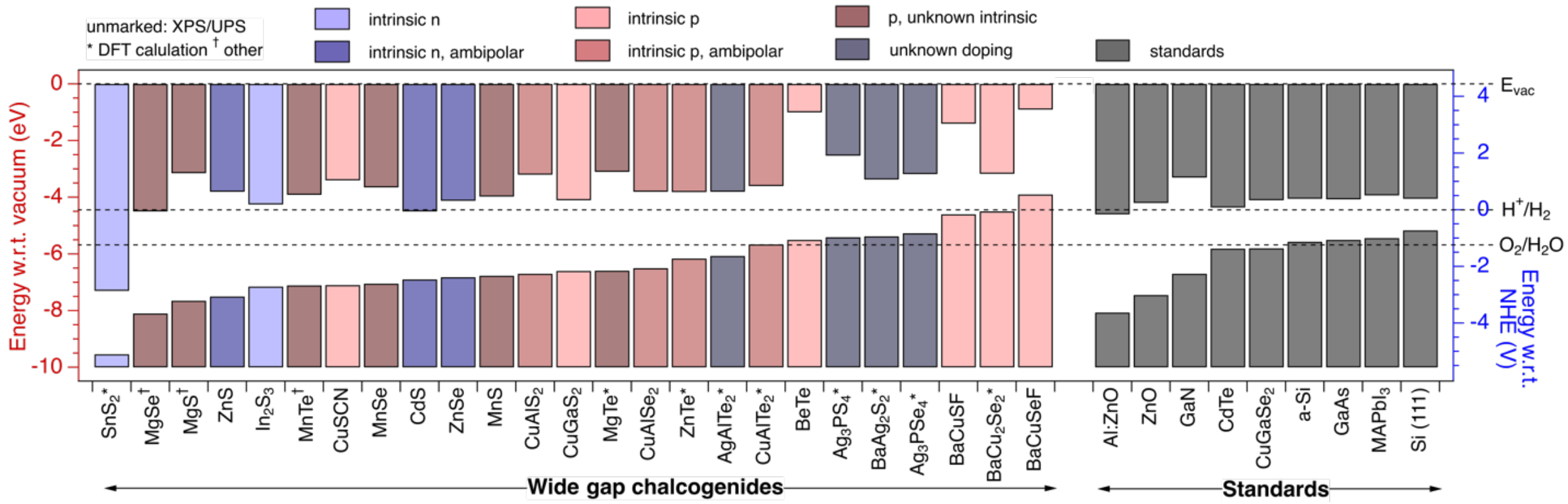

**Figure 24**. Literature reported band offsets of the wide band gap chalcogenides from this review, referenced to vacuum energy $E_{vac}$ and to the normal hydrogen electron (NHE). HER and OER potentials are labeled as "$H^+/H_2$" and "$O_2/H_2O$", respectively. Values, measurement or calculation details, and corresponding references are tabulated in the Supporting Information. We include several PV absorbers and n-type TCOs as standards (grey). All values are determined from x-ray and ultraviolet electron spectroscopy (XPS/UPS), except those with * are from DFT calculations (see S.I. for DFT functional used), and those with † are from another technique. Colors indicate the doping type reported in the literature. We note that modifications due to surface dipoles, band-bending, polar mismatch, and strain are not included.

# 5. Summary and outlook

In this review, we have explored the current state of materials research in wide band gap ($E_G$ > 2 eV) chalcogenide (*Ch* = S, Se, Te) semiconductors, demonstrating that this is an important space for ongoing and future research. In Section 1, we introduced the existing material space of wide-gap chalcogenides in comparison to oxides and other chemistries. In Section 2, we gave an overview of computational and experimental design criteria and methodology to investigate functional wide-gap chalcogenide semiconductors. In Section 3, we focused on distinctive chemical bonding and properties of chalcogenides and then covered general classes of wide band gap chalcogenide semiconductors that have been experimentally synthesized, including binary chalcogenides *MCh* (e.g. Zn*Ch*), ternaries (represented by chalcopyrites e.g. $CuAlCh_2$, sulvanites e.g. $Cu_3TaCh_4$, α-$BaCu_2S_2$ and related structures), layered mixed-anion oxychalcogenides (e.g. LaCuO*Ch*, BaCu*Ch*F, etc.), and touched on multinary compounds as well as 2D chalcogenides. We focused on their respective optical properties as well as the mobilities and conductivities they have achieved (see Tables 1–3), and commented on the influence of structure and chemistry on these properties. And finally in Section 4, we elaborated upon the use of wide-gap chalcogenides as functional layers in optoelectronic devices. Specifically, we looked at solar cells, thin film transistors, water splitting devices, and light emitting diodes. Most of these devices use simple binaries (e.g. CdS, ZnTe), so we suggest incorporation of more complex chalcogenide materials into these devices to address technological challenges. A conclusion from this review is that wide band gap chalcogenide material and device spaces are still underexplored. We discussed the paradigm of materials exploration as it applies to searching for new wide-gap chalcogenide semiconductors, including several classes of computationally predicted compounds, and proposed potential future directions in new device applications of known materials.

We have discussed both n-type and p-type materials (see Figure 18 and Tables 1–4), but in particular have covered dozens of wide-gap *p-type* chalcogenide compounds and demonstrated how computational methods enable future discoveries. Prospective guidelines for computational discovery of new chalcogenide (semi)transparent p-type conductors are (1) to further investigate previously synthesized materials in databases for their properties, (2) to take structural analogs of oxides or other counterparts (e.g. chalcogenide delafossites or perovskites, see Sections 3.4.3 and 3.4.4) in order to find new materials, or (3) to propose new stoichiometries and chemistries altogether (e.g. $Cs_2Zn_3Se_4$). As noted, only a portion of these materials have been calculated to be p-type dopable; only a fraction of those have specific dopants proposed, and even fewer have been synthesized with proposed dopants. As mentioned previously, tools are now available (e.g. PyCDT, PyLADA) to ease implementation of defect calculations in high throughput computational screenings. Defect calculations and corresponding experimental exploration of dopants will be essential to understand physical properties and potential optoelectronic applications. In addition to further studies of dopability, defects, and synthesizability, it is also important to consider the band offsets and alignments of these materials to justify their use in specific optoelectronic devices.

What are some next steps that can be taken through experimental materials engineering and design? First, it will be important to synthese and characterize computationally predicted wide-gap conducting p-type materials (e.g. ZrOS, $Cs_2Zn_3Se_4$, etc.). Second, we suggest continued research into combining known compounds with complementary properties to tune desired functionality. This involves continued investigation into anion and cation alloying in the materials discussed in this paper (e.g. $Cd_{1-x}Zn_xS$, $Cu_xAg_{1-x}AlS_2$, $LaCuOS_{1-x}Se_x$), and in others we have yet to discover. As shown in this review, as well as in the n-type TCO literature, many material breakthroughs have originated from tuning properties in solid solutions and alloys. In addition, exploring composite material systems (e.g. $Cu_xS$:ZnS, Section 3.1.7) and using "$\delta$-doping" (e.g. Zn*Ch* system, see Section 3.1.1) to combine different systems could leverage the desired properties of each subsystem. Third, this review promotes clever engineering of known chalcogenide compounds, including new processing techniques, non-equilibrium synthesis methods, post-processing, etc., to promote carrier transport and high doping in chalcogenide host materials. Additionally, the community could engineer quasi-molecular structures such as CuSCN, where the VBM is more disperse than the CBM, and explore band engineering by thinning materials to tune band gaps and mobilities (see Section 2.6). These suggested approaches are inspired by the progress so far in the wide-gap chalcogenide community, as discussed in this review.

Finally, this review motivates continued investigation of interfaces of wide band gap chalcogenides and their device applications. Developing a theoretical understanding of stability, degradation, and interface modeling in chalcogenides is essential to deciding which materials can be used in devices. In particular, the band alignment of many of the systems we have discussed remain undetermined, and understanding them will be critical to tailoring materials to a particular device. Experimentally, this can be realized by synthesis of heterostructures and measurement of their interfacial band offsets. Incorporating both n-type and p-type chalcogenide materials into existing devices, as well as designing new device architectures using these unique materials, could potentially lead to significantly higher performances and efficiencies. Together, this suite of future advances into wide band gap chalcogenide materials and devices containing such materials could be a key component in accelerating society's renewable energy transformation and enabling future technological advances.

# Acknowledgements

This work was authored at the National Renewable Energy Laboratory, operated by the Alliance for Sustainable Energy, LLC, for the U.S. Department of Energy (DOE) under Contract No. DEAC36-08GO28308. Funding is provided by the U.S. DOE, Office of Science, Basic Energy Sciences, as part of an Energy Frontier Research Center "Center for Next Generation Materials by Design (CNGMD)". R.W.R. acknowledges her Ph.D. funding from the UC Berkeley Chancellor's Fellowship and the National Science Foundation (NSF) Graduate Research Fellowship Program (GRFP) under Grant Numbers DGE1106400 and DGE1752814, and thanks Dr. Shyam Dwaraknath and Dr. Matthew Horton for discussion. Y.H. acknowledges his stipend from China Scholarship Council, and in particular thanks Yan Wang for her inspiration and discussion. H.Z. acknowledges funding support from Office of Basic Energy Sciences, Division of Chemical Sciences, Geosciences and Biosciences. The views expressed in the article do not necessarily represent the views of the DOE or the U.S. Government.

# Author contributions

A.Z, R.W.R., and Y.H. conceptualized the review and edited the manuscript. R.W.R. and Y.H. co-wrote Section 2 and Sections 3.1–3.5. R.W.R. wrote Section 1, Section 4.3, Section 5, and introductions to all sections. H.Z. wrote Section 3.6. T.A. and R.W.R. wrote Section 4.1. Y.H. wrote section 4.2.1 and 4.2.3. I.K. wrote section 4.2.2. R.W.R. created original Figures 1, 2, 3, 4, 6, 7c, 11, 17c, 18, 19, and 24, and Y.H. created original Figures 5 and 10d. A.Z. and K.P. supervised the work.

# Author bios

**Rachel Woods-Robinson** is a Ph.D. Candidate at UC Berkeley, researching at the Lawrence Berkeley National Laboratory and National Renewable Energy Laboratory under Prof. Kristin Persson and Dr. Andriy Zakutayev. She received a Physics B.S. from UCLA, and is an NSF Graduate Research Fellow and UC Chancellors Fellow, and. Her research focuses on combining computational and experimental materials discovery of p-type transparent chalcogenide semiconductors for photovoltaics. She is also the co-founder of Cycle for Science, a STEM outreach organization.

**Yanbing Han** is currently a researcher at Zhengzhou University. He received his Ph.D. in Physical Electronics in 2019 from Fudan University. During his Ph.D., he spent 2016–2018 at the National Renewable Energy Laboratory as a visiting student, researching chalcogenide semiconductors and stabilization of metastable polymorphs under the supervision of Dr. Andriy Zakutayev. His research interests include novel oxide and chalcogenide semiconductors and their applications.

**Hanyu Zhang** obtained a Ph.D. in 2016 from the School of Mechanical Engineering at Purdue University as well as his BSME and MSME. He currently is working as a postdoctoral researcher at the National Renewable Energy Laboratory. His research focuses on investigating 2D transition metal dichalcogenides for energy conversion applications via surface modulation and phase engineering.

**Tursun Ablekim** is a postdoctoral research scientist within the thin film photovoltaics group at the National Renewable Energy Laboratory (NREL) in Golden, CO. He joined NREL in 2017 after receiving his Ph.D. in Materials Science and Engineering from Washington State University with Prof. Kelvin Lynn as his advisor. His graduate work was focused on synthesis and defect analysis of CdTe bulk crystals. Tursun's current research interests include II-VI semiconductor materials and their applications for energy conversion.

**Imran Khan** is a postdoctoral researcher at the National Renewable Energy Laboratory. His current work involves the synthesis and characterization of novel oxide and nitride thin films for photovoltaic and photoelectrochemical applications. He obtained his B.S. from Bangladesh University of Engineering & Technology. He received his M.S. and Ph.D. at the University of South Florida, focusing on the development of dye sensitized solar cells and cadmium telluride solar cells, respectively.

**Kristin Persson** is an Associate Professor at UC Berkeley and a Senior Faculty Scientist at Lawrence Berkeley National Laboratory. She is the Director of the Materials Project (www.materialsproject.org), a leading Materials Genome Initiative program. She is an Associate Editor for Chemistry of Materials, and has published over 160 papers in peer-reviewed journals. She has received the DOE Secretary of Energy's Achievement Award, the TMS Faculty Early Career Award, and she is a Kavli Fellow.

**Andriy Zakutayev** is a scientist at National Renewable Energy Laboratory, working on discovery and design of new materials for energy applications. To achieve these goals, he applies high-throughput combinatorial thin film methods in close iterative coupling with first principles calculations. Dr. Zakutayev received his Ph.D. in Physics from Oregon State University, where he studied mixed-anion p-type transparent conductors.

# Supporting information

S1. Atomic and molecular orbital diagram details
S2. Band gap vs. lattice constant data

S3. Band offset data
S4. Other wide band gap chalcogenides
S5. Computed wide-gap p-type chalcogenides